\documentclass[aps,superscriptaddress,showpacs,amsmath,amssymb]{revtex4-2}

\usepackage{natbib}
\usepackage{subfigure}
\usepackage{graphicx} 
\usepackage{hyperref}
\usepackage{cancel}

\usepackage[mathlines]{lineno}

\newcommand{\aap}{Astron. Astrophys.}   % Astronomy and Astrophysics
\newcommand{\jcap}{J. Cosmol. Astropart. Phys.}   % Journal of Cosmology and Astroparticle Physics
\newcommand{\mnras}{Mon. Not. R. Astron. Soc.}   % Monthly Notices of the RAS
\graphicspath{{./}{figures/}}

\begin{document}

\title{An indirect search for dark matter with a combined analysis of dwarf spheroidal galaxies from VERITAS}

% \coror{Conor McGrath (\href{mailto:conor.mcgrath2@ucdconnect.ie}{conor.mcgrath2@ucdconnect.ie}), Donggeun Tak (\href{mailto:donggeun.tak@gmail.com}{donggeun.tak@gmail.com})}

\author{A.~Acharyya}\affiliation{CP3-Origins, University of Southern Denmark, Campusvej 55, 5230 Odense M, Denmark}
\author{C.~B.~Adams}\affiliation{Physics Department, Columbia University, New York, NY 10027, USA}
\author{P.~Bangale}\affiliation{Department of Physics and Astronomy and the Bartol Research Institute, University of Delaware, Newark, DE 19716, USA}
\author{J.~T.~Bartkoske}\affiliation{Department of Physics and Astronomy, University of Utah, Salt Lake City, UT 84112, USA}
\author{P.~Batista}\affiliation{DESY, Platanenallee 6, 15738 Zeuthen, Germany}
\author{W.~Benbow}\affiliation{Center for Astrophysics $|$ Harvard \& Smithsonian, Cambridge, MA 02138, USA}
\author{J.~L.~Christiansen}\affiliation{Physics Department, California Polytechnic State University, San Luis Obispo, CA 94307, USA}
\author{A.~J.~Chromey}\affiliation{Center for Astrophysics $|$ Harvard \& Smithsonian, Cambridge, MA 02138, USA}
\author{A.~Duerr}\affiliation{Department of Physics and Astronomy, University of Utah, Salt Lake City, UT 84112, USA}
\author{M.~Errando}\affiliation{Department of Physics, Washington University, St. Louis, MO 63130, USA}
\author{A.~Falcone}\affiliation{Department of Astronomy and Astrophysics, 525 Davey Lab, Pennsylvania State University, University Park, PA 16802, USA}
\author{Q.~Feng}\affiliation{Department of Physics and Astronomy, University of Utah, Salt Lake City, UT 84112, USA}
\author{G.~M.~Foote}\affiliation{Department of Physics and Astronomy and the Bartol Research Institute, University of Delaware, Newark, DE 19716, USA}
\author{L.~Fortson}\affiliation{School of Physics and Astronomy, University of Minnesota, Minneapolis, MN 55455, USA}
\author{A.~Furniss}\affiliation{Department of Physics, California State University - East Bay, Hayward, CA 94542, USA}
\author{W.~Hanlon}\affiliation{Center for Astrophysics $|$ Harvard \& Smithsonian, Cambridge, MA 02138, USA}
\author{D.~Hanna}\affiliation{Physics Department, McGill University, Montreal, QC H3A 2T8, Canada}
\author{O.~Hervet}\affiliation{Santa Cruz Institute for Particle Physics and Department of Physics, University of California, Santa Cruz, CA 95064, USA}
\author{C.~E.~Hinrichs}\affiliation{Center for Astrophysics $|$ Harvard \& Smithsonian, Cambridge, MA 02138, USA and Department of Physics and Astronomy, Dartmouth College, 6127 Wilder Laboratory, Hanover, NH 03755 USA}
\author{J.~Holder}\affiliation{Department of Physics and Astronomy and the Bartol Research Institute, University of Delaware, Newark, DE 19716, USA}
\author{T.~B.~Humensky}\affiliation{Department of Physics, University of Maryland, College Park, MD, USA and NASA GSFC, Greenbelt, MD 20771, USA}
\author{W.~Jin}\affiliation{Department of Physics and Astronomy, University of California, Los Angeles, CA 90095, USA}
\author{M.~N.~Johnson}\affiliation{Santa Cruz Institute for Particle Physics and Department of Physics, University of California, Santa Cruz, CA 95064, USA}
\author{P.~Kaaret}\affiliation{Department of Physics and Astronomy, University of Iowa, Van Allen Hall, Iowa City, IA 52242, USA}
\author{M.~Kertzman}\affiliation{Department of Physics and Astronomy, DePauw University, Greencastle, IN 46135-0037, USA}
\author{D.~Kieda}\affiliation{Department of Physics and Astronomy, University of Utah, Salt Lake City, UT 84112, USA}
\author{T.~K.~Kleiner}\affiliation{DESY, Platanenallee 6, 15738 Zeuthen, Germany}
\author{N.~Korzoun}\affiliation{Department of Physics and Astronomy and the Bartol Research Institute, University of Delaware, Newark, DE 19716, USA}
\author{S.~Kumar}\affiliation{Department of Physics, University of Maryland, College Park, MD, USA }
\author{M.~J.~Lang}\affiliation{School of Natural Sciences, University of Galway, University Road, Galway, H91 TK33, Ireland}
\author{M.~Lundy}\affiliation{Physics Department, McGill University, Montreal, QC H3A 2T8, Canada}
\author{G.~Maier}\affiliation{DESY, Platanenallee 6, 15738 Zeuthen, Germany}
\author{C.~E~McGrath}
\altaffiliation{Conor McGrath, \href{mailto:mcgrathconor94@gmail.com}{mcgrathconor94@gmail.com}}
\affiliation{School of Physics, University College Dublin, Belfield, Dublin 4, Ireland}
\author{M.~J.~Millard}\affiliation{Department of Physics and Astronomy, University of Iowa, Van Allen Hall, Iowa City, IA 52242, USA}
\author{C.~L.~Mooney}\affiliation{Department of Physics and Astronomy and the Bartol Research Institute, University of Delaware, Newark, DE 19716, USA}
\author{P.~Moriarty}\affiliation{School of Natural Sciences, University of Galway, University Road, Galway, H91 TK33, Ireland}
\author{R.~Mukherjee}\affiliation{Department of Physics and Astronomy, Barnard College, Columbia University, NY 10027, USA}
\author{W.~Ning}\affiliation{Department of Physics and Astronomy, University of California, Los Angeles, CA 90095, USA}
\author{S.~O'Brien}\affiliation{Physics Department, McGill University, Montreal, QC H3A 2T8, Canada and Arthur B. McDonald Canadian Astroparticle Physics Research Institute, 64 Bader Lane, Queen's University, Kingston, ON Canada, K7L 3N6}
\author{R.~A.~Ong}\affiliation{Department of Physics and Astronomy, University of California, Los Angeles, CA 90095, USA}
\author{N.~Park}\affiliation{Department of Physics, Engineering Physics and Astronomy, Queen's University, Kingston, ON K7L 3N6, Canada}
\author{M.~Pohl}\affiliation{Institute of Physics and Astronomy, University of Potsdam, 14476 Potsdam-Golm, Germany and DESY, Platanenallee 6, 15738 Zeuthen, Germany}
\author{E.~Pueschel}\affiliation{Fakult\"at f\"ur Physik \& Astronomie, Ruhr-Universit\"at Bochum, D-44780 Bochum, Germany}
\author{J.~Quinn}\affiliation{School of Physics, University College Dublin, Belfield, Dublin 4, Ireland}
\author{P.~L.~Rabinowitz}\affiliation{Department of Physics, Washington University, St. Louis, MO 63130, USA}
\author{K.~Ragan}\affiliation{Physics Department, McGill University, Montreal, QC H3A 2T8, Canada}
\author{P.~T.~Reynolds}\affiliation{Department of Physical Sciences, Munster Technological University, Bishopstown, Cork, T12 P928, Ireland}
\author{D.~Ribeiro}\affiliation{School of Physics and Astronomy, University of Minnesota, Minneapolis, MN 55455, USA}
\author{E.~Roache}\affiliation{Center for Astrophysics $|$ Harvard \& Smithsonian, Cambridge, MA 02138, USA}
\author{J.~L.~Ryan}\affiliation{Department of Physics and Astronomy, University of California, Los Angeles, CA 90095, USA}
\author{I.~Sadeh}\affiliation{DESY, Platanenallee 6, 15738 Zeuthen, Germany}
\author{L.~Saha}\affiliation{Center for Astrophysics $|$ Harvard \& Smithsonian, Cambridge, MA 02138, USA}
\author{G.~H.~Sembroski}\affiliation{Department of Physics and Astronomy, Purdue University, West Lafayette, IN 47907, USA}
\author{R.~Shang}\affiliation{Department of Physics and Astronomy, Barnard College, Columbia University, NY 10027, USA}
\author{M.~Splettstoesser}\affiliation{Santa Cruz Institute for Particle Physics and Department of Physics, University of California, Santa Cruz, CA 95064, USA}
\author{D.~Tak}
\altaffiliation{Donggeun Tak, \href{mailto:donggeun.tak@gmail.com}{donggeun.tak@gmail.com}}
\affiliation{SNU Astronomy Research Center, Seoul National University, Seoul 08826, Republic of Korea.}
\author{A.~K.~Talluri}\affiliation{School of Physics and Astronomy, University of Minnesota, Minneapolis, MN 55455, USA}
\author{J.~V.~Tucci}\affiliation{Department of Physics, Indiana University-Purdue University Indianapolis, Indianapolis, IN 46202, USA}
\author{V.~V.~Vassiliev}\affiliation{Department of Physics and Astronomy, University of California, Los Angeles, CA 90095, USA}
\author{A.~Weinstein}\affiliation{Department of Physics and Astronomy, Iowa State University, Ames, IA 50011, USA}
\author{D.~A.~Williams}\affiliation{Santa Cruz Institute for Particle Physics and Department of Physics, University of California, Santa Cruz, CA 95064, USA}
\author{S.~L.~Wong}\affiliation{Physics Department, McGill University, Montreal, QC H3A 2T8, Canada}

\collaboration{VERITAS Collaboration}

\date{\today}
\begin{abstract}
Understanding the nature and identity of dark matter is a key goal in the physics community. In the case that TeV-scale dark matter particles decay or annihilate into standard model particles, very-high-energy (VHE) gamma rays (greater than 100 GeV) will be present in the final state. The Very Energetic Radiation Imaging Telescope Array System (VERITAS) is an imaging atmospheric Cherenkov telescope array that can indirectly detect VHE gamma rays in an energy range of 100 GeV to $>$ 30 TeV. Dwarf spheroidal galaxies (dSphs) are ideal candidates in the search for dark matter due to their high dark matter content, high mass-to-light ratios, and their low gamma-ray fluxes from astrophysical processes. This study uses a legacy data set of 638 hours collected on 17 dSphs, built over 11 years with an observing strategy optimized according to the dark matter content of the targets. The study addresses a broad dark matter particle mass range, extending from 200 GeV to 30 PeV. In the absence of a detection, we set the upper limits on the dark matter velocity-weighted annihilation cross section.
\end{abstract}

\maketitle

\section{\label{sec:intro} Introduction} 
%\textit{Elisa}
%\textit{General dark matter background, indirect detection, appropriateness of IACTs for indirect detection searches. Mention recent indirect searches from Fermi, IACTs, HAWC, mentioning systems of interest - Galactic center, dSphs, galaxy clusters, sub-halo targets. Include a brief discussion of the motivation for UHDM interpretation.}

Cold dark matter is a key component of the current cosmological picture, comprising $\sim$85\% of the matter of the universe~\cite{Planck:2018vyg}. Indirect astrophysical dark matter searches provide a means of probing the nature of dark matter, complementary to direct detection and collider searches. Assuming that dark matter is made up of particles, and that these particles interact with standard baryonic matter, it is possible to search for annihilation or decay of dark matter particles to Standard Model particles. The focus of this paper is searching for dark matter self-annihilation signatures, assuming self-conjugate dark matter. 

Gamma rays are typically present in the final state following dark matter annihilation, whether as mono-energetic ``line" signatures from promptly produced photons, or as a continuum gamma-ray spectrum with a cut-off at the dark matter particle mass. For the weakly interacting massive particle (WIMP) class of dark matter candidates, the preferred dark matter mass lies in the GeV to TeV range. The annihilation of such particles leads to gamma rays in the sensitive energy range of ground- and space-based gamma-ray instruments. Indirect searches for dark matter annihilation with gamma rays have been conducted and limits set on the dark matter velocity-weighted annihilation cross section with the space telescope $Fermi$-LAT~\cite{dm_fermi, dm_fermiGCline, dm_fermi2}, the imaging atmospheric Cherenkov telescopes (IACTs) H.E.S.S.~\citep{dm_hess, dm_hessGCline, dm_hess2, dm_hessGC}, MAGIC~\citep{dm_magic2, dm_magicGCline} and VERITAS~\citep{dm_veritas}, and the water Cherenkov observatory HAWC~\citep{dm_hawc, dm_hawcGH, dm_hawc2}; the cited publications highlight the most recent results.

The Galactic Center and the Milky Way satellite dwarf spheroidal galaxies (dSphs) are canonical targets for indirect searches with gamma rays from dark matter annihilation. While the Galactic Center is the nearest large repository of dark matter and hence has the largest predicted annihilation signal, dSphs offer a set of targets with low astrophysical backgrounds and modest angular extensions compared to the point-spread functions of the gamma-ray instruments. The dSphs can be divided into classical and ultra-faint objects. The former contain on the order of hundreds of stars, while the latter contain on the order of tens of stars. As stellar motions are used to constrain the dark matter content of these systems, estimates of the dark matter content of classical dSphs tend to be more robust, as their stellar populations are well-measured. This paper analyzes the complete set of observations of dSphs, including classical and ultra-faint targets, collected by VERITAS over an 11 year period, and it utilizes the lack of a detected signal to derive upper limits on the dark matter velocity-weighted annihilation cross section in the standard WIMP mass range below $\sim$100 TeV.

A recent work by \cite{Tak_2022} has highlighted the theoretical motivation for dark matter annihilation searches outside of the standard WIMP mass range. %and the capability of  ground-based gamma-ray instruments to extend their search range above 100 TeV. 
The unitarity limit at $\sim$194 TeV \cite{Griest1990, Smirnov:2019ngs} is violated by point-like thermal-relic dark matter. However, dark matter composed of composite states or non-thermal relics can evade this limit \citep[e.g.,][]{Harigaya2016, Geller2018}. Also, \cite{Tak_2022} showed that the annihilation of ultra-heavy dark matter particles (UHDM) up to 30 PeV can produce a sufficient number of photons below 100 TeV to enable VERITAS to constrain the velocity-averaged cross section for PeV dark matter. This UHDM mass range has been probed with gamma-ray and neutrino searches \cite{Tak_2023, ANTARES2022}, but a broad range of dark matter annihilation cross sections remains unconstrained. Consequently, we also present limits on annihilation of UHDM, with particle masses between 194 TeV and 30 PeV. 

\section{Predicted gamma-ray flux and target selection} \label{sec:targets}

The predicted observable gamma-ray flux from a dSph, $\phi_{s}$, is given by the product of two terms; the ``particle physics factor'' ($\Phi_{PP}$) which is based on theoretical models of the annihilation of dark matter particles producing gamma-ray products, and the ``astrophysical factor'' or ``J-factor'', which is determined by the dark matter content of the region observed:
\begin{equation} \label{eq:dm_flux}
\phi_{s}(\Delta\Omega)=\underbrace{\frac{1}{4\pi}\frac{\left<\sigma v\right>}{2M^2_{\chi}}\int^{E_{max}}_{E_{min}}\frac{dN}{dE}dE}_{\Phi_{PP}}\times\underbrace{\int_{\Delta\Omega}\frac{dJ}{d\Omega}d\Omega}_{\text{J-factor}},
\end{equation}
where $M_{\chi}$ is the dark matter particle mass, $\left< \sigma v \right>$ is the thermally averaged, velocity-weighted annihilation cross-section of the dark matter particles, $dN/dE$ is the final-state photon spectrum, and $dJ/d\Omega$ is the differential annihilation J-factor with respect to solid angle $\Omega$ (Equation~\ref{eq:jfactor}).

The photon spectrum from dark matter annihilation depends on the annihilation channel of the dark matter pair into Standard Model particles. Here we assume a 100\% branching ratio of dark matter particles into the following channels: $e^{+}e^{-}$, $\mu^{+}\mu^{-}$, $\tau^{+}\tau^{-}$, $t\bar{t}$, $b\bar{b}$, $W^{+}W^{-}$, $ZZ$, $\gamma\gamma$, and $\nu_e \bar{\nu}_e$. For each of the annihilation channels, we exploit the spectrum provided by {\tt PPPC4DMID}, which offers trustworthy spectra in the 5 GeV to 100 TeV energy range \citep{Cirelli_2011}, and {\tt HDMSpectra}, which provides reliable spectra from 1 TeV to the Planck energy \citep{Bauer_2021}. The latter includes many of the interactions in the full unbroken Standard Model that are relevant for estimating accurate UHDM annihilation spectra. Thus, we adopt the spectra from {\tt PPPC4DMID} for analyzing masses from 200 GeV to $<$ 100 TeV, and {\tt HDMSpectra} for analyzing masses from 100 TeV to 30 PeV. We stress that the spectra from the two packages are consistent within the mass range from 1 TeV to 100 TeV \citep{Tak_2022}, implying a low systematic uncertainty in the dark matter photon spectra. We also note that another dark matter photon-spectrum calculation method, CosmiXs (energy range: 5 GeV to 100 TeV) \citep{Arina2024}, also shows differences for indirect dark matter searches with IACTs that are smaller than the uncertainties due to the dwarf geometrical factors and the Poissonian fluctuations \citep[e.g.,][]{Circiello2024}.

The differential annihilation J-factor with respect to solid angle is obtained by integrating the dark matter density profile squared ($\rho^2$) along the line-of-sight ($l$) through the dSph for a given observation angle: 

\begin{equation}\label{eq:jfactor}
    \frac{dJ}{d\Omega}=\int_{los}\rho^2(l,\Omega)\,dl
\end{equation}

Using the commonly employed simplifying assumption that the dark matter density profiles are spherically symmetric, we can write the angular J-profile as
a function of the observation angle ($\theta$) from the center of the dSph as:

\begin{equation}
    \frac{dJ}{d\theta}=2\pi\sin\theta\,\frac{dJ}{d\Omega}
\end{equation}

The above J-profile can be integrated over a given angular range to determine the J-factor for a given dSph analysis.

%\begin{equation}
%    J(\Delta\Omega)=\int_{\Delta\Omega}\int_{los}\rho^2(l,\Omega)\,dl\,d\Omega. 
%\end{equation}
%We also utilise the J-profile, $dJ/d\theta$, for which the radial integral in the equation above is not performed.

For the dark matter density profile we utilize the Navarro-Frenk-White (NFW) model: 
\begin{equation}\label{eq:dm_NFW}
\rho(r) = \frac{\rho_s}{(r/r_s)(1+(r/r_s))^2}, \quad \textrm{if $r<r_t$, otherwise $\rho$ = 0,}
\end{equation}
where $\rho_s$ is a characteristic density, and $r_s$ and $r_t$ are a scale radius and a tidal truncation radius, respectively. 

The set of NFW parameters ($\rho_s$, $r_s$, and $r_t$) may be estimated from dynamical analyses of kinematic data of member stars. For this work we make use of the publication by \cite{Ando2020} (Ando+20) and the associated Python package\footnote{\url{https://github.com/shinichiroando/dwarf_params}} that provides the probability density functions (PDFs) for those parameters. With their Bayesian approach, they estimated the posterior distributions of the parameters by adopting either a physically motivated prior on a circular velocity of a satellite in a host subhalo \citep{Graus2019} for the ultra-faint galaxies, or a well-known J-factor prior constrained by the large number of member stars \citep{Pace2019} for the classical dSphs. With the use of physically informative priors, Ando+20 takes into account the current understanding of the role of structure formation on dark matter subhalos in dSphs, such as the extended Press-Schechter formalism combined with tidal effects on subhalo evolution \citep[e.g.,][]{Strigari2007, Bullock2001, Graus2019, Pace2019}. Ando+20 cites the use of informative priors as important for improving the accuracy of the J-factor calculation for ultra-faint dwarfs with scarce kinematic data. As the majority of the targets included in our analysis are ultra-faint, we consider the use of Ando+20 J-factors appropriate. The impact of this choice is elaborated on when assessing the results.

This work presents the analysis of the complete VERITAS observational data set obtained on dSphs and covers the time period of 2007 - 2018. The VERITAS observation strategy evolved with time to focus on deep observations of those objects with the largest predicted J-factors, but to also obtain exposures on additional dSphs, while taking into account uncertainties on the published J-factors and the possibility that J-factors could have been underestimated. VERITAS devoted deep exposures to three dark-matter-dominated dSphs\footnote{We note that Draco II is one of the most dark-matter rich dSphs, but is not included in our deep-exposure sample because, during the data collection phase, Ando+20 had not yet been published.}: Segue 1, Ursa Minor, and Ursa Major II, with more than 100 hours obtained on each target. To supplement these deep observations, survey observations were made on an additional 14 dSphs. 

\begin{table*}[t!]
	\centering 
	\begin{tabular}{c | c c c | c c c | c c}
    \hline\hline
    Dwarf & $\rho_{s, \text{ref}}$ & $r_{s, \text{ref}}$ & $r_{t, \text{ref}}$ & $\rho_s$ & $r_s$ & $r_t$ & log$_{10}J(\theta_{\rm cut})$ & log$_{10}J(0.5^o)$ \\ 
    &[M$_\odot/{\rm pc}^3$] & [pc] & [kpc] & [M$_\odot/{\rm pc}^3$] & [pc] & [kpc] & [GeV$^2$/cm$^5$] & [GeV$^2$/cm$^5$] \\ \hline
    Bo\"otes  &  0.045  &  425  &  3.50  &  7.0$^{+11.2}_{-4.3}\times10^{-2}$  &  3.1$^{+2.9}_{-1.4}\times10^{2}$  & 2.1$^{+4.0}_{-1.2}$  &  $17.67_{-0.25}^{+0.26}$ & $17.77_{-0.24}^{+0.23}$  \\ 
    Coma Berenices  &  0.067  &  397  &  2.15  &  7.9$^{+12.6}_{-4.9}\times10^{-2}$  &  3.5$^{+3.9}_{-1.8}\times10^{2}$  &  2.7$^{+5.4}_{-1.7}$  &  $18.16_{-0.32}^{+0.29}$& $18.37_{-0.33}^{+0.30}$  \\ 
    CVn I  &  0.042  &  702  &  11.92  &  4.7$^{+6.5}_{-2.7}\times10^{-2}$  &  6.4$^{+4.7}_{-2.6}\times10^{2}$  &  6.0$^{+6.9}_{-3.1}$  &  $17.31_{-0.12}^{+0.15}$& $17.38_{-0.11}^{+0.11}$  \\ 
    CVn II  &  0.062  &  381  &  2.28  &  8.4$^{+14.0}_{-5.2}\times10^{-2}$  &  3.1$^{+3.7}_{-1.7}\times10^{2}$  &  2.3$^{+5.1}_{-1.5}$  &  $17.16_{-0.44}^{+0.38}$&$17.19_{-0.47}^{+0.37}$  \\ 
    Draco II  &  0.054  &  741  &  8.19  &  5.9$^{+7.4}_{-3.2}\times10^{-2}$  &  7.0$^{+5.2}_{-3.0}\times10^{2}$  &  7.7$^{+8.5}_{-4.2}$  &  $19.14_{-0.25}^{+0.25}$&$19.49_{-0.25}^{+0.20}$  \\ 
    Hercules I  &  0.055  &  300  &  1.15  &  9.2$^{+15.7}_{-6.0}\times10^{-2}$  &  2.1$^{+2.5}_{-1.1}\times10^{2}$  &  1.2$^{+2.8}_{-0.7}$  &  $16.92_{-0.36}^{+0.31}$&$16.93_{-0.39}^{+0.34}$  \\ 
    Leo I  &  0.046  &  946  &  11.40  &  5.7$^{+6.7}_{-2.9}\times10^{-2}$  &  8.2$^{+5.2}_{-3.3}\times10^{2}$  &  10.0$^{+8.8}_{-4.8}$  &  $17.63_{-0.08}^{+0.09}$&$17.70_{-0.08}^{+0.07}$  \\ 
    Leo II  &  0.044  &  808  &  11.42  &  5.5$^{+7.5}_{-3.1}\times10^{-2}$  &  6.7$^{+5.5}_{-2.9}\times10^{2}$  &  6.8$^{+8.1}_{-3.6}$  &  $17.48_{-0.09}^{+0.09}$&$17.54_{-0.10}^{+0.10}$  \\ 
    Leo IV  &  0.093  &  181  &  0.53  &  1.2$^{+2.3}_{-0.8}\times10^{-1}$  &  1.5$^{+2.4}_{-0.9}\times10^{2}$  &  0.8$^{+2.4}_{-0.6}$  &  $16.57_{-0.64}^{+0.52}$&$16.56_{-0.66}^{+0.57}$  \\ 
    Leo V  &  0.102  &  181  &  0.87  &  1.1$^{+2.3}_{-0.8}\times10^{-1}$  &  1.7$^{+3.0}_{-1.1}\times10^{2}$  &  0.9$^{+3.4}_{-0.7}$  &  $16.57_{-0.69}^{+0.60}$&$16.58_{-0.69}^{+0.60}$  \\ 
    Segue 1  &  0.086  &  344  &  3.18  &  1.1$^{+1.7}_{-0.7}\times10^{-1}$  &  2.9$^{+3.7}_{-1.7}\times10^{2}$  &  2.2$^{+5.2}_{-1.6}$  &  $18.61_{-0.41}^{+0.34}$&$18.91_{-0.48}^{+0.39}$  \\ 
    Segue 2  &  0.186  &  65  &  0.25  &  1.9$^{+3.6}_{-1.3}\times10^{-1}$  &  0.6$^{+1.3}_{-0.4}\times10^{2}$  &  0.2$^{+0.9}_{-0.2}$  &  $17.15_{-0.79}^{+0.48}$&$17.23_{-0.99}^{+0.58}$  \\ 
    Sextans I  &  0.043  &  645  &  11.07  &  5.0$^{+7.0}_{-2.9}\times10^{-2}$  &  5.6$^{+4.6}_{-2.4}\times10^{2}$  &  5.1$^{+6.5}_{-2.9}$  &  $17.87_{-0.28}^{+0.28}$&$18.05_{-0.29}^{+0.25}$  \\ 
    Triangulum II  &  0.100  &  134  &  1.64  &  1.4$^{+3.0}_{-1.0}\times10^{-1}$  &  1.0$^{+2.4}_{-0.7}\times10^{2}$  &  0.5$^{+2.1}_{-0.3}$  &  $17.56_{-0.77}^{+0.60}$&$17.65_{-0.90}^{+0.68}$  \\ 
    Ursa Major I  &  0.056  &  638  &  3.43  &  7.4$^{+11.0}_{-4.4}\times10^{-2}$  &  4.2$^{+4.4}_{-2.2}\times10^{2}$  &  3.6$^{+6.5}_{-2.3}$  &  $18.05_{-0.24}^{+0.21}$&$18.19_{-0.25}^{+0.22}$  \\ 
    Ursa Major II  &  0.065  &  452  &  2.41  &  4.9$^{+6.5}_{-2.7}\times10^{-2}$  &  7.5$^{+5.5}_{-3.2}\times10^{2}$  &  8.1$^{+8.5}_{-4.3}$  &  $18.43_{-0.38}^{+0.35}$&$18.79_{-0.48}^{+0.36}$  \\ 
    Ursa Minor  & 0.049  &  765  &  9.48 &  6.0$^{+8.0}_{-3.4}\times10^{-2}$  &6.0$^{+5.1}_{-2.7}\times10^{2}$&5.9$^{+7.7}_{-3.4}$&$18.22_{-0.24}^{+0.25}$&$18.47_{-0.22}^{+0.20}$ \\
    \hline\hline
    \end{tabular}
    \caption{Table of the NFW parameters of the 17 dSphs used in this work. The first three columns show the reference NFW profile parameters, which approximate the median of the distribution of viable J-profiles \citep[described in Section~\ref{sec:targets}; adopted from][]{Ando2020}. The next three columns give the medians with 68\% confidence intervals of the NFW profile parameters. The last two columns show the logarithm of the median J-factors with 68\% confidence intervals integrated to $\theta_{\rm cut}$ and 0.5$^\circ$ from the centers of the galaxies, respectively. Each dSph was analyzed with an appropriate value of $\theta_{\rm cut}$: 0.089, 0.110, or 0.141 degrees (see Section~\ref{sec:obs} and Table~\ref{tab:dSphs}).} 
    \label{tab:j_pars}
\end{table*}

In Table~\ref{tab:j_pars} we present the reference NFW parameters $\rho_s$, $r_s$, and $r_t$, as well as the median and 68\% confidence interval for the parameters, as provided by Ando+20. The median and interval are calculated from a random sample of 1,000 viable parameter sets for each dSph. The parameter sets constitute the 1,000 randomly-drawn profiles, providing the value and errors of the J-factor integrated out to an observation angle of 0.5$^\circ$, as well as $\theta_{\rm cut}$ (see Section~\ref{sec:obs}). Since a strong correlation exists among three NFW parameters \citep[$\rho_s$, $r_s$, and $r_t$; see][]{Song2024}, the use of the median parameters does not guarantee the median value of the J-profile. To provide a reference J-profile for each dSph, we select the individual randomly-drawn profile that most closely matches the median J-profile calculated from 1,000 randomly drawn profiles for each object. 

%To obtain a feasible parameter set for each dSph, we make use of the work by \cite{Ando2020} (Ando+20) \footnote{\url{https://github.com/shinichiroando/dwarf_params}}, which provides the probability density functions (PDFs) for those parameters. With the Bayesian approach, they estimated the posterior distributions of such parameters by adopting either a physically motivated prior on a circular velocity of a satellite in a host subhalo \citep{Graus2019} or a well-known J-factor prior constrained by the large number of member stars \citep{Pace2019}. The former prior is applied to the ultra-faint dSphs, and the latter is the classical dSphs\footnote{In our study, Leo I, Leo II, Sextans, and Ursa Minor are a classical dSph, and the others are classified as a ultra-faint dSph}. From the PDFs, we can randomly sample the set of the J profile parameters.

\section{VERITAS Instrument, Observations and Analysis} \label{sec:obs}
\subsection{Instrument}
The Very Energetic Radiation Imaging Telescope Array System (VERITAS) is a four 12~m-diameter IACT array, located at the Fred Lawrence Whipple Observatory in southern Arizona. It has been in full operation since 2007. VERITAS indirectly detects gamma rays in the energy range 85 GeV to $>30$ TeV through imaging the fast (ns) flashes of Cherenkov light produced by extensive air showers in the atmosphere.
Each of the four VERITAS telescopes has a reflector comprised of 350 hexagonal mirrors, mounted on a Davies-Cotton optical support structure, which images the Cherenkov light onto a 499-pixel photomultiplier-tube (PMT) camera mounted in the focal plane~\citep{VERITASInstrument}. Stereoscopic analysis of the images allows showers initiated by gamma rays to be preferentially selected over showers initiated by hadronic cosmic ray particles and the energies and directions of the primary gamma rays to be estimated. 

The performance of VERITAS varies with energy; for 1~TeV photons VERITAS has an energy resolution of 17$\%$, an
angular resolution of 0.08 deg (68\% containment radius) and an effective gamma-ray collection area on the order of 10$^5$ m$^2$.  The VERITAS source location accuracy is $<$50 arcseconds. The sensitivity of VERITAS  allows a statistically significant detection (5 standard deviations, 5$\sigma$, above background) of a point source with flux equivalent to 1$\%$ that of the Crab Nebula in $\sim$25 hours~\citep{Park_2015}. 

VERITAS has undergone two major configuration changes since it began full four-telescope operations in 2007; in 2009 Telescope 1 was moved to a new location in order to improve sensitivity, while in 2012 the cameras were upgraded with higher quantum-efficiency PMTs and a revised trigger system, resulting in an improved low-energy response~\citep{Kieda2013}. In addition to the upgrades discussed above, the VERITAS instrument response has changed with time due to mirror reflectivity changes and PMT gain changes. We perform regular calibration measurements \citep{NievasRosillo_2021} that are used to produce time-dependent instrument response functions (IRFs), which are used in this study since the data were taken over a time period of roughly 11 years (2007-2018).

%\subsection{Observations} \label{sec:obs}
%Since the commencement of operations, VERITAS has amassed over 600 hours of quality-selected data on 17 dSphs. Of these data, $\sim$220 hours was previously analysed and published \citep{dm_veritas}. In this work, we include both the previously published data as well as including all remaining data taken to date. A breakdown of the exposure time per dSph is shown in Table~\ref{tab:dSphs}. 

\subsection{Observations}

All observations were made in the ``wobble" observing mode, whereby the target being observed is offset in the field of view by 0.5$^\circ$ to allow for simultaneous measurement of background at the same radial distance as the target from the center of the field of view \citep{Fomin1994}. Observing runs are typically 20-30 minutes in duration, and the direction of the offset is cycled through the four cardinal directions to reduce systematic effects.

\begin{table*}[t!]
	\centering 
	\begin{tabular}{c | c c | c c | c | c }
    \hline\hline
    Dwarf & $N_{ON}$ & $N_{OFF}$ & Exposure & $\theta_{\rm cut}$ & Significance & $\Phi^{95\%}$ \\ 
    &[counts] & [\textbf{total} counts] & [hours] & [deg] & [$\sigma$]&[10$^{-12}$cm$^{-2}$s$^{-1}$] \\ \hline
    Bo\"otes  &  446  &  2569  &  13.98  &  0.141  &  0.8  &  0.91  \\ 
    Coma Berenices  &  1122  &  6770  &  39.76  &  0.110  &  -0.2  &  0.66  \\ 
    CVn I  &  411  &  2430  &  9.72  &  0.141  &  0.3  &  0.98  \\ 
    CVn II  &  335  &  1822  &  8.14  &  0.141  &  1.6  &  1.69  \\ 
    Draco II  &  223  &  1335  &  8.02  &  0.141  &  0.0  &  1.62  \\ 
    Hercules I  &  369  &  2187  &  9.46  &  0.141  &  0.2  &  1.18  \\ 
    Leo I  &  196  &  1182  &  5.66  &  0.141  &  -0.1  &  1.98  \\ 
    Leo II  &  550  &  3275  &  11.31  &  0.141  &  0.2  &  0.97  \\ 
    Leo IV  &  7  &  65  &  0.48  &  0.141  &  -1.2  &  2.58  \\ 
    Leo V  &  33  &  218  &  1.38  &  0.141  &  -0.5  &  1.55  \\ 
    Segue 1  &  3070  &  18336  &  126.29  &  0.110  &  0.2  &  0.30  \\ 
    Segue 2  &  487  &  3000  &  12.51  &  0.110  &  -0.5  &  0.80  \\ 
    Sextans I  &  213  &  1262  &  7.45  &  0.141  &  0.2  &  0.95  \\ 
    Triangulum II  &  751  &  4870  &  29.51  &  0.110  &  -2.0  &  0.36  \\ 
    Ursa Major I  &  358  &  2073  &  6.63  &  0.141  &  0.6  &  0.99  \\ 
    Ursa Major II  &  2266  &  13855  &  212.32  &  0.089  &  -0.8  &  0.20  \\ 
    Ursa Minor  &  2253 & 13608 & 135.3 & 0.110  &  -0.3 & 0.28 \\ 
    \hline\hline
    \end{tabular}
    \caption{Table of the VERITAS observations of the 17 dwarf spheroidal galaxies. The first and second columns give the counts of gamma-like events in the ON and OFF regions, respectively. The next two columns show the exposure times and radial cuts defining the ON-region, respectively. The detection significance is given in the following column in terms of standard deviations above the background. The final column gives the 95\% confidence level upper limit on the flux above 300 GeV, assuming a spectral index of -2.4. In all 17 cases, the background normalisation value $\alpha$ is 0.167.}
    \label{tab:dSphs}
\end{table*}

Overall, VERITAS obtained a total quality-selected exposure of 638 hours on the 17 dSphs. The targets and exposures are shown in Table~\ref{tab:dSphs}. Note that four dSphs---namely Leo I, Leo II, Sextans, and Ursa Minor---are classical dSphs, and the rest are classified as ultra-faint dSphs. This study uses approximately 3--4 times more data than in the previous VERITAS study \citep[total 230 hours on four dSphs;][]{dm_veritas}, and the previous data set is included in this work.

\subsection{Analysis}

VERITAS scientific results are produced and validated with two independent software packages: {\tt VEGAS} \citep{VEGAS} and {\tt EventDisplay} \citep{eventdisplay}. Since the previous VERITAS dark matter publication, both packages have been improved by the addition of updated analysis techniques; specifically, boosted decision trees \citep[BDTs;][]{BDTs} in {\tt EventDisplay} and an image template model \citep[ITM;][]{ITM} in {\tt VEGAS}. Studies have shown a $\sim$20$\%$ - 25$\%$ increase in sensitivity from the addition of BDTs and a $\sim$30$\%$ increase in sensitivity from the addition of ITM analysis. In this work, we present the results from the {\tt EventDisplay} package, although they are verified using {\tt VEGAS} and an independent analysis pipeline for performing the dark matter likelihood analysis.

The gamma-ray-selection/hadron-rejection parameters chosen for this analysis are optimized for a moderate energy threshold, which differs from the previous VERITAS dark matter search where the parameters used were optimized for the lowest possible energy threshold. The decision to raise the energy threshold for this analysis is made in order to focus the dark matter search on the energy range where VERITAS is most sensitive, while avoiding systematic effects associated with deep exposures that affect data with less restrictive gamma/hadron selection. 

While VERITAS has the capability to detect events with energies as low as 85 GeV, we apply a low-energy threshold to avoid events with poorly reconstructed energy and angular information. The energy threshold for each run is determined by taking into account the VERITAS IRFs.\footnote{Specifically, a run's low-energy threshold is established as the higher of two values: the energy corresponding to 15\% of the maximum point in the effective area curve or the energy at which the deviation between the true energy and the reconstructed energy is less than 20\%.} This threshold typically ranges from 200 GeV to 300 GeV, depending on the observing conditions. From a pilot study comparing results with and without applying the low-energy threshold, we found that considering a threshold effectively mitigates systematic effects without compromising sensitivity at energies where VERITAS is most effective in the dark matter search.

The data were analyzed to select candidate gamma-ray events, and a reflected-region background estimation was performed \citep{Berge_2007} to search for an excess of events from the direction of each target. The ON-source region was defined to be a circular region centered on the target.  
Multiple OFF regions are defined as circular regions of identical size to the ON region again offset by 0.5$^ \circ$ from the center of the field-of-view. Note that for {\tt EventDisplay}, the number of OFF regions is set to six for each run, giving the relative exposure time between the ON and OFF regions to be 1/6 ($\alpha \simeq 0.167$) for all dSph observations.

The radius of the ON-source region is normally chosen be comparable to the VERITAS point-spread-function (PSF) for a point source, or larger for a spatially extended target.  All of the dSph objects in this study have angular extensions greater than the VERITAS PSF. However, when one expands the ON-source region size, there are fewer available regions for background estimation. Furthermore, for large data sets the analysis is more susceptible to systematic effects such as gradients in the number of events recorded across the cameras due to the varying zenith angle across the field of view. This effect is exacerbated for a larger ON-source region. The expected field-of-view significance distributions for an empty field is a Gaussian distribution of unit width centered at zero. Gradients across the camera can dramatically broaden the distribution, resulting in an unreliable assessment of the signal significance at the target position. 

Prior to this analysis being conducted, a study was performed to optimize the radial cut used to define the ON-source region for dSphs with deep and shallow exposures to maximize the J-factor enclosed while preserving an acceptable field-of-view significance distribution on a control data set. Due to the large number of targets, all with differing exposures, we considered three possible exposure times, corresponding to targets with 16 hours of exposure, 16-150 hours of exposure, and more than 150 hours of exposure. We iteratively increased the radial cut defining the ON-source region, and selected the maximum radial cut for which no systematic effects in the field-of-view significance distributions were visible on a test data set.

Based on the results of the study on the control sample, we use radial cuts to define the ON and OFF regions of 0.089$^ \circ$ ($\theta^2 = 0.008~\text{deg}^2$) for the deepest exposure target (Ursa Major II), 0.110$^ \circ$ ($\theta^2 = 0.012~\text{deg}^2$) for the remaining targets with more than 16 hours of exposure, and 0.141$^ \circ$ ($\theta^2 = 0.02~\text{deg}^2$) for targets with less than 16 hours of exposure (see Table~\ref{tab:dSphs}). Note that the data for Segue 2 were analyzed with a radial cut appropriate for a longer exposure, as its raw exposure of more than 16 hours becomes less than 16 hours after application of time cuts to remove periods with weather or hardware issues.

\section{Maximum Likelihood Estimation}\label{sec:mle}
%\textit{Conor}
%Give formula for the maximum likelihood method. Probably should give some explanation of why we don't use the event-weighting method of the previous analysis.\\
\subsection{Predicted signal}

For a given VERITAS analysis we can calculate the number of gamma-ray events expected ($\mathcal{S}$) for a given data set by taking the predicted flux (Equation \ref{eq:dm_flux}) and folding it with the VERITAS IRFs: 

\begin{equation}
\mathcal{S}=\frac{\left<\sigma v \right>T_{obs}}{8\pi M^2_{\chi}} \int_0^{\infty} \int_0^{\infty} \frac{dN}{dE'}J(E')A(E')D(E|E')dE'dE,
\label{equation:expectedEvnts}
\end{equation}
where $T_{obs}$ is the total exposure time, $J(E')$ is the J-profile convolved with the VERITAS 
energy-dependent PSF and integrated to the appropriate value of $\theta$, $A(E')$ is the VERITAS effective area, 
and $D(E|E')$ is the energy migration matrix, in which $E$ is the reconstructed photon energy and $E'$ is the true 
photon energy. Time-averaged IRFs are used in this analysis for each dSph. Using a maximum likelihood 
analysis we can test for a significant excess of gamma rays and 
place constraints on the value of $\left<\sigma v \right>$ as a function of dark matter particle mass.

%We perform the above calculation for a set of nine final-state photon spectra, assuming a 100\% branching ratio of dark matter particles in nine different annihilation channels: $e^{+}e^{-}$, $\mu^{+}\mu^{-}$, $\tau^{+}\tau^{-}$, $t\bar{t}$, $b\bar{b}$, $W^{+}W^{-}$, $ZZ$, $\gamma\gamma$, and $\nu_e \bar{\nu}_e$.

\subsection{Likelihood}

The use of the maximum likelihood estimation (MLE) method has been shown to maximize the sensitivity to gamma-rays produced from dark matter interactions \citep[e.g.][]{Aleksi__2012}. We adopt the MLE method as described in \cite{Aleksi__2014}. Source analysis with an IACT such as VERITAS is performed by comparing the observed number of ON-region (source region) events with the observed number of OFF-region (background region) events. As the ON and OFF counts are Poisson-distributed, we construct the full likelihood function by multiplying two Poisson likelihood functions with the model-dependent likelihood function $P_i(E_i|M_\chi,\langle \sigma v \rangle)$,

 \begin{equation}
     \mathcal{L}=\frac{(\mathcal{S}+\alpha \mathcal{B})^{N_{on}} e^{-(\mathcal{S}+\alpha \mathcal{B})}} {N_{on}!} \frac{\mathcal{B}^{N_{off}} e^{-\mathcal{B}}}{N_{off}!} \prod^{N_{on}}_{i=1} P_i(E_i|M_\chi,\langle \sigma v \rangle),
 \end{equation}
where $\alpha$ is a background normalisation factor that accounts for the ratio of the number of ON regions to OFF regions, $\mathcal{B}$ is a nuisance parameter that describes the total expected number of background counts from multiple OFF regions, and $\mathcal{S}$ is the expected number of events from dark matter annihilation at a given dark matter mass and velocity-weighted cross section within the ON region (Equation~\ref{equation:expectedEvnts}). Finally, $P_i(E_i|M_\chi,\langle \sigma v \rangle)$ is the predicted energy-dependent counts distribution in the ON region, where these counts can be from the dark matter self-annihilation or cosmic-ray background. The distribution is given by

\begin{equation}
    P_i(E_i|M_{\chi},\left<\sigma v\right>)=\frac{\mathcal{S} p_{s}(E_i)+\alpha \mathcal{B} p_{b}(E_i)}{\mathcal{S}+\alpha \mathcal{B}},
\end{equation}
where $p_{s}$ is from the normalised signal spectrum at a given energy, and $p_{b}$ is generated from a normalised histogram of the energies of all background events. In other words, they are the probability density functions for the dark matter signal and the background, respectively. %\textbf{We note that, at both low and high energies, the values of $p_b$ and thus the limits are sensitive to the binning of the background histograms due to low-count statistics. We choose an intermediate value for the number of bins used.} 

We can then maximize the log likelihood function, neglecting constant terms,
 \begin{equation}
     \log{\mathcal{L}(\langle\sigma v\rangle; \mathbf{\mathcal{B}})} = N_{off} \log{\mathcal{B}}-\mathcal{S}-(1+\alpha)\mathcal{B}+\sum_{i=1}^{N_{on}}{\log{(\alpha \mathcal{B}\,p_{b}(E_i) + \mathcal{S}\,p_{s}(E_i))}},
 \end{equation}
 with respect to $\left<\sigma v\right>$ and $\mathcal{B}$, giving us a constraint on the dark matter velocity-weighted annihilation cross-section.

For the joint-fit MLE analysis, we combine data from the 17 dSphs to maximize statistical power using the joint likelihood function
\begin{equation}
    \log{\mathcal{L}_{\rm joint}(\langle\sigma v\rangle; \mathbf{\mathcal{B}})} = \sum_{i = 1}^{N_{\text{target}}} \log{\mathcal{L}(\langle\sigma v\rangle; \mathcal{B}_i)}.
\end{equation}

To determine the significance of the dark matter signal over the background, we compare two likelihoods (null and alternative hypotheses) using the following equation, 
\begin{equation}\label{eq:ts}
    \lambda = -2 \log \left(\frac{\mathcal{L}_{\mathcal{S} \equiv 0}}{\mathcal{L}_{\mathcal{S} \neq 0}} \right)\!.
\end{equation}

When the resulting value of $\lambda$ is below the detection threshold of 25 (equivalent to $5\sigma$ detection), we calculate an upper limit (UL) on the velocity-weighted annihilation cross section by utilizing the likelihood profile.
The UL is obtained by computing $\Delta \log\mathcal{L}$ of 1.35 compared to the likelihood maximum, corresponding to the one-sided 95\% confidence level.

\begin{figure*}[t!]
    \centering
    \includegraphics[width=\textwidth]{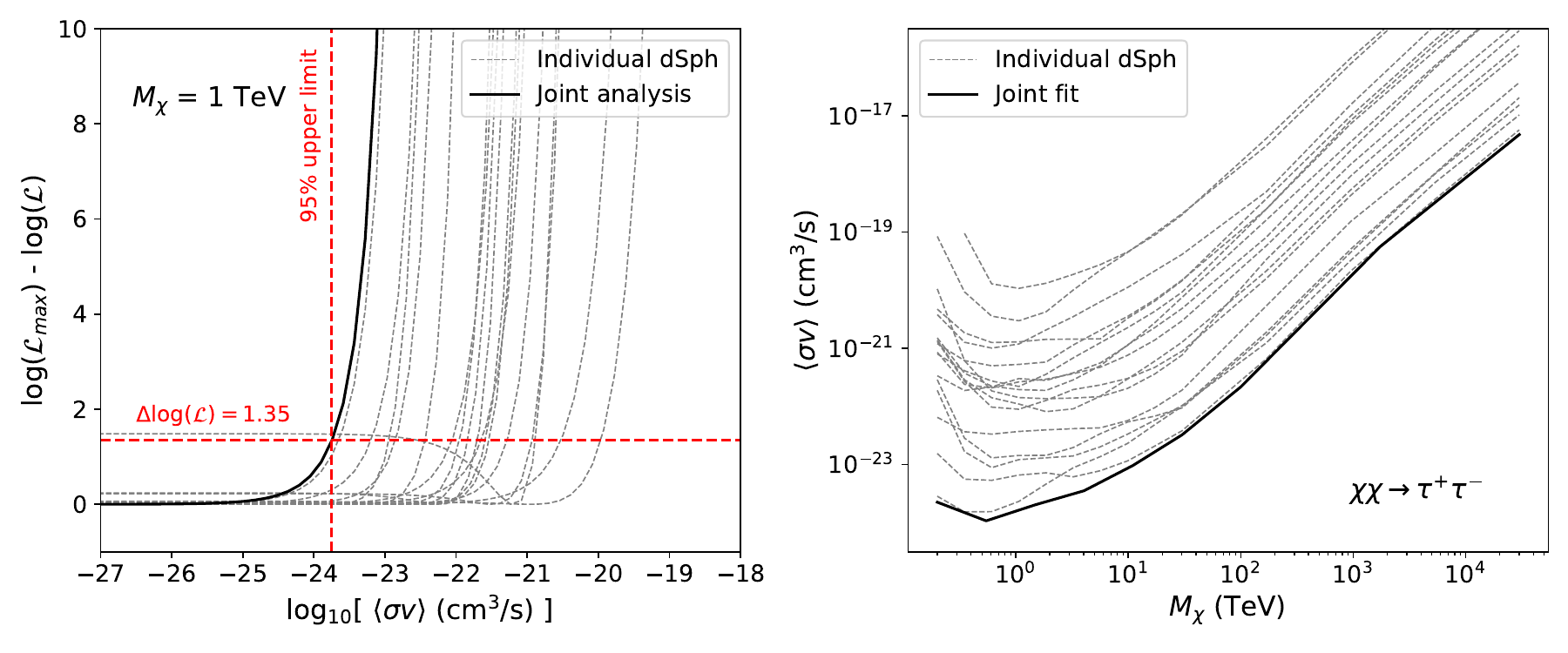}
    \caption{Profile likelihood for a joint-fit analysis with 17 dSphs for a dark matter particle mass of 1 TeV (left) and the VERITAS 95\% upper limit on the dark matter cross section (right), considering the $\tau^+\tau^-$ annihilation channel. The solid black line shows the joint-fit analysis result, and the dashed gray lines are from the individual dSph analyses.}
    \label{fig:method}
\end{figure*}

\section{Results} \label{sec:result}
\subsection{Non-detection with Li \& Ma analysis}
The significance of a signal above background in the ON region is estimated based on the Li \& Ma method~\citep{LiMa}, using the counts in the ON and OFF regions and the ratio of areas between the ON and OFF regions ($\alpha$=0.167 for this analysis). A summary of the counts, the detection significance, and the 95\% confidence level upper limits on the flux is shown in Table~\ref{tab:dSphs}. No dSph shows a significant signal. 

\subsection{Non-detection with maximum likelihood estimation}
We test nine annihilation channels ($e^{+}e^{-}$, $\mu^{+}\mu^{-}$, $\tau^{+}\tau^{-}$, $t\bar{t}$, $b\bar{b}$, $W^{+}W^{-}$, $ZZ$, $\gamma\gamma$, and $\nu_e \bar{\nu}_e$) using the MLE analysis and find that there is no evidence of dark matter annihilation signals from the 17 dSph observations; i.e., the flux and energy spectrum of observed events from the source region is consistent with that of the background fluctuations. The following sections describe our constraint on the velocity-weighted annihilation cross section of dark matter in various aspects. Note that in the case of the $\nu_e \bar{\nu}_e$ annihilation channel, production of final-state gamma rays is enabled by radiation of a $Z$ boson by an off-shell neutrino at sufficiently high energies, or decay of an off-shell neutrino to a $W$ boson and an electron or positron.

\subsection{Upper limits on the dark matter velocity-weighted annihilation cross section}\label{sec:ul_w_sys}

\begin{figure*}[t!]
    \centering
    \subfigure{\includegraphics[width=0.329\textwidth]{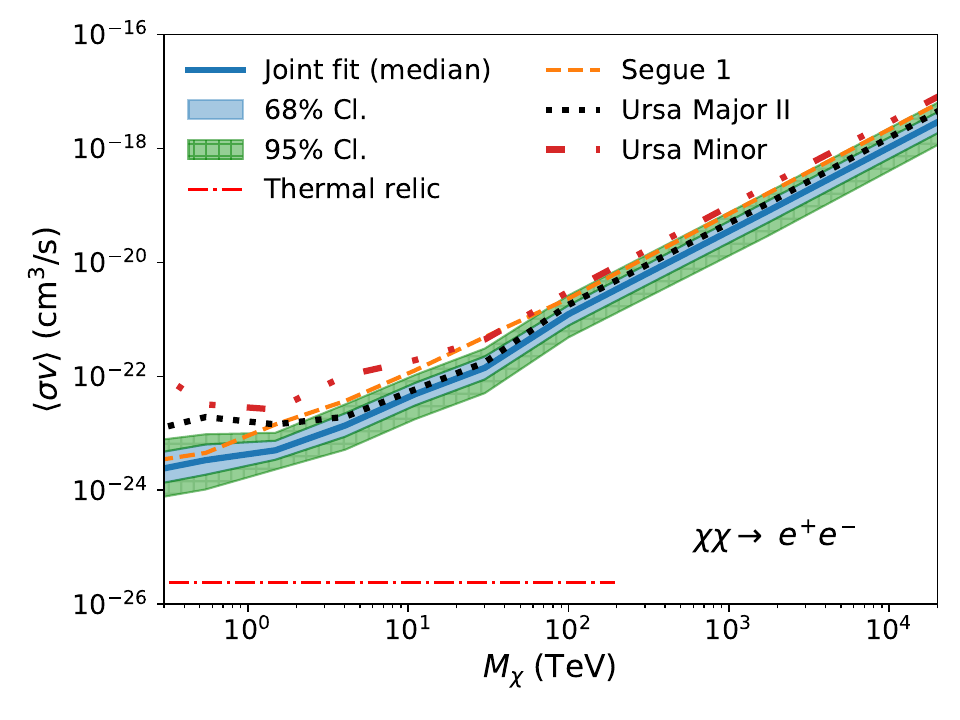}}
    \subfigure{\includegraphics[width=0.329\textwidth]{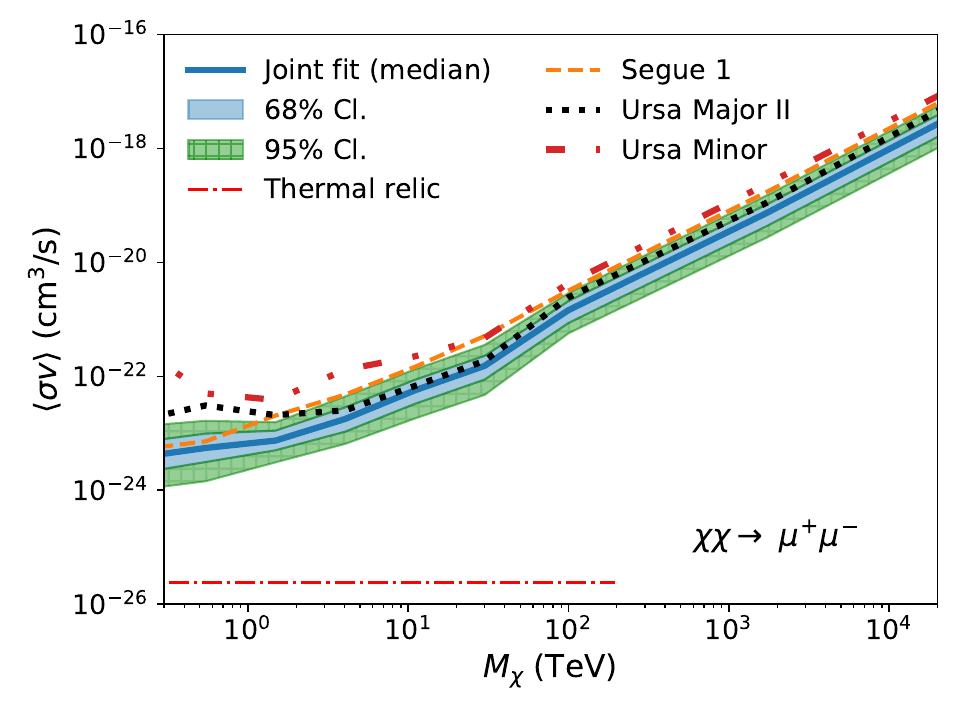}}
    \subfigure{\includegraphics[width=0.329\textwidth]{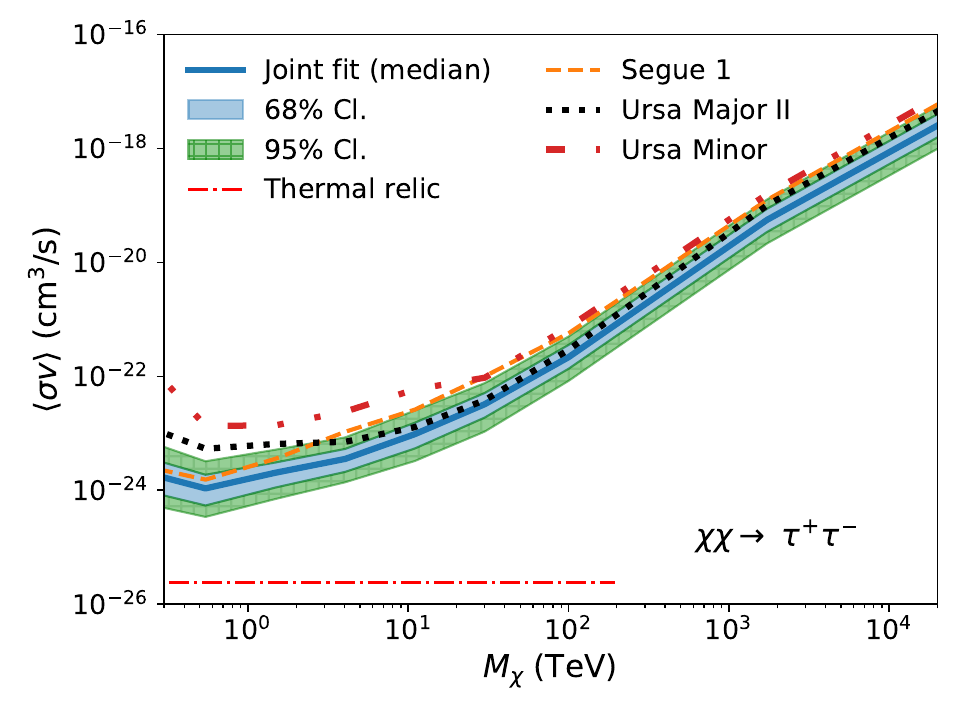}}\\
    \subfigure{\includegraphics[width=0.329\textwidth]{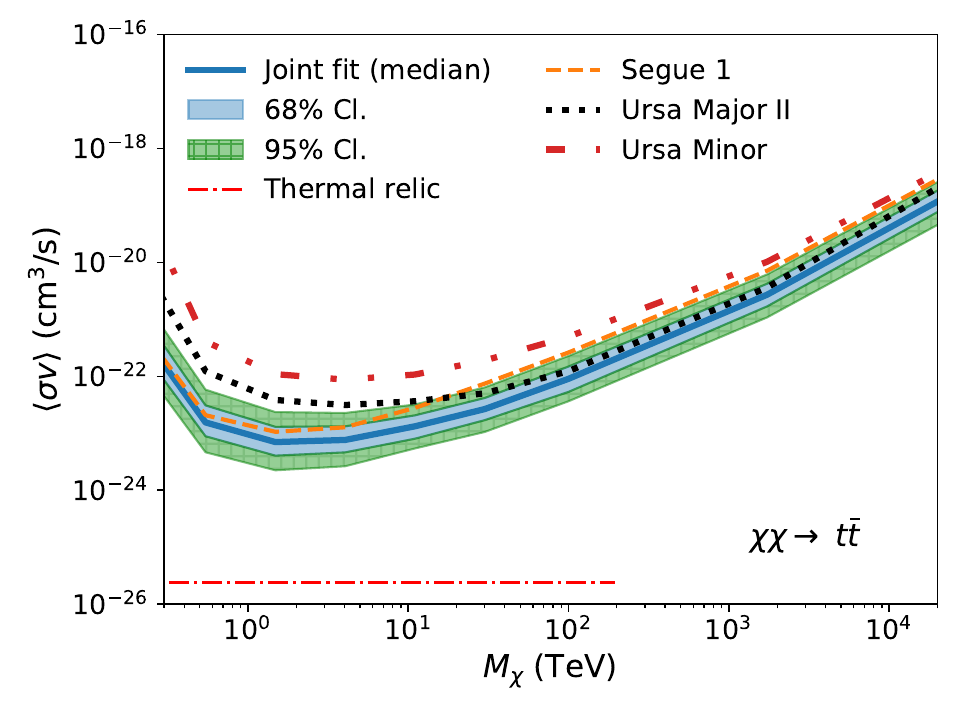}}
    \subfigure{\includegraphics[width=0.329\textwidth]{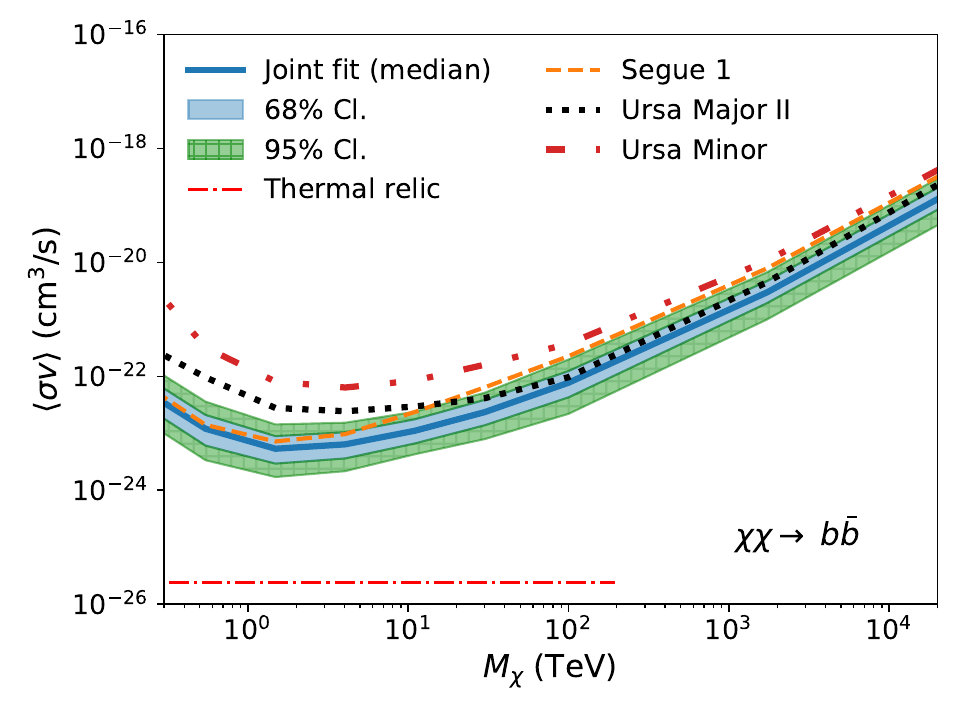}}
    \subfigure{\includegraphics[width=0.329\textwidth]{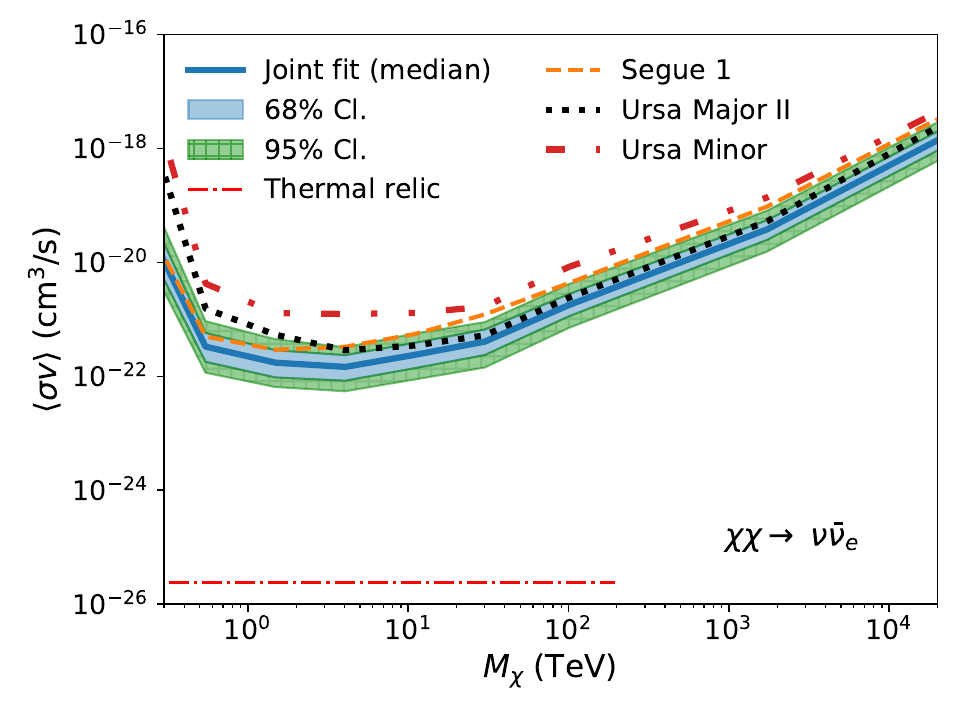}}\\
    \subfigure{\includegraphics[width=0.329\textwidth]{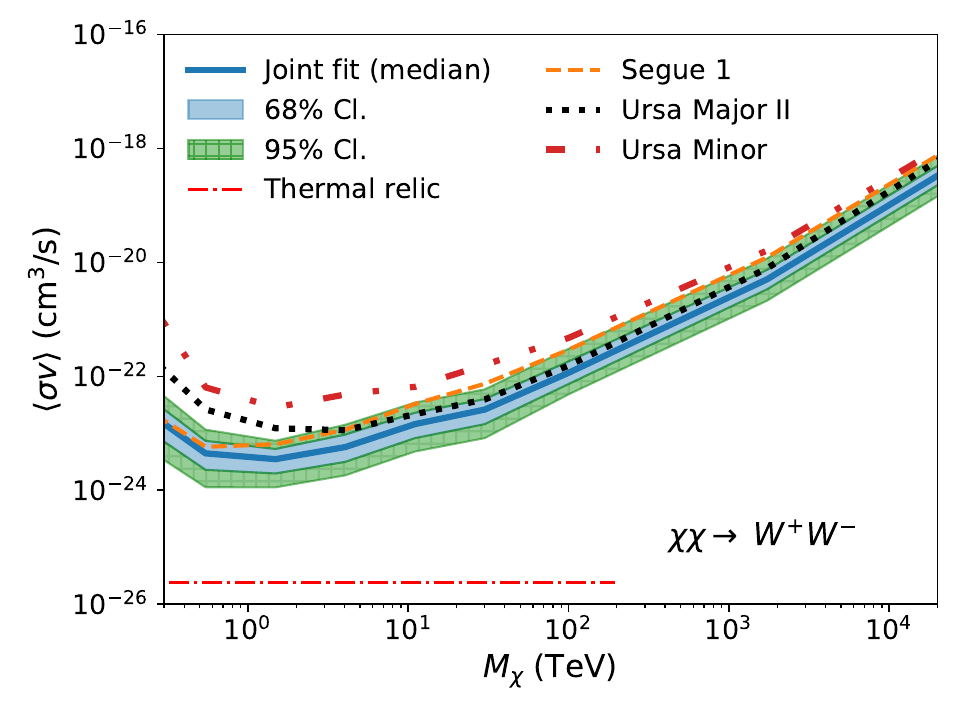}}
    \subfigure{\includegraphics[width=0.329\textwidth]{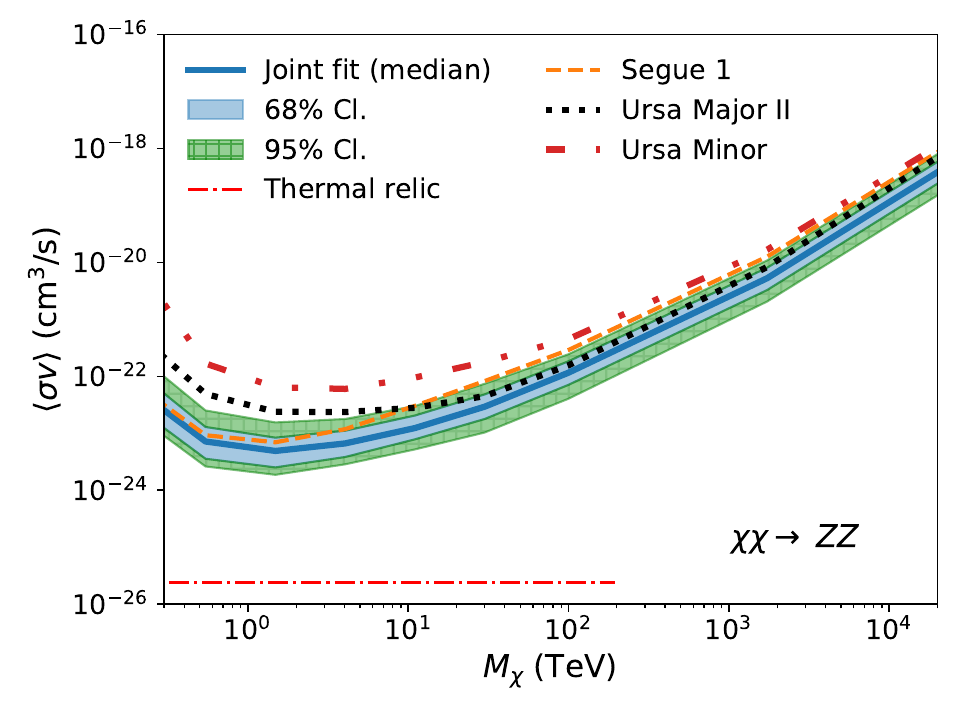}}
    \subfigure{\includegraphics[width=0.329\textwidth]{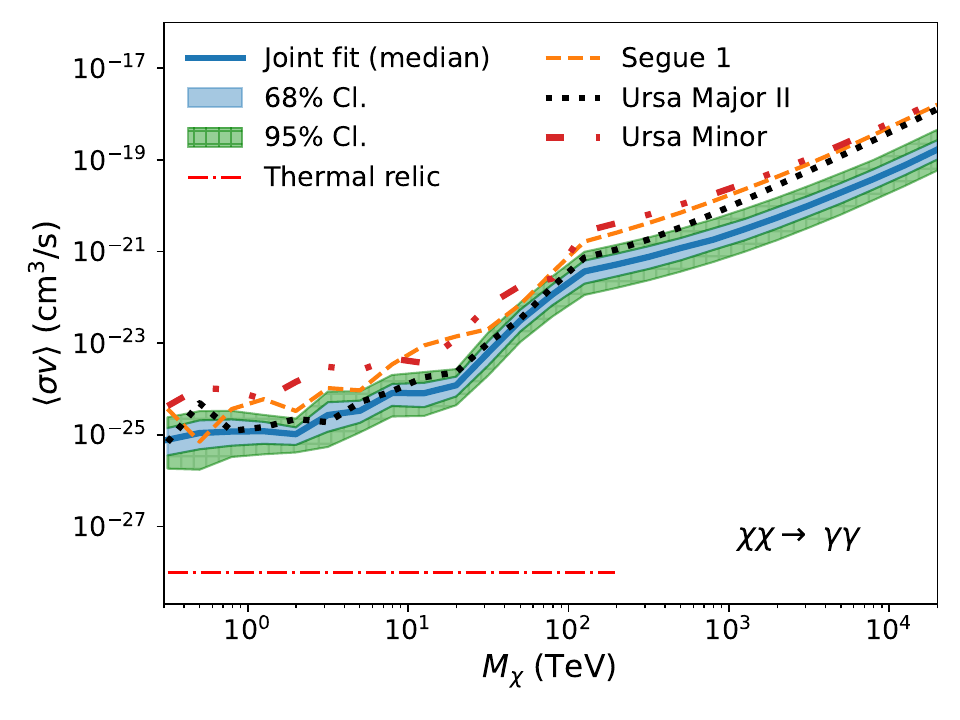}}
    \caption{Velocity-weighted annihilation cross section upper limits produced from 17 dSphs observations, by annihilation channel. The blue (green hatched) shaded uncertainty band depicts the 68\% (95\%) containment obtained from 300 realizations of viable dark matter density profiles. The blue solid curve denotes the median of the band. The orange, black, and red lines represent the upper limits derived from individual observations of Segue 1, Ursa Major II, and Ursa Minor, respectively. These limits result from a reference J-profile for each dSph (Table~\ref{tab:j_pars}). The red dotted-dashed line is the expected velocity-weighted annihilation cross section for a thermal-relic dark matter scenario.}
    \label{fig:ul_sys}
\end{figure*}

Fig.~\ref{fig:method} describes the method of obtaining the dark matter annihilation cross section from the joint-fit MLE analysis and the results. The left panel shows an example of the profile likelihood from individual MLE analyses (dotted or dashed lines) and the joint-fit MLE analysis (solid black line) for the dark matter mass of 1 TeV. In this case, the joint-fit profile likelihood has $\Delta \log(\mathcal{L})$ of 1.35 at $\langle\sigma v\rangle = 1.76\times 10^{-24}$, corresponding to the one-sided 95\% confidence UL. This process is repeated for other dark matter masses so that we can get the UL curve as seen in the right panel of Fig.~\ref{fig:method}. Generally, the joint-fit result provides a stronger constraint on the velocity-weighted dark matter annihilation cross section compared to those from the individual fits. 

Since the J-factors have significant uncertainties, we compute for every dSph a set of ULs for each annihilation channel using 300 possible J-factors. Each J-factor is randomly generated by drawing the three NFW parameters from the posterior distributions provided by Ando+20 (see Table~\ref{tab:dSphs}) and integrating over the line of sight and solid angle. From the 300 realizations, we obtain the median as well as 68\% and 95\% containments for ULs on the dark matter annihilation cross section, which reflects the systematic uncertainty due to the limited understanding of the dark matter distribution.

Fig.~\ref{fig:ul_sys} shows the UL band from the joint-fit analysis, overplotted with UL curves from the three dSphs with the deepest exposures (Segue 1, Ursa Major II, and Ursa Minor) as a reference (see also Appendix~\ref{app:deepexp}). The red dotted-dashed line corresponds to the theoretical expectation of the velocity-weighted annihilation cross section of thermal-relic dark matter, $\langle\sigma v \rangle \sim 2.4\times10^{-26} $cm$^3$/s, extending to the unitarity bound at $\sim$100 TeV \citep{Smirnov:2019ngs}, except for the $\gamma\gamma$ channel. For the loop-suppressed $\gamma\gamma$ channel, we use $\langle\sigma v \rangle \sim 1\times10^{-28} $cm$^3$/s \citep{Bergstrom1994}. Note that for the individual dSph analyses, we take the median of the J-factor distribution and compute a single UL. In general, Segue 1 primarily influences the joint-fit result in the low mass range (up to about 10 TeV), while the joint-fit result in the high mass range is predominantly determined by the observations of Ursa Major II. Ursa Minor and other dSphs have minimal impact on the joint-fit result. %\textbf{We note that, in the case of Segue 1, the uncertainty in choosing member stars has a substantial impact on the dark matter distribution, which can lead to an additional systematic uncertainty of up to two orders of magnitude in the J-factor \citep{Bonnivard2016}. Since Segue 1 plays a crucial role in determining the upper limits in the low mass range, excluding it from the joint-fit analysis significantly impacts the upper limits in that mass range, typically by a factor of 2 to 6, depending on the annihilation channel and mass.}

The discontinuity observed in the limit for $M_\chi$ at  $\sim100$~TeV for the $\gamma\gamma$ channel is expected. We do not analyze photon events with energies above $100$~TeV, as they are beyond the energy range in which events can be reliably reconstructed. Hence, for $M_\chi > 100$~TeV we only consider the continuum $\gamma\gamma$ signal that produces $\lesssim$100 TeV gamma rays, rather than the more easily detectable line signature located above 100 TeV.

\subsection{Comparison with the null hypothesis}\label{subsec:null}

\begin{figure*}[t!]
    \centering
    \subfigure{\includegraphics[width=0.329\linewidth]{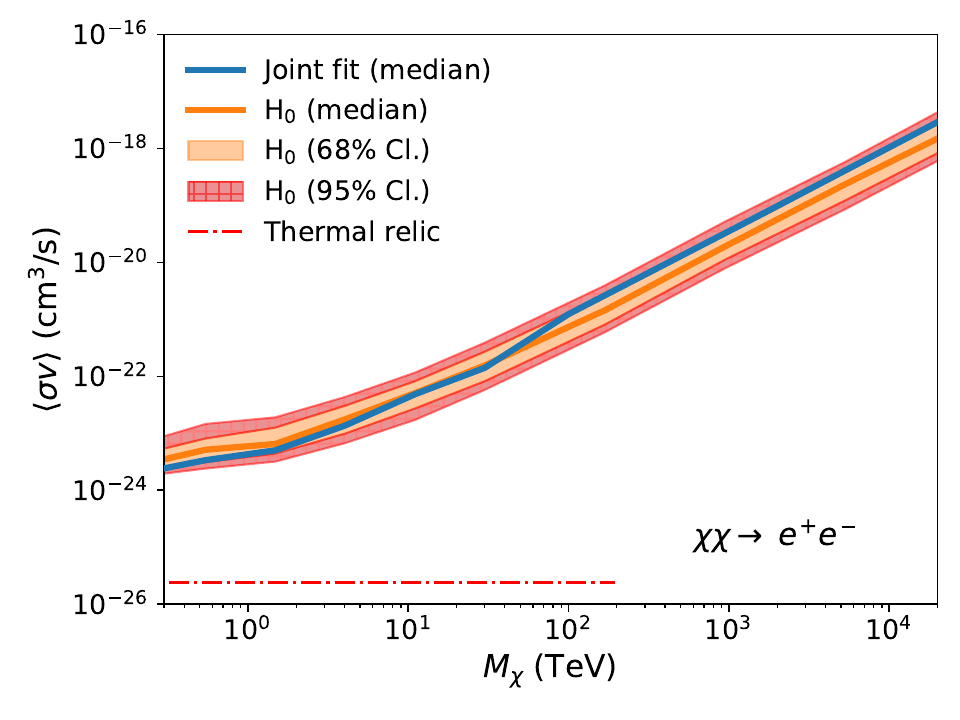}}
    \subfigure{\includegraphics[width=0.329\textwidth]{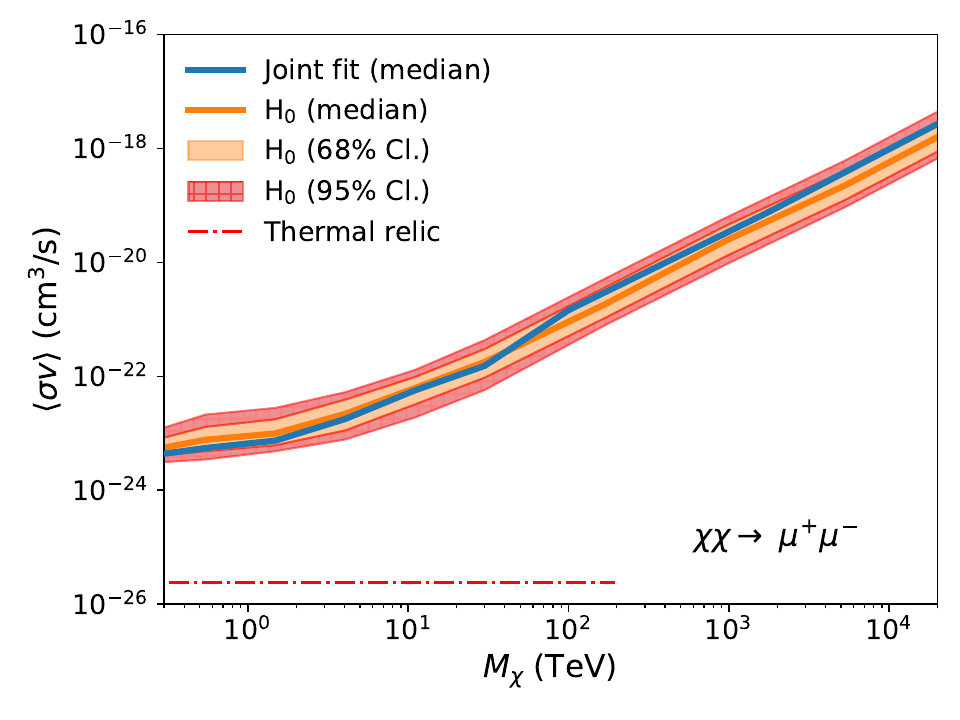}}
    \subfigure{\includegraphics[width=0.329\textwidth]{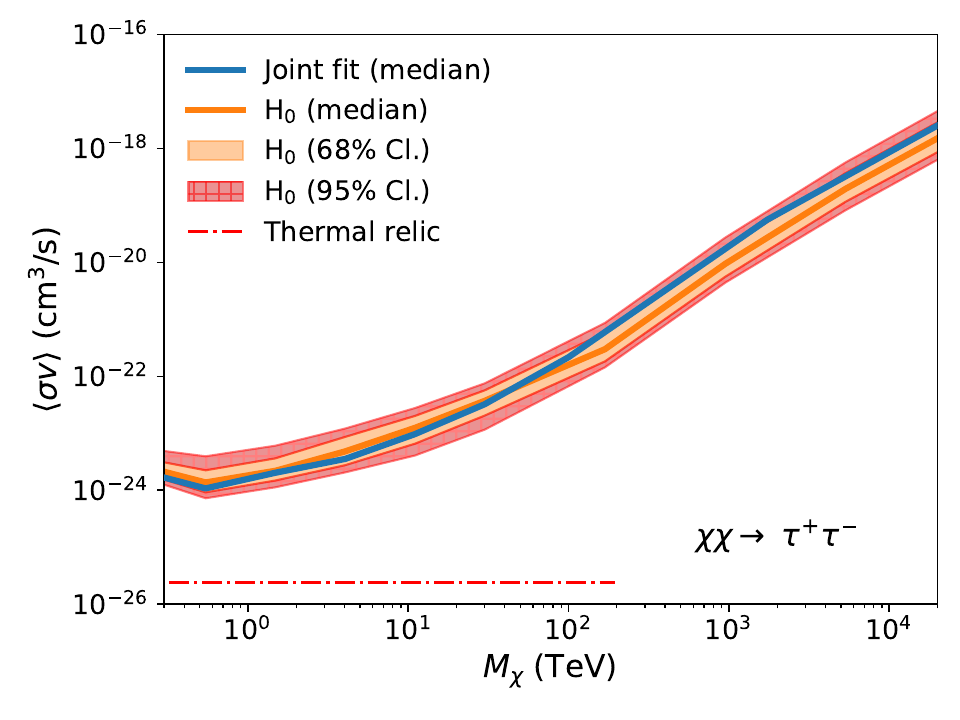}}\\
    \subfigure{\includegraphics[width=0.329\textwidth]{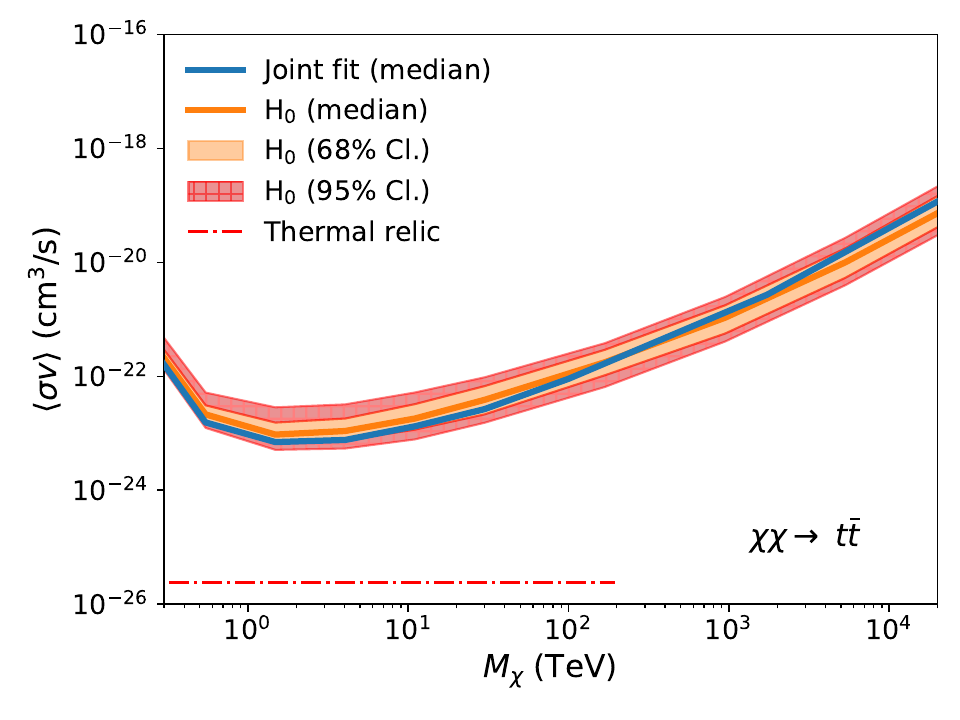}}
    \subfigure{\includegraphics[width=0.329\textwidth]{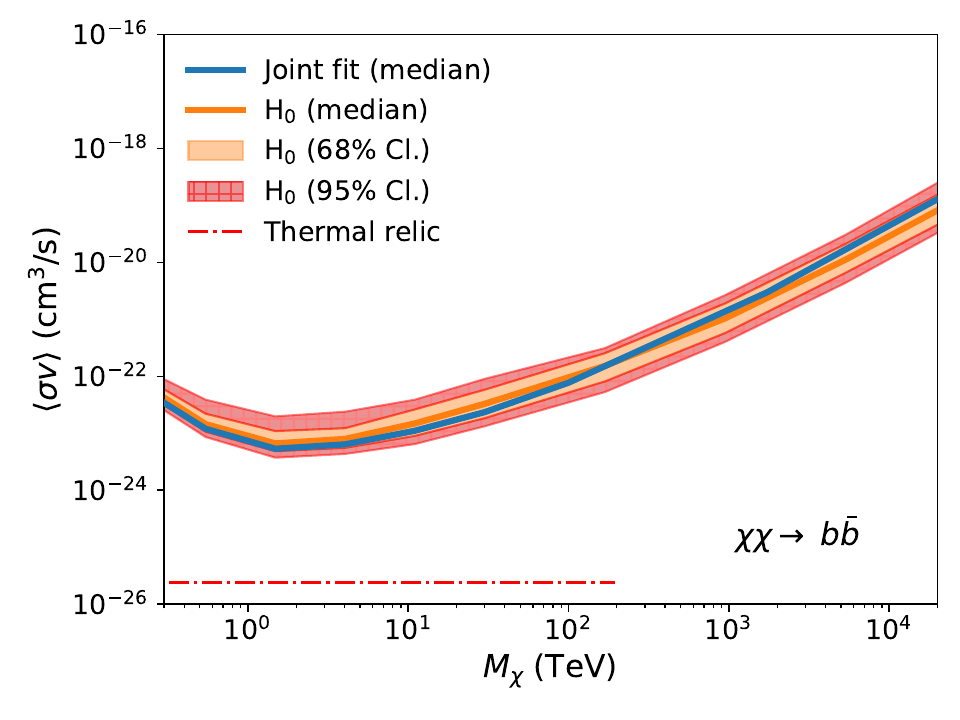}}
    \subfigure{\includegraphics[width=0.329\textwidth]{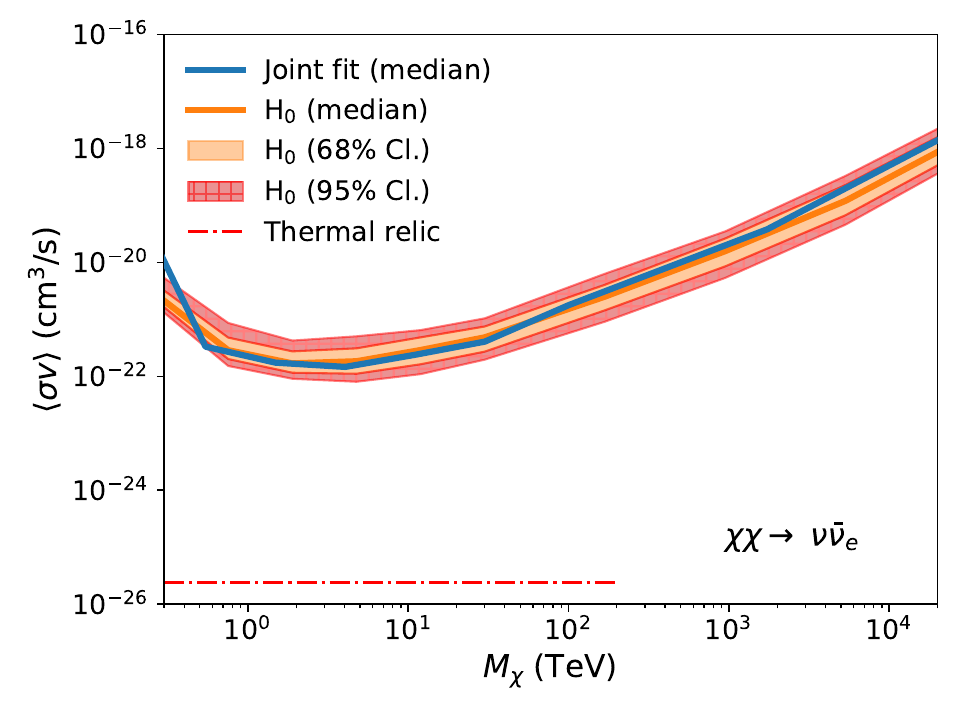}}\\
    \subfigure{\includegraphics[width=0.329\textwidth]{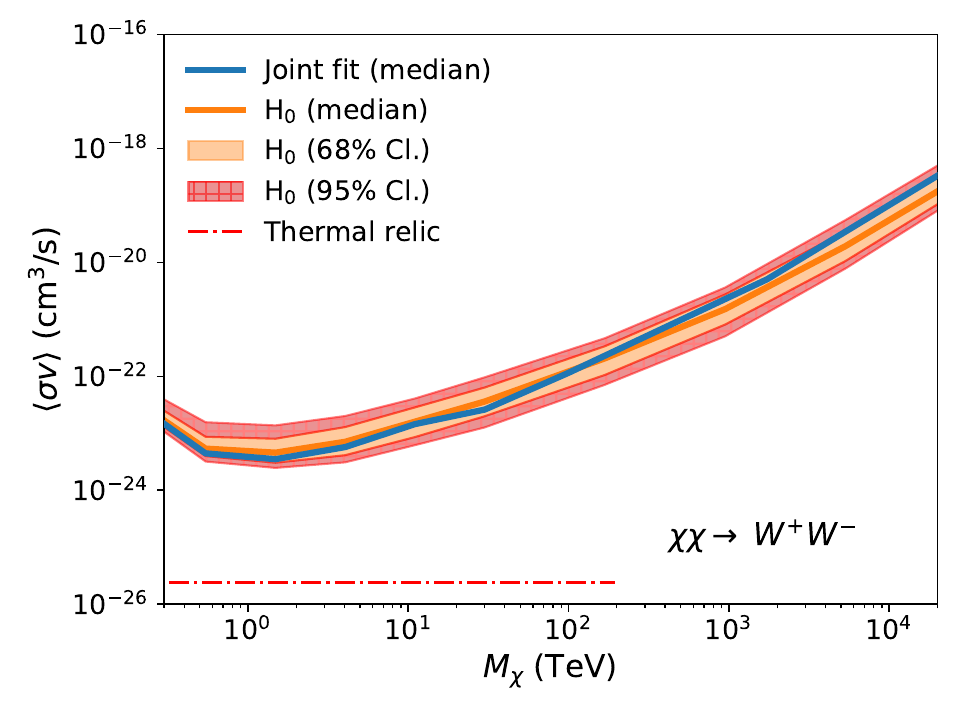}}
    \subfigure{\includegraphics[width=0.329\textwidth]{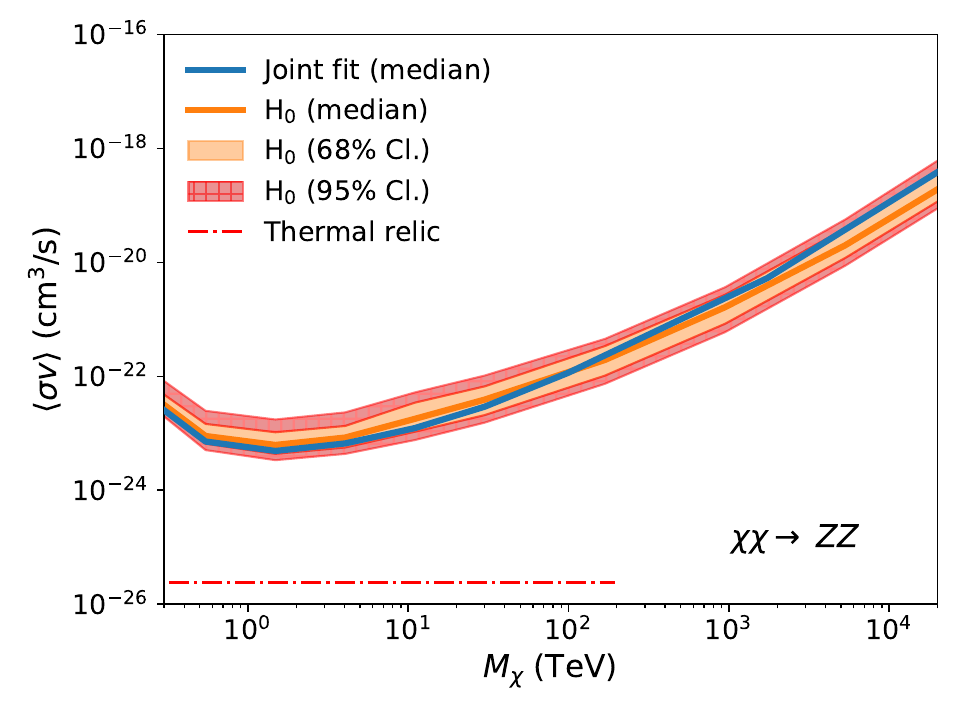}}
    \subfigure{\includegraphics[width=0.329\textwidth]{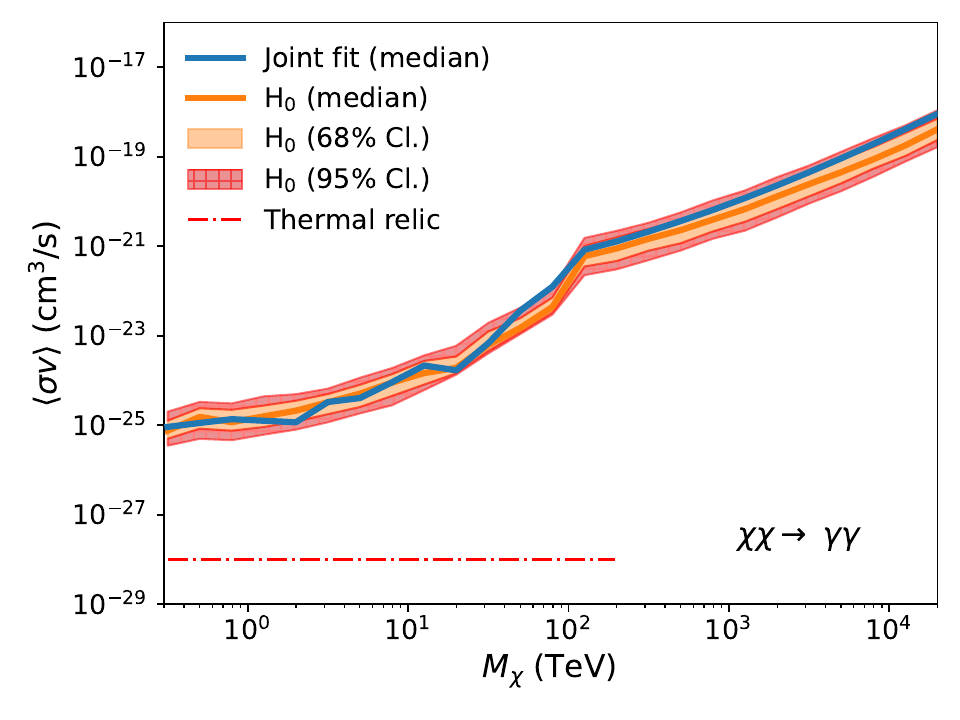}}
    \caption{Velocity-weighted annihilation cross section upper limit curves produced from VERITAS observations by channel compared with their null-hypothesis bands (H$_0$; background-only hypothesis). We present upper limits derived from the joint-fit dSph observations as Fig~\ref{fig:ul_sys} (blue) and upper limits with the Poisson background fluctuation (orange). The orange (red hatched) uncertainty band depicts the 68\% (95\%) containment obtained from 300 realizations of random fluctuations of the background. The red dotted-dashed line is the expected velocity-weighted annihilation cross section for a thermal-relic dark matter scenario.}
    \label{fig:ul_exp}
\end{figure*}

%We assess whether the distribution of observed ON-region events includes the DM annihilation signal above the background level (alternative hypothesis; H$_1$). The null hypothesis (H$_0$) is that the distribution of observed ON-region events is consistent with the background fluctuation (i.e., $\langle\sigma v \rangle = 0$). We perform a simulation study for testing these hypotheses. This study involves estimating an expected upper limit, based on a simulated (or synthesized) ON region. The simulated ON region is 

In addition to calculating ``observed" ULs, we calculate ``expected" ULs assuming the background-only (null) hypothesis. Simulated ON regions are generated through a procedure where events (and their corresponding energies) are randomly selected (with replacement) from the observed OFF-region events, which are assumed to be pure background. The events are sampled  according to a Poisson distribution, with the mean equal to the observed number of OFF-region events scaled by the ratio of the areas of the ON and OFF regions; i.e., $N_{\rm on,sim} = {\rm Pois}(\alpha N_{\rm off,obs})$. Simulated OFF-regions are obtained in the same manner, considering Poisson fluctuations through random sampling. We repeat this process 300 times for each channel, resulting in an expected upper limit band. The width of this band is determined by the magnitude of the Poisson fluctuations of ON and OFF regions. We use the same J-profile for both the expected and observed limits (Section~\ref{sec:ul_w_sys})

Fig.~\ref{fig:ul_exp} shows the comparison between the expected UL band (orange) and the observed UL (blue solid line). For each annihilation channel, the expected upper limit band shows the 68\% containment (orange shaded region) and 95\% containment (red hatched region). The observed upper limits remain consistent with the expected upper limits within the 95\% confidence level across all annihilation channels. This result indicates that a dark matter annihilation signal has not been observed, while also quantifying the influence of statistical uncertainty on the derived ULs. 

\subsection{Ultra-heavy dark matter search}

\begin{figure*}[t!]
    \centering
    \subfigure{\includegraphics[width=0.495\textwidth]{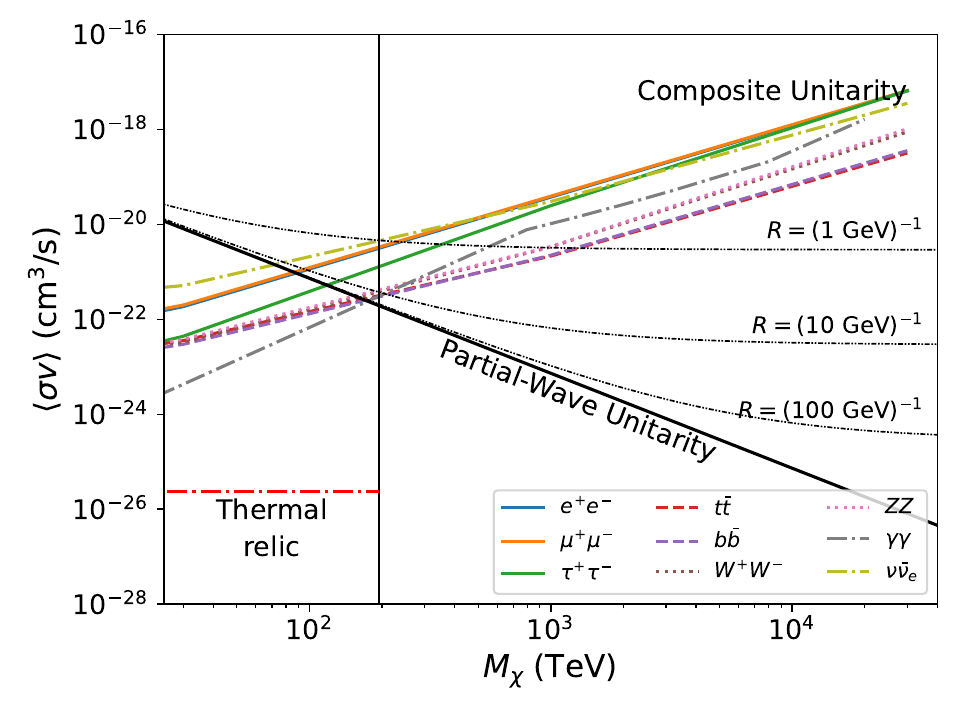}}
\subfigure{\includegraphics[width=0.495\textwidth]{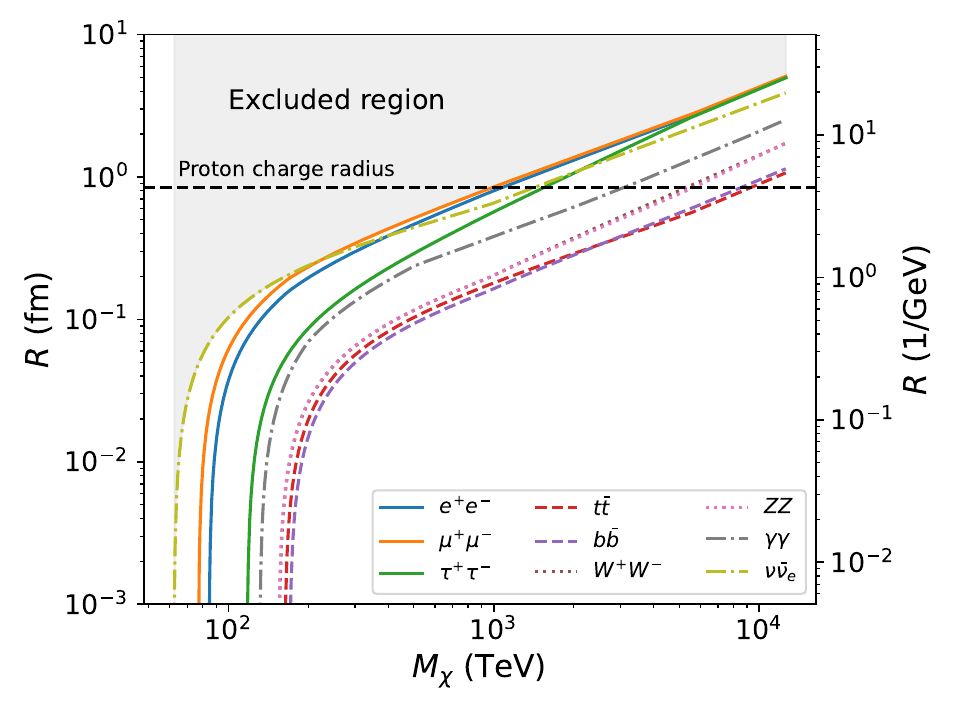}}
    \caption{Left panel: VERITAS 95\% confidence upper limits curves on the dark matter velocity-weighted annihilation cross section compared with benchmark theoretical models. The black solid line shows the partial-wave unitarity bound for a point-like dark matter particle with angular momentum $\textbf{J}$ = 0, while the dashed black lines show the unitarity bounds for composite dark matter particles of different radii. Right panel: VERITAS 95\% confidence upper limits curves on the dark matter particle radius (units of femtometers) as a function of mass, for the nine annihilation channels considered. The dashed black line refers the proton charge radius as a reference \citep{Lin2022}. The shaded areas denote exclusion regions.}
    \label{fig:uhdmR}
\end{figure*}

For a dark matter mass exceeding 194 TeV (UHDM regime), the observed data did not show significant deviation between the observed and expected limits (Fig.~\ref{fig:ul_exp}). In Fig.~\ref{fig:uhdmR}, we interpret this non-detection in terms of theoretical benchmark models \citep[for details on the benchmark models, see][]{Tak_2022}. The left panel of Fig.~\ref{fig:uhdmR} shows both the unitarity limit for a point-like particle with angular momentum $\textbf{J}$ = 0 (partial-wave unitarity limit; $\langle \sigma v \rangle_{unitarity} \propto 1/v_{rel}$) and a set of unitarity bounds for composite dark matter particles of different radii, $\langle \sigma v \rangle_{unitarity}(v_{rel}, R)$. Note that, when computing the partial-wave and composite unitarity bounds, we adopt $v_{rel}/c = 2 \times 10^{-5}$, where $v_{rel}$ represents the relative velocity between dark matter particles in dSph galaxies \citep{Martinez:2010xn,McGaugh:2021tyj}. This velocity $v_{rel}$ is much slower than that of the thermal-relic dark matter particles in the early universe. While the limits from this dataset are unable to probe below the limit for a point-like particle, it is possible to rule out models with large dark matter particle radii.
The right panel shows that we are able to exclude a certain range of values for the radius of the composite particle. If the mass of the dark matter particle is less than 1 PeV, its radius should be smaller than the proton charge radius, 0.84 fm \citep{Lin2022}. 

\subsection{Comparison with previous works}\label{sec:comp}

\begin{figure*}[t!]
    \centering
    \subfigure{\includegraphics[width=0.495\textwidth]{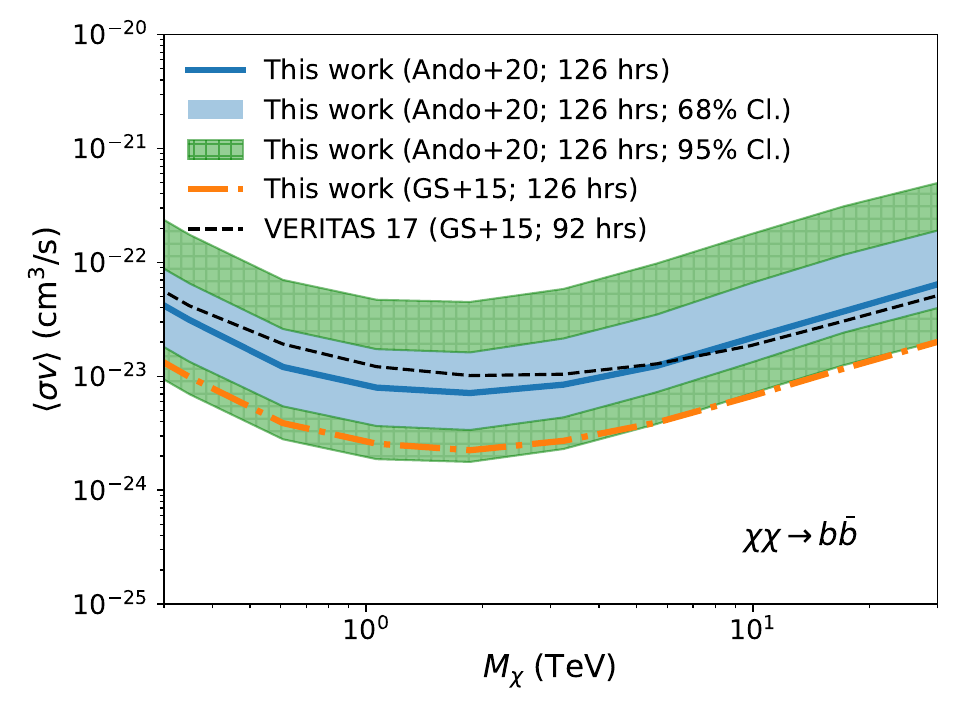}}
    \subfigure{\includegraphics[width=0.495\textwidth]{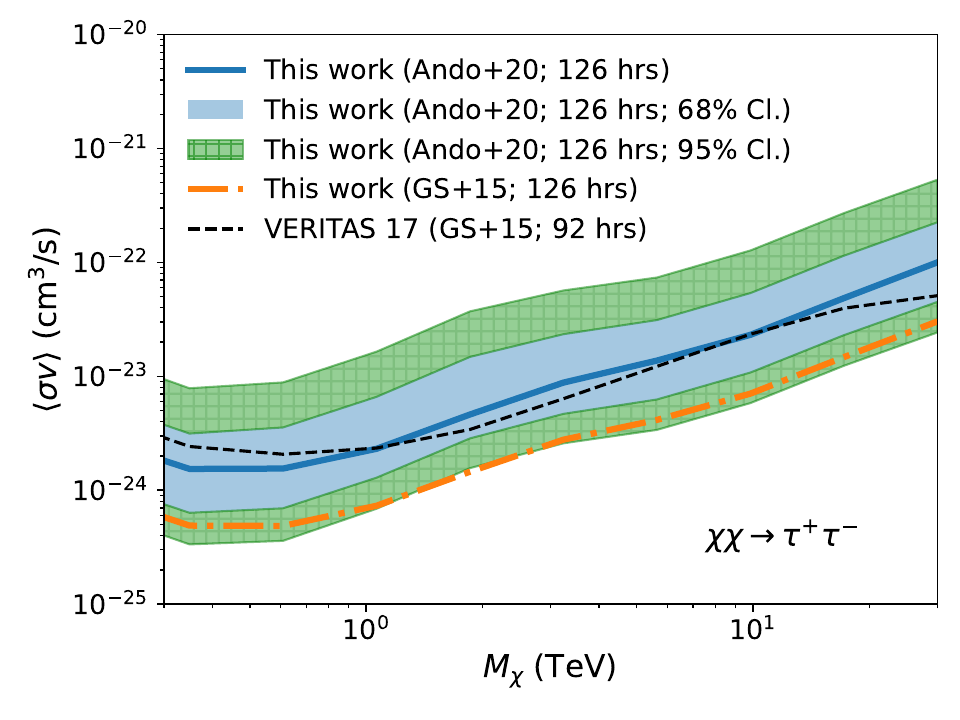}}
    \caption{Comparison of 95\% confidence upper limit curves of the dark matter cross section with various conditions. The blue and orange lines result from the 126-hour Segue 1 observation with the J-profile adopted from \cite{Ando2020} and \cite{GS2015}, respectively. The blue (green hatched) shaded uncertainty band depicts the 68\% (95\%) containment obtained from 300 realizations of viable dark matter density profiles. The black dashed line denotes the previous VERITAS publication \citep{dm_veritas}, which is from the 92-hour Segue 1 observation.}
    \label{fig:compare_with_old}
\end{figure*}

We compare the derived ULs with those from the previous VERITAS work \citep[VERITAS\,17;][]{dm_veritas}. As described in Section~\ref{sec:obs}, in this work, we analyzed a larger dataset with improved analysis techniques. Another significant difference in this study is the utilization of updated J-factors from Ando+20 (Section~\ref{sec:targets}), whereas the previous VERITAS work relied on J-factors estimated using uniform (or uninformative) priors \cite[GS+15;][]{GS2015}. We note that the J-factors calculated using the NFW parameters of Ando+20 result in a lower J-factor on average than the parameters used in the previous VERITAS study using the NFW parameters from GS+15. This is discussed in more detail in Appendix \ref{app:comp}. To quantify the improvement, we compute ULs for Segue 1 for the published dataset and current analysis tools, and compare against the published Segue 1 ULs. We consider the $b^+b^-$ and $\tau^+\tau^-$ annihilation channels. 

Fig.~\ref{fig:compare_with_old} shows UL curves from the extended Segue 1 observation for the two J-factors, GS+15 (orange dotted line) and Ando+20 (blue solid line), and the UL curve from VERITAS\,17 (black dashed line). The limits indicate that our enhanced techniques, such as BDTs, ITM, and an optimized $\theta^2$ cut, along with the increased exposure time, contribute to constraints on the dark matter annihilation cross-section that are more stringent by a factor of 1.7 to 5.1. This is evident when we compare the limits obtained from the same J factor, represented by the orange dotted line (this work) and the black dashed line (the previous work) in Fig.~\ref{fig:compare_with_old}. Note that while extending the exposure time on Segue 1 by 37\% (126~h/92~h) lowers the upper limit slightly, this enhancement is negligible compared to the significant improvement achieved by the enhanced techniques, which lower the limits by an average factor of 3. In contrast, utilizing Ando+20 leads to a less restrictive limit on the dark matter annihilation cross section, as seen by the upper limit (blue solid line) increasing by approximately three times compared to the UL curve with GS+15 (orange dotted line). Note that the J-factor of Segue 1 from Ando+20 is smaller than that of GS+15 by a factor of about 3 (see Appendix~\ref{app:comp}). Overall, taking into account both the positive and negative effects, we arrive at a UL curve similar to that of VERITAS\,17.

\section{Discussion and Conclusions}\label{sec:discussion}
\begin{figure*}[t!]
    \centering
    \subfigure{\includegraphics[width=0.495\textwidth]{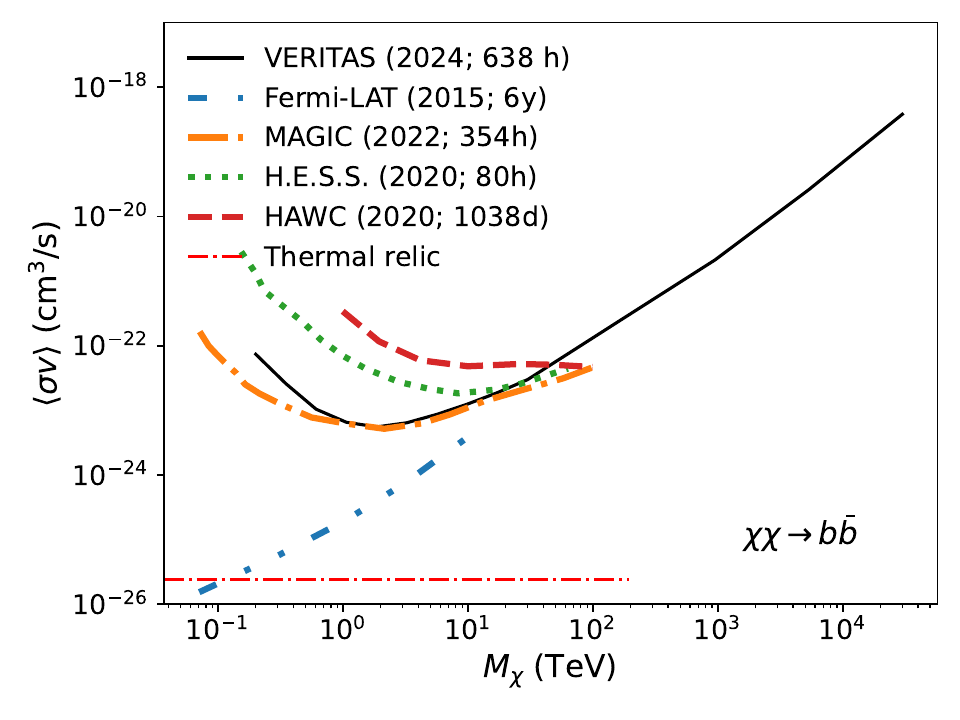}}
    \subfigure{\includegraphics[width=0.495\textwidth]{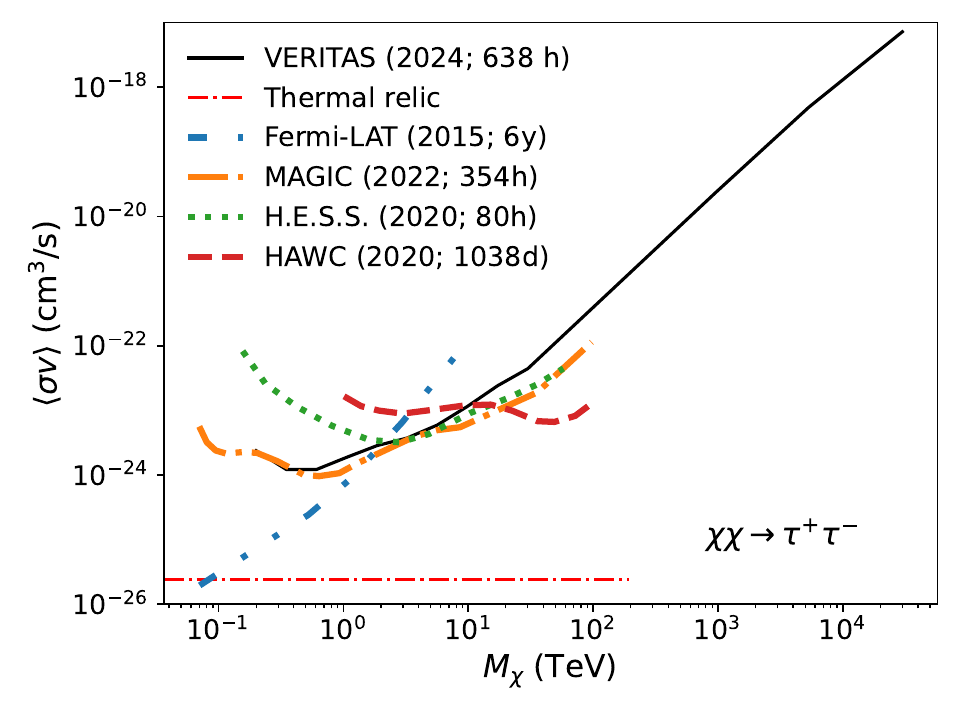}}
    \caption{VERITAS upper limit curves obtained from 17 dSphs compared with other published upper limit curves. All curves show 95\% confidence upper limits on the dark matter velocity-weighted annihilation cross section for the $b\bar{b}$ 
    (left) and $\tau^{+}\tau^{-}$ (right) annihilation channels. This work is represented with the solid black line. %To take into consideration the J-factor ratio between \cite{Ando2020} and \cite{GS2015} (see Appendix~\ref{app:comp}), the \textbf{limit} is lowered by a factor of about three \textbf{to give the black dotted curve}. 
    Other results are adapted from \cite{dm_fermi} (Fermi-LAT; blue dashed line), \cite{dm_magic2} (MAGIC; orange dashed line), \cite{dm_hess2} (H.E.S.S.; green dashed line), and \cite{dm_hawc2} (HAWC; red dashed line). The red dotted-dashed line is the expected velocity-weighted annihilation cross section for a thermal-relic dark matter scenario.}
    \label{fig:compare_with_others}
\end{figure*}

%We note that previous studies on estimating the DM density in dSphs had relied on uninformative priors or neglected relevant physical information. 

%\donggeun{The J factors of GS+15 and Ando+20, calculated at a 0.5-degree offset from the dSph center, exhibit a noticeable distinction, depending on dSphs (see Appendix~\ref{app:comp}). For example, in GS+15, the J factors attributed to Segue 1, Ursa Major II, and Ursa Minor (dSphs subjected to deep exposure in this work) are approximately 2.8 to 4.3 times higher than those presented in Ando+20. As we lack access to the J-factor parameters from GS+15 for our set of 17 dSphs, an accurate computation of the extent to which this discrepancy impacts the upper limit curve for the DM cross section cannot be achieved. Nonetheless, assuming a constant factor by which the two J factors vary across all degrees for each dSph, a rough estimation of the DM upper limit curve can be obtained; i.e., for each dSph, the number of expected events from the DM signal (as described in Equation~\ref{equation:expectedEvnts}) is multiplied by a constant factor based on Table~\ref{tab:j_comp}. Note that for Draco II and Triangulum II, we assume the ratio is 1 because GS+15 did not present the J factors for those dSphs.} 

Fig.~\ref{fig:compare_with_others} shows our upper limits for two annihilation channels (black lines) in comparison to results utilizing dSph observations from the \textit{Fermi}-LAT \citep[blue dashed;][]{dm_fermi}, MAGIC \citep[orange dashed;][]{dm_magic2}, H.E.S.S. \citep[greeen dashed;][]{dm_hess2}, and HAWC \citep[red dashed;][]{dm_hawc2} collaborations. Using the Ando+20 J-factors (black solid), this work demonstrates results on par with the most stringent results from other collaborations. Note that the results from the other collaborations were derived from a range of sources for the J-factor, including but not limited to GS+15.

Due to the significant influence of the J-factor on the constraint of the dark matter velocity-weighted annihilation cross-section, it is crucial to utilize an accurate J-factor when conducting indirect dark matter searches with gamma rays. In this work, we adopt the recent J-factor study by Ando+20, which employs physically motivated informative priors (Section~\ref{sec:targets}). We stress that this is in contrast to previous J-factor estimates (e.g., GS+15, used in the previous VERITAS work) which rely on uniform priors or neglect relevant physical information. If we account for the differences in the J factor between GS+15 and Ando+20, our upper limits are expected to be lowered by a factor of about three (see Appendix~\ref{app:comp}). This effect arises from the inverse proportionality between the upper limit and the J-factor, $\langle\sigma v\rangle_{95} \propto 1/J(\Delta \Omega)$. %; for dSphs subjected to deep exposure, the J factors differ by about 3. With Ando+20, this work, therefore, provides the most conservative but stringent constraints on the DM upper limits.

In summary, this paper conducts an indirect search for a dark matter annihilation signal in gamma rays. We analyzed 638 hours of observations taken by the VERITAS array, spanning from 2007 to 2018. Seventeen dSphs were observed, employing two observational strategies: deep exposures on dark matter-dense dSphs and survey observations of dark matter-sparse dSphs. Our search targeted the detection of the final state gamma rays resulting from nine annihilation channels. Although no significant signals were detected, we derived upper limits on the dark matter velocity-weighted annihilation cross section for the nine annihilation channels. These limits were obtained using a joint-fit MLE analysis and cover a range of dark matter particle masses, extending from 200 GeV to 30 PeV. With the extended data set and the improved techniques, we were able to provide competitive ULs above about 1 TeV, even using a set of smaller but physically motivated J-factors. In addition, we examined the ultra-heavy dark matter scenario and constrained the radius of a composite ultra-heavy dark matter particle.

\section{Acknowledgments}
This research is supported by grants from the U.S. Department of Energy Office of Science, the U.S. National Science Foundation and the Smithsonian Institution, by NSERC in Canada, and by the Helmholtz Association in Germany. This research used resources provided by the Open Science Grid, which is supported by the National Science Foundation and the U.S. Department of Energy's Office of Science, and resources of the National Energy Research Scientific Computing Center (NERSC), a U.S. Department of Energy Office of Science User Facility operated under Contract No. DE-AC02-05CH11231. We acknowledge the excellent work of the technical support staff at the Fred Lawrence Whipple Observatory and at the collaborating institutions in the construction and operation of the instrument.  D. Tak is supported by the National Research Foundation of Korea (NRF) grant, No.2020R1A2C3011091, and No.2021M3F7A1084525, and RS-2024-00343729, funded by the Korea government (MSIT). 

\bibliography{references}

\begin{thebibliography}{52}%
\makeatletter
\providecommand \@ifxundefined [1]{%
 \@ifx{#1\undefined}
}%
\providecommand \@ifnum [1]{%
 \ifnum #1\expandafter \@firstoftwo
 \else \expandafter \@secondoftwo
 \fi
}%
\providecommand \@ifx [1]{%
 \ifx #1\expandafter \@firstoftwo
 \else \expandafter \@secondoftwo
 \fi
}%
\providecommand \natexlab [1]{#1}%
\providecommand \enquote  [1]{``#1''}%
\providecommand \bibnamefont  [1]{#1}%
\providecommand \bibfnamefont [1]{#1}%
\providecommand \citenamefont [1]{#1}%
\providecommand \href@noop [0]{\@secondoftwo}%
\providecommand \href [0]{\begingroup \@sanitize@url \@href}%
\providecommand \@href[1]{\@@startlink{#1}\@@href}%
\providecommand \@@href[1]{\endgroup#1\@@endlink}%
\providecommand \@sanitize@url [0]{\catcode `\\12\catcode `\$12\catcode
  `\&12\catcode `\#12\catcode `\^12\catcode `\_12\catcode `\%12\relax}%
\providecommand \@@startlink[1]{}%
\providecommand \@@endlink[0]{}%
\providecommand \url  [0]{\begingroup\@sanitize@url \@url }%
\providecommand \@url [1]{\endgroup\@href {#1}{\urlprefix }}%
\providecommand \urlprefix  [0]{URL }%
\providecommand \Eprint [0]{\href }%
\providecommand \doibase [0]{https://doi.org/}%
\providecommand \selectlanguage [0]{\@gobble}%
\providecommand \bibinfo  [0]{\@secondoftwo}%
\providecommand \bibfield  [0]{\@secondoftwo}%
\providecommand \translation [1]{[#1]}%
\providecommand \BibitemOpen [0]{}%
\providecommand \bibitemStop [0]{}%
\providecommand \bibitemNoStop [0]{.\EOS\space}%
\providecommand \EOS [0]{\spacefactor3000\relax}%
\providecommand \BibitemShut  [1]{\csname bibitem#1\endcsname}%
\let\auto@bib@innerbib\@empty
%</preamble>
\bibitem [{\citenamefont {Aghanim}\ \emph {et~al.}(2020)\citenamefont {Aghanim}
  \emph {et~al.}}]{Planck:2018vyg}%
  \BibitemOpen
  \bibfield  {author} {\bibinfo {author} {\bibfnamefont {N.}~\bibnamefont
  {Aghanim}} \emph {et~al.} (\bibinfo {collaboration} {Planck}),\ }\href
  {https://doi.org/10.1051/0004-6361/201833910} {\bibfield  {journal} {\bibinfo
   {journal} {Astron. Astrophys.}\ }\textbf {\bibinfo {volume} {641}},\
  \bibinfo {pages} {A6} (\bibinfo {year} {2020})},\ \bibinfo {note} {[Erratum:
  Astron.Astrophys. 652, C4 (2021)]},\ \Eprint
  {https://arxiv.org/abs/1807.06209} {arXiv:1807.06209 [astro-ph.CO]}
  \BibitemShut {NoStop}%
\bibitem [{\citenamefont {Ackermann}\ \emph
  {et~al.}(2015{\natexlab{a}})\citenamefont {Ackermann} \emph
  {et~al.}}]{dm_fermi}%
  \BibitemOpen
  \bibfield  {author} {\bibinfo {author} {\bibfnamefont {M.}~\bibnamefont
  {Ackermann}} \emph {et~al.},\ }\href
  {https://doi.org/10.1103/PhysRevLett.115.231301} {\bibfield  {journal}
  {\bibinfo  {journal} {\prl}\ }\textbf {\bibinfo {volume} {115}},\ \bibinfo
  {eid} {231301} (\bibinfo {year} {2015}{\natexlab{a}})},\ \Eprint
  {https://arxiv.org/abs/1503.02641} {arXiv:1503.02641 [astro-ph.HE]}
  \BibitemShut {NoStop}%
\bibitem [{\citenamefont {Ackermann}\ \emph
  {et~al.}(2015{\natexlab{b}})\citenamefont {Ackermann}, \citenamefont
  {Ajello}, \citenamefont {Albert}, \citenamefont {Anderson}, \citenamefont
  {Atwood}, \citenamefont {Baldini}, \citenamefont {Barbiellini}, \citenamefont
  {Bastieri}, \citenamefont {Bellazzini}, \citenamefont {Bissaldi},
  \citenamefont {Blandford}, \citenamefont {Bloom}, \citenamefont {Bonino},
  \citenamefont {Bottacini}, \citenamefont {Brandt}, \citenamefont {Bregeon},
  \citenamefont {Bruel}, \citenamefont {Buehler}, \citenamefont {Buson},
  \citenamefont {Caliandro}, \citenamefont {Cameron}, \citenamefont {Caputo},
  \citenamefont {Caragiulo}, \citenamefont {Caraveo}, \citenamefont {Cecchi},
  \citenamefont {Charles}, \citenamefont {Chekhtman}, \citenamefont {Chiang},
  \citenamefont {Chiaro}, \citenamefont {Ciprini}, \citenamefont {Claus},
  \citenamefont {Cohen-Tanugi}, \citenamefont {Conrad}, \citenamefont {Cuoco},
  \citenamefont {Cutini}, \citenamefont {D'Ammando}, \citenamefont
  {de~Angelis}, \citenamefont {de~Palma}, \citenamefont {Desiante},
  \citenamefont {Digel}, \citenamefont {Di~Venere}, \citenamefont {Drell},
  \citenamefont {Drlica-Wagner}, \citenamefont {Favuzzi}, \citenamefont
  {Fegan}, \citenamefont {Franckowiak}, \citenamefont {Fukazawa}, \citenamefont
  {Funk}, \citenamefont {Fusco}, \citenamefont {Gargano}, \citenamefont
  {Gasparrini}, \citenamefont {Giglietto}, \citenamefont {Giordano},
  \citenamefont {Giroletti}, \citenamefont {Godfrey}, \citenamefont
  {Gomez-Vargas}, \citenamefont {Grenier}, \citenamefont {Grove}, \citenamefont
  {Guiriec}, \citenamefont {Gustafsson}, \citenamefont {Hewitt}, \citenamefont
  {Hill}, \citenamefont {Horan}, \citenamefont {J\'ohannesson}, \citenamefont
  {Johnson}, \citenamefont {Kuss}, \citenamefont {Larsson}, \citenamefont
  {Latronico}, \citenamefont {Li}, \citenamefont {Li}, \citenamefont {Longo},
  \citenamefont {Loparco}, \citenamefont {Lovellette}, \citenamefont {Lubrano},
  \citenamefont {Malyshev}, \citenamefont {Mayer}, \citenamefont {Mazziotta},
  \citenamefont {McEnery}, \citenamefont {Michelson}, \citenamefont {Mizuno},
  \citenamefont {Moiseev}, \citenamefont {Monzani}, \citenamefont {Morselli},
  \citenamefont {Murgia}, \citenamefont {Nuss}, \citenamefont {Ohsugi},
  \citenamefont {Orienti}, \citenamefont {Orlando}, \citenamefont {Ormes},
  \citenamefont {Paneque}, \citenamefont {Pesce-Rollins}, \citenamefont
  {Piron}, \citenamefont {Pivato}, \citenamefont {Rain\`o}, \citenamefont
  {Rando}, \citenamefont {Razzano}, \citenamefont {Reimer}, \citenamefont
  {Reposeur}, \citenamefont {Ritz}, \citenamefont {S\'anchez-Conde},
  \citenamefont {Schulz}, \citenamefont {Sgr\`o}, \citenamefont {Siskind},
  \citenamefont {Spada}, \citenamefont {Spandre}, \citenamefont {Spinelli},
  \citenamefont {Tajima}, \citenamefont {Takahashi}, \citenamefont {Thayer},
  \citenamefont {Tibaldo}, \citenamefont {Torres}, \citenamefont {Tosti},
  \citenamefont {Troja}, \citenamefont {Vianello}, \citenamefont {Werner},
  \citenamefont {Winer}, \citenamefont {Wood}, \citenamefont {Wood},
  \citenamefont {Zaharijas},\ and\ \citenamefont {Zimmer}}]{dm_fermiGCline}%
  \BibitemOpen
  \bibfield  {author} {\bibinfo {author} {\bibfnamefont {M.}~\bibnamefont
  {Ackermann}}, \bibinfo {author} {\bibfnamefont {M.}~\bibnamefont {Ajello}},
  \bibinfo {author} {\bibfnamefont {A.}~\bibnamefont {Albert}}, \bibinfo
  {author} {\bibfnamefont {B.}~\bibnamefont {Anderson}}, \bibinfo {author}
  {\bibfnamefont {W.~B.}\ \bibnamefont {Atwood}}, \bibinfo {author}
  {\bibfnamefont {L.}~\bibnamefont {Baldini}}, \bibinfo {author} {\bibfnamefont
  {G.}~\bibnamefont {Barbiellini}}, \bibinfo {author} {\bibfnamefont
  {D.}~\bibnamefont {Bastieri}}, \bibinfo {author} {\bibfnamefont
  {R.}~\bibnamefont {Bellazzini}}, \bibinfo {author} {\bibfnamefont
  {E.}~\bibnamefont {Bissaldi}}, \bibinfo {author} {\bibfnamefont {R.~D.}\
  \bibnamefont {Blandford}}, \bibinfo {author} {\bibfnamefont {E.~D.}\
  \bibnamefont {Bloom}}, \bibinfo {author} {\bibfnamefont {R.}~\bibnamefont
  {Bonino}}, \bibinfo {author} {\bibfnamefont {E.}~\bibnamefont {Bottacini}},
  \bibinfo {author} {\bibfnamefont {T.~J.}\ \bibnamefont {Brandt}}, \bibinfo
  {author} {\bibfnamefont {J.}~\bibnamefont {Bregeon}}, \bibinfo {author}
  {\bibfnamefont {P.}~\bibnamefont {Bruel}}, \bibinfo {author} {\bibfnamefont
  {R.}~\bibnamefont {Buehler}}, \bibinfo {author} {\bibfnamefont
  {S.}~\bibnamefont {Buson}}, \bibinfo {author} {\bibfnamefont {G.~A.}\
  \bibnamefont {Caliandro}}, \bibinfo {author} {\bibfnamefont {R.~A.}\
  \bibnamefont {Cameron}}, \bibinfo {author} {\bibfnamefont {R.}~\bibnamefont
  {Caputo}}, \bibinfo {author} {\bibfnamefont {M.}~\bibnamefont {Caragiulo}},
  \bibinfo {author} {\bibfnamefont {P.~A.}\ \bibnamefont {Caraveo}}, \bibinfo
  {author} {\bibfnamefont {C.}~\bibnamefont {Cecchi}}, \bibinfo {author}
  {\bibfnamefont {E.}~\bibnamefont {Charles}}, \bibinfo {author} {\bibfnamefont
  {A.}~\bibnamefont {Chekhtman}}, \bibinfo {author} {\bibfnamefont
  {J.}~\bibnamefont {Chiang}}, \bibinfo {author} {\bibfnamefont
  {G.}~\bibnamefont {Chiaro}}, \bibinfo {author} {\bibfnamefont
  {S.}~\bibnamefont {Ciprini}}, \bibinfo {author} {\bibfnamefont
  {R.}~\bibnamefont {Claus}}, \bibinfo {author} {\bibfnamefont
  {J.}~\bibnamefont {Cohen-Tanugi}}, \bibinfo {author} {\bibfnamefont
  {J.}~\bibnamefont {Conrad}}, \bibinfo {author} {\bibfnamefont
  {A.}~\bibnamefont {Cuoco}}, \bibinfo {author} {\bibfnamefont
  {S.}~\bibnamefont {Cutini}}, \bibinfo {author} {\bibfnamefont
  {F.}~\bibnamefont {D'Ammando}}, \bibinfo {author} {\bibfnamefont
  {A.}~\bibnamefont {de~Angelis}}, \bibinfo {author} {\bibfnamefont
  {F.}~\bibnamefont {de~Palma}}, \bibinfo {author} {\bibfnamefont
  {R.}~\bibnamefont {Desiante}}, \bibinfo {author} {\bibfnamefont {S.~W.}\
  \bibnamefont {Digel}}, \bibinfo {author} {\bibfnamefont {L.}~\bibnamefont
  {Di~Venere}}, \bibinfo {author} {\bibfnamefont {P.~S.}\ \bibnamefont
  {Drell}}, \bibinfo {author} {\bibfnamefont {A.}~\bibnamefont
  {Drlica-Wagner}}, \bibinfo {author} {\bibfnamefont {C.}~\bibnamefont
  {Favuzzi}}, \bibinfo {author} {\bibfnamefont {S.~J.}\ \bibnamefont {Fegan}},
  \bibinfo {author} {\bibfnamefont {A.}~\bibnamefont {Franckowiak}}, \bibinfo
  {author} {\bibfnamefont {Y.}~\bibnamefont {Fukazawa}}, \bibinfo {author}
  {\bibfnamefont {S.}~\bibnamefont {Funk}}, \bibinfo {author} {\bibfnamefont
  {P.}~\bibnamefont {Fusco}}, \bibinfo {author} {\bibfnamefont
  {F.}~\bibnamefont {Gargano}}, \bibinfo {author} {\bibfnamefont
  {D.}~\bibnamefont {Gasparrini}}, \bibinfo {author} {\bibfnamefont
  {N.}~\bibnamefont {Giglietto}}, \bibinfo {author} {\bibfnamefont
  {F.}~\bibnamefont {Giordano}}, \bibinfo {author} {\bibfnamefont
  {M.}~\bibnamefont {Giroletti}}, \bibinfo {author} {\bibfnamefont
  {G.}~\bibnamefont {Godfrey}}, \bibinfo {author} {\bibfnamefont {G.~A.}\
  \bibnamefont {Gomez-Vargas}}, \bibinfo {author} {\bibfnamefont {I.~A.}\
  \bibnamefont {Grenier}}, \bibinfo {author} {\bibfnamefont {J.~E.}\
  \bibnamefont {Grove}}, \bibinfo {author} {\bibfnamefont {S.}~\bibnamefont
  {Guiriec}}, \bibinfo {author} {\bibfnamefont {M.}~\bibnamefont {Gustafsson}},
  \bibinfo {author} {\bibfnamefont {J.~W.}\ \bibnamefont {Hewitt}}, \bibinfo
  {author} {\bibfnamefont {A.~B.}\ \bibnamefont {Hill}}, \bibinfo {author}
  {\bibfnamefont {D.}~\bibnamefont {Horan}}, \bibinfo {author} {\bibfnamefont
  {G.}~\bibnamefont {J\'ohannesson}}, \bibinfo {author} {\bibfnamefont {R.~P.}\
  \bibnamefont {Johnson}}, \bibinfo {author} {\bibfnamefont {M.}~\bibnamefont
  {Kuss}}, \bibinfo {author} {\bibfnamefont {S.}~\bibnamefont {Larsson}},
  \bibinfo {author} {\bibfnamefont {L.}~\bibnamefont {Latronico}}, \bibinfo
  {author} {\bibfnamefont {J.}~\bibnamefont {Li}}, \bibinfo {author}
  {\bibfnamefont {L.}~\bibnamefont {Li}}, \bibinfo {author} {\bibfnamefont
  {F.}~\bibnamefont {Longo}}, \bibinfo {author} {\bibfnamefont
  {F.}~\bibnamefont {Loparco}}, \bibinfo {author} {\bibfnamefont {M.~N.}\
  \bibnamefont {Lovellette}}, \bibinfo {author} {\bibfnamefont
  {P.}~\bibnamefont {Lubrano}}, \bibinfo {author} {\bibfnamefont
  {D.}~\bibnamefont {Malyshev}}, \bibinfo {author} {\bibfnamefont
  {M.}~\bibnamefont {Mayer}}, \bibinfo {author} {\bibfnamefont {M.~N.}\
  \bibnamefont {Mazziotta}}, \bibinfo {author} {\bibfnamefont {J.~E.}\
  \bibnamefont {McEnery}}, \bibinfo {author} {\bibfnamefont {P.~F.}\
  \bibnamefont {Michelson}}, \bibinfo {author} {\bibfnamefont {T.}~\bibnamefont
  {Mizuno}}, \bibinfo {author} {\bibfnamefont {A.~A.}\ \bibnamefont {Moiseev}},
  \bibinfo {author} {\bibfnamefont {M.~E.}\ \bibnamefont {Monzani}}, \bibinfo
  {author} {\bibfnamefont {A.}~\bibnamefont {Morselli}}, \bibinfo {author}
  {\bibfnamefont {S.}~\bibnamefont {Murgia}}, \bibinfo {author} {\bibfnamefont
  {E.}~\bibnamefont {Nuss}}, \bibinfo {author} {\bibfnamefont {T.}~\bibnamefont
  {Ohsugi}}, \bibinfo {author} {\bibfnamefont {M.}~\bibnamefont {Orienti}},
  \bibinfo {author} {\bibfnamefont {E.}~\bibnamefont {Orlando}}, \bibinfo
  {author} {\bibfnamefont {J.~F.}\ \bibnamefont {Ormes}}, \bibinfo {author}
  {\bibfnamefont {D.}~\bibnamefont {Paneque}}, \bibinfo {author} {\bibfnamefont
  {M.}~\bibnamefont {Pesce-Rollins}}, \bibinfo {author} {\bibfnamefont
  {F.}~\bibnamefont {Piron}}, \bibinfo {author} {\bibfnamefont
  {G.}~\bibnamefont {Pivato}}, \bibinfo {author} {\bibfnamefont
  {S.}~\bibnamefont {Rain\`o}}, \bibinfo {author} {\bibfnamefont
  {R.}~\bibnamefont {Rando}}, \bibinfo {author} {\bibfnamefont
  {M.}~\bibnamefont {Razzano}}, \bibinfo {author} {\bibfnamefont
  {A.}~\bibnamefont {Reimer}}, \bibinfo {author} {\bibfnamefont
  {T.}~\bibnamefont {Reposeur}}, \bibinfo {author} {\bibfnamefont
  {S.}~\bibnamefont {Ritz}}, \bibinfo {author} {\bibfnamefont {M.}~\bibnamefont
  {S\'anchez-Conde}}, \bibinfo {author} {\bibfnamefont {A.}~\bibnamefont
  {Schulz}}, \bibinfo {author} {\bibfnamefont {C.}~\bibnamefont {Sgr\`o}},
  \bibinfo {author} {\bibfnamefont {E.~J.}\ \bibnamefont {Siskind}}, \bibinfo
  {author} {\bibfnamefont {F.}~\bibnamefont {Spada}}, \bibinfo {author}
  {\bibfnamefont {G.}~\bibnamefont {Spandre}}, \bibinfo {author} {\bibfnamefont
  {P.}~\bibnamefont {Spinelli}}, \bibinfo {author} {\bibfnamefont
  {H.}~\bibnamefont {Tajima}}, \bibinfo {author} {\bibfnamefont
  {H.}~\bibnamefont {Takahashi}}, \bibinfo {author} {\bibfnamefont {J.~B.}\
  \bibnamefont {Thayer}}, \bibinfo {author} {\bibfnamefont {L.}~\bibnamefont
  {Tibaldo}}, \bibinfo {author} {\bibfnamefont {D.~F.}\ \bibnamefont {Torres}},
  \bibinfo {author} {\bibfnamefont {G.}~\bibnamefont {Tosti}}, \bibinfo
  {author} {\bibfnamefont {E.}~\bibnamefont {Troja}}, \bibinfo {author}
  {\bibfnamefont {G.}~\bibnamefont {Vianello}}, \bibinfo {author}
  {\bibfnamefont {M.}~\bibnamefont {Werner}}, \bibinfo {author} {\bibfnamefont
  {B.~L.}\ \bibnamefont {Winer}}, \bibinfo {author} {\bibfnamefont {K.~S.}\
  \bibnamefont {Wood}}, \bibinfo {author} {\bibfnamefont {M.}~\bibnamefont
  {Wood}}, \bibinfo {author} {\bibfnamefont {G.}~\bibnamefont {Zaharijas}},\
  and\ \bibinfo {author} {\bibfnamefont {S.}~\bibnamefont {Zimmer}},\ }\href
  {https://doi.org/10.1103/PhysRevD.91.122002} {\bibfield  {journal} {\bibinfo
  {journal} {Phys. Rev. D}\ }\textbf {\bibinfo {volume} {91}},\ \bibinfo
  {pages} {122002} (\bibinfo {year} {2015}{\natexlab{b}})}\BibitemShut
  {NoStop}%
\bibitem [{\citenamefont {{Albert}}\ \emph {et~al.}(2017)\citenamefont
  {{Albert}}, \citenamefont {{Anderson}}, \citenamefont {{Bechtol}},
  \citenamefont {{Drlica-Wagner}}, \citenamefont {{Meyer}}, \citenamefont
  {{S{\'a}nchez-Conde}}, \citenamefont {{Strigari}}, \citenamefont {{Wood}},
  \citenamefont {{Abbott}}, \citenamefont {{Abdalla}}, \citenamefont
  {{Benoit-L{\'e}vy}}, \citenamefont {{Bernstein}}, \citenamefont
  {{Bernstein}}, \citenamefont {{Bertin}}, \citenamefont {{Brooks}},
  \citenamefont {{Burke}}, \citenamefont {{Carnero Rosell}}, \citenamefont
  {{Carrasco Kind}}, \citenamefont {{Carretero}}, \citenamefont {{Crocce}},
  \citenamefont {{Cunha}}, \citenamefont {{D'Andrea}}, \citenamefont {{da
  Costa}}, \citenamefont {{Desai}}, \citenamefont {{Diehl}}, \citenamefont
  {{Dietrich}}, \citenamefont {{Doel}}, \citenamefont {{Eifler}}, \citenamefont
  {{Evrard}}, \citenamefont {{Fausti Neto}}, \citenamefont {{Finley}},
  \citenamefont {{Flaugher}}, \citenamefont {{Fosalba}}, \citenamefont
  {{Frieman}}, \citenamefont {{Gerdes}}, \citenamefont {{Goldstein}},
  \citenamefont {{Gruen}}, \citenamefont {{Gruendl}}, \citenamefont
  {{Honscheid}}, \citenamefont {{James}}, \citenamefont {{Kent}}, \citenamefont
  {{Kuehn}}, \citenamefont {{Kuropatkin}}, \citenamefont {{Lahav}},
  \citenamefont {{Li}}, \citenamefont {{Maia}}, \citenamefont {{March}},
  \citenamefont {{Marshall}}, \citenamefont {{Martini}}, \citenamefont
  {{Miller}}, \citenamefont {{Miquel}}, \citenamefont {{Neilsen}},
  \citenamefont {{Nord}}, \citenamefont {{Ogando}}, \citenamefont {{Plazas}},
  \citenamefont {{Reil}}, \citenamefont {{Romer}}, \citenamefont {{Rykoff}},
  \citenamefont {{Sanchez}}, \citenamefont {{Santiago}}, \citenamefont
  {{Schubnell}}, \citenamefont {{Sevilla-Noarbe}}, \citenamefont {{Smith}},
  \citenamefont {{Soares-Santos}}, \citenamefont {{Sobreira}}, \citenamefont
  {{Suchyta}}, \citenamefont {{Swanson}}, \citenamefont {{Tarle}},
  \citenamefont {{Vikram}}, \citenamefont {{Walker}}, \citenamefont
  {{Wechsler}}, \citenamefont {{Fermi-LAT Collaboration}},\ and\ \citenamefont
  {{DES Collaboration}}}]{dm_fermi2}%
  \BibitemOpen
  \bibfield  {author} {\bibinfo {author} {\bibfnamefont {A.}~\bibnamefont
  {{Albert}}}, \bibinfo {author} {\bibfnamefont {B.}~\bibnamefont
  {{Anderson}}}, \bibinfo {author} {\bibfnamefont {K.}~\bibnamefont
  {{Bechtol}}}, \bibinfo {author} {\bibfnamefont {A.}~\bibnamefont
  {{Drlica-Wagner}}}, \bibinfo {author} {\bibfnamefont {M.}~\bibnamefont
  {{Meyer}}}, \bibinfo {author} {\bibfnamefont {M.}~\bibnamefont
  {{S{\'a}nchez-Conde}}}, \bibinfo {author} {\bibfnamefont {L.}~\bibnamefont
  {{Strigari}}}, \bibinfo {author} {\bibfnamefont {M.}~\bibnamefont {{Wood}}},
  \bibinfo {author} {\bibfnamefont {T.~M.~C.}\ \bibnamefont {{Abbott}}},
  \bibinfo {author} {\bibfnamefont {F.~B.}\ \bibnamefont {{Abdalla}}}, \bibinfo
  {author} {\bibfnamefont {A.}~\bibnamefont {{Benoit-L{\'e}vy}}}, \bibinfo
  {author} {\bibfnamefont {G.~M.}\ \bibnamefont {{Bernstein}}}, \bibinfo
  {author} {\bibfnamefont {R.~A.}\ \bibnamefont {{Bernstein}}}, \bibinfo
  {author} {\bibfnamefont {E.}~\bibnamefont {{Bertin}}}, \bibinfo {author}
  {\bibfnamefont {D.}~\bibnamefont {{Brooks}}}, \bibinfo {author}
  {\bibfnamefont {D.~L.}\ \bibnamefont {{Burke}}}, \bibinfo {author}
  {\bibfnamefont {A.}~\bibnamefont {{Carnero Rosell}}}, \bibinfo {author}
  {\bibfnamefont {M.}~\bibnamefont {{Carrasco Kind}}}, \bibinfo {author}
  {\bibfnamefont {J.}~\bibnamefont {{Carretero}}}, \bibinfo {author}
  {\bibfnamefont {M.}~\bibnamefont {{Crocce}}}, \bibinfo {author}
  {\bibfnamefont {C.~E.}\ \bibnamefont {{Cunha}}}, \bibinfo {author}
  {\bibfnamefont {C.~B.}\ \bibnamefont {{D'Andrea}}}, \bibinfo {author}
  {\bibfnamefont {L.~N.}\ \bibnamefont {{da Costa}}}, \bibinfo {author}
  {\bibfnamefont {S.}~\bibnamefont {{Desai}}}, \bibinfo {author} {\bibfnamefont
  {H.~T.}\ \bibnamefont {{Diehl}}}, \bibinfo {author} {\bibfnamefont {J.~P.}\
  \bibnamefont {{Dietrich}}}, \bibinfo {author} {\bibfnamefont
  {P.}~\bibnamefont {{Doel}}}, \bibinfo {author} {\bibfnamefont {T.~F.}\
  \bibnamefont {{Eifler}}}, \bibinfo {author} {\bibfnamefont {A.~E.}\
  \bibnamefont {{Evrard}}}, \bibinfo {author} {\bibfnamefont {A.}~\bibnamefont
  {{Fausti Neto}}}, \bibinfo {author} {\bibfnamefont {D.~A.}\ \bibnamefont
  {{Finley}}}, \bibinfo {author} {\bibfnamefont {B.}~\bibnamefont
  {{Flaugher}}}, \bibinfo {author} {\bibfnamefont {P.}~\bibnamefont
  {{Fosalba}}}, \bibinfo {author} {\bibfnamefont {J.}~\bibnamefont
  {{Frieman}}}, \bibinfo {author} {\bibfnamefont {D.~W.}\ \bibnamefont
  {{Gerdes}}}, \bibinfo {author} {\bibfnamefont {D.~A.}\ \bibnamefont
  {{Goldstein}}}, \bibinfo {author} {\bibfnamefont {D.}~\bibnamefont
  {{Gruen}}}, \bibinfo {author} {\bibfnamefont {R.~A.}\ \bibnamefont
  {{Gruendl}}}, \bibinfo {author} {\bibfnamefont {K.}~\bibnamefont
  {{Honscheid}}}, \bibinfo {author} {\bibfnamefont {D.~J.}\ \bibnamefont
  {{James}}}, \bibinfo {author} {\bibfnamefont {S.}~\bibnamefont {{Kent}}},
  \bibinfo {author} {\bibfnamefont {K.}~\bibnamefont {{Kuehn}}}, \bibinfo
  {author} {\bibfnamefont {N.}~\bibnamefont {{Kuropatkin}}}, \bibinfo {author}
  {\bibfnamefont {O.}~\bibnamefont {{Lahav}}}, \bibinfo {author} {\bibfnamefont
  {T.~S.}\ \bibnamefont {{Li}}}, \bibinfo {author} {\bibfnamefont {M.~A.~G.}\
  \bibnamefont {{Maia}}}, \bibinfo {author} {\bibfnamefont {M.}~\bibnamefont
  {{March}}}, \bibinfo {author} {\bibfnamefont {J.~L.}\ \bibnamefont
  {{Marshall}}}, \bibinfo {author} {\bibfnamefont {P.}~\bibnamefont
  {{Martini}}}, \bibinfo {author} {\bibfnamefont {C.~J.}\ \bibnamefont
  {{Miller}}}, \bibinfo {author} {\bibfnamefont {R.}~\bibnamefont {{Miquel}}},
  \bibinfo {author} {\bibfnamefont {E.}~\bibnamefont {{Neilsen}}}, \bibinfo
  {author} {\bibfnamefont {B.}~\bibnamefont {{Nord}}}, \bibinfo {author}
  {\bibfnamefont {R.}~\bibnamefont {{Ogando}}}, \bibinfo {author}
  {\bibfnamefont {A.~A.}\ \bibnamefont {{Plazas}}}, \bibinfo {author}
  {\bibfnamefont {K.}~\bibnamefont {{Reil}}}, \bibinfo {author} {\bibfnamefont
  {A.~K.}\ \bibnamefont {{Romer}}}, \bibinfo {author} {\bibfnamefont {E.~S.}\
  \bibnamefont {{Rykoff}}}, \bibinfo {author} {\bibfnamefont {E.}~\bibnamefont
  {{Sanchez}}}, \bibinfo {author} {\bibfnamefont {B.}~\bibnamefont
  {{Santiago}}}, \bibinfo {author} {\bibfnamefont {M.}~\bibnamefont
  {{Schubnell}}}, \bibinfo {author} {\bibfnamefont {I.}~\bibnamefont
  {{Sevilla-Noarbe}}}, \bibinfo {author} {\bibfnamefont {R.~C.}\ \bibnamefont
  {{Smith}}}, \bibinfo {author} {\bibfnamefont {M.}~\bibnamefont
  {{Soares-Santos}}}, \bibinfo {author} {\bibfnamefont {F.}~\bibnamefont
  {{Sobreira}}}, \bibinfo {author} {\bibfnamefont {E.}~\bibnamefont
  {{Suchyta}}}, \bibinfo {author} {\bibfnamefont {M.~E.~C.}\ \bibnamefont
  {{Swanson}}}, \bibinfo {author} {\bibfnamefont {G.}~\bibnamefont {{Tarle}}},
  \bibinfo {author} {\bibfnamefont {V.}~\bibnamefont {{Vikram}}}, \bibinfo
  {author} {\bibfnamefont {A.~R.}\ \bibnamefont {{Walker}}}, \bibinfo {author}
  {\bibfnamefont {R.~H.}\ \bibnamefont {{Wechsler}}}, \bibinfo {author}
  {\bibnamefont {{Fermi-LAT Collaboration}}},\ and\ \bibinfo {author}
  {\bibnamefont {{DES Collaboration}}},\ }\href
  {https://doi.org/10.3847/1538-4357/834/2/110} {\bibfield  {journal} {\bibinfo
   {journal} {\apj}\ }\textbf {\bibinfo {volume} {834}},\ \bibinfo {eid} {110}
  (\bibinfo {year} {2017})},\ \Eprint {https://arxiv.org/abs/1611.03184}
  {arXiv:1611.03184 [astro-ph.HE]} \BibitemShut {NoStop}%
\bibitem [{\citenamefont {Abdalla}\ \emph {et~al.}(2018)\citenamefont {Abdalla}
  \emph {et~al.}}]{dm_hess}%
  \BibitemOpen
  \bibfield  {author} {\bibinfo {author} {\bibfnamefont {H.}~\bibnamefont
  {Abdalla}} \emph {et~al.},\ }\href
  {https://doi.org/10.1088/1475-7516/2018/11/037} {\bibfield  {journal}
  {\bibinfo  {journal} {JCAP}\ }\textbf {\bibinfo {volume} {2018}}\bibfield
  {number} {\bibinfo  {number} { (11)},\ \bibinfo {eid} {037}},\ }\Eprint
  {https://arxiv.org/abs/1810.00995} {arXiv:1810.00995 [astro-ph.HE]}
  \BibitemShut {NoStop}%
\bibitem [{\citenamefont {Abdallah}\ \emph {et~al.}(2018)\citenamefont
  {Abdallah}, \citenamefont {Abramowski}, \citenamefont {Aharonian},
  \citenamefont {Ait~Benkhali}, \citenamefont {Ang\"uner}, \citenamefont
  {Arakawa}, \citenamefont {Arrieta}, \citenamefont {Aubert}, \citenamefont
  {Backes}, \citenamefont {Balzer}, \citenamefont {Barnard}, \citenamefont
  {Becherini}, \citenamefont {Becker~Tjus}, \citenamefont {Berge},
  \citenamefont {Bernhard}, \citenamefont {Bernl\"ohr}, \citenamefont
  {Blackwell}, \citenamefont {B\"ottcher}, \citenamefont {Boisson},
  \citenamefont {Bolmont}, \citenamefont {Bonnefoy}, \citenamefont {Bordas},
  \citenamefont {Bregeon}, \citenamefont {Brun}, \citenamefont {Brun},
  \citenamefont {Bryan}, \citenamefont {B\"uchele}, \citenamefont {Bulik},
  \citenamefont {Capasso}, \citenamefont {Caroff}, \citenamefont {Carosi},
  \citenamefont {Carr}, \citenamefont {Casanova}, \citenamefont {Cerruti},
  \citenamefont {Chakraborty}, \citenamefont {Chaves}, \citenamefont {Chen},
  \citenamefont {Chevalier}, \citenamefont {Colafrancesco}, \citenamefont
  {Condon}, \citenamefont {Conrad}, \citenamefont {Davids}, \citenamefont
  {Decock}, \citenamefont {Deil}, \citenamefont {Devin}, \citenamefont
  {deWilt}, \citenamefont {Dirson}, \citenamefont {Djannati-Ata\"{\i}},
  \citenamefont {Domainko}, \citenamefont {Donath}, \citenamefont {Drury},
  \citenamefont {Dutson}, \citenamefont {Dyks}, \citenamefont {Edwards},
  \citenamefont {Egberts}, \citenamefont {Eger}, \citenamefont {Emery},
  \citenamefont {Ernenwein}, \citenamefont {Eschbach}, \citenamefont {Farnier},
  \citenamefont {Fegan}, \citenamefont {Fernandes}, \citenamefont {Fiasson},
  \citenamefont {Fontaine}, \citenamefont {F\"orster}, \citenamefont {Funk},
  \citenamefont {F\"u\ss{}ling}, \citenamefont {Gabici}, \citenamefont
  {Gallant}, \citenamefont {Garrigoux}, \citenamefont {Gat\'e}, \citenamefont
  {Giavitto}, \citenamefont {Giebels}, \citenamefont {Glawion}, \citenamefont
  {Glicenstein}, \citenamefont {Gottschall}, \citenamefont {Grondin},
  \citenamefont {Hahn}, \citenamefont {Haupt}, \citenamefont {Hawkes},
  \citenamefont {Heinzelmann}, \citenamefont {Henri}, \citenamefont {Hermann},
  \citenamefont {Hinton}, \citenamefont {Hofmann}, \citenamefont {Hoischen},
  \citenamefont {Holch}, \citenamefont {Holler}, \citenamefont {Horns},
  \citenamefont {Ivascenko}, \citenamefont {Iwasaki}, \citenamefont
  {Jacholkowska}, \citenamefont {Jamrozy}, \citenamefont {Janiak},
  \citenamefont {Jankowsky}, \citenamefont {Jankowsky}, \citenamefont {Jingo},
  \citenamefont {Jouvin}, \citenamefont {Jung-Richardt}, \citenamefont
  {Kastendieck}, \citenamefont {Katarzy\ifmmode~\acute{n}\else \'{n}\fi{}ski},
  \citenamefont {Katsuragawa}, \citenamefont {Katz}, \citenamefont {Kerszberg},
  \citenamefont {Khangulyan}, \citenamefont {Kh\'elifi}, \citenamefont {King},
  \citenamefont {Klepser}, \citenamefont {Klochkov}, \citenamefont
  {Klu\ifmmode~\acute{z}\else \'{z}\fi{}niak}, \citenamefont {Komin},
  \citenamefont {Kosack}, \citenamefont {Krakau}, \citenamefont {Kraus},
  \citenamefont {Kr\"uger}, \citenamefont {Laffon}, \citenamefont {Lamanna},
  \citenamefont {Lau}, \citenamefont {Lees}, \citenamefont {Lefaucheur},
  \citenamefont {Lemi\`ere}, \citenamefont {Lemoine-Goumard}, \citenamefont
  {Lenain}, \citenamefont {Leser}, \citenamefont {Liu}, \citenamefont {Lohse},
  \citenamefont {Lorentz}, \citenamefont {L\'opez-Coto}, \citenamefont
  {Lypova}, \citenamefont {Malyshev}, \citenamefont {Marandon}, \citenamefont
  {Marcowith}, \citenamefont {Mariaud}, \citenamefont {Marx}, \citenamefont
  {Maurin}, \citenamefont {Maxted}, \citenamefont {Mayer}, \citenamefont
  {Meintjes}, \citenamefont {Meyer}, \citenamefont {Mitchell}, \citenamefont
  {Moderski}, \citenamefont {Mohamed}, \citenamefont {Mohrmann}, \citenamefont
  {Mor\aa{}}, \citenamefont {Moulin}, \citenamefont {Murach}, \citenamefont
  {Nakashima}, \citenamefont {de~Naurois}, \citenamefont {Ndiyavala},
  \citenamefont {Niederwanger}, \citenamefont {Niemiec}, \citenamefont {Oakes},
  \citenamefont {O'Brien}, \citenamefont {Odaka}, \citenamefont {Ohm},
  \citenamefont {Ostrowski}, \citenamefont {Oya}, \citenamefont {Padovani},
  \citenamefont {Panter}, \citenamefont {Parsons}, \citenamefont {Pekeur},
  \citenamefont {Pelletier}, \citenamefont {Perennes}, \citenamefont
  {Petrucci}, \citenamefont {Peyaud}, \citenamefont {Piel}, \citenamefont
  {Pita}, \citenamefont {Poireau}, \citenamefont {Poon}, \citenamefont
  {Prokhorov}, \citenamefont {Prokoph}, \citenamefont {P\"uhlhofer},
  \citenamefont {Punch}, \citenamefont {Quirrenbach}, \citenamefont {Raab},
  \citenamefont {Rauth}, \citenamefont {Reimer}, \citenamefont {Reimer},
  \citenamefont {Renaud}, \citenamefont {de~los Reyes}, \citenamefont {Rieger},
  \citenamefont {Rinchiuso}, \citenamefont {Romoli}, \citenamefont {Rowell},
  \citenamefont {Rudak}, \citenamefont {Rulten}, \citenamefont {Sahakian},
  \citenamefont {Saito}, \citenamefont {Sanchez}, \citenamefont {Santangelo},
  \citenamefont {Sasaki}, \citenamefont {Schandri}, \citenamefont
  {Schlickeiser}, \citenamefont {Sch\"ussler}, \citenamefont {Schulz},
  \citenamefont {Schwanke}, \citenamefont {Schwemmer}, \citenamefont
  {Seglar-Arroyo}, \citenamefont {Settimo}, \citenamefont {Seyffert},
  \citenamefont {Shafi}, \citenamefont {Shilon}, \citenamefont {Shiningayamwe},
  \citenamefont {Simoni}, \citenamefont {Sol}, \citenamefont {Spanier},
  \citenamefont {Spir-Jacob}, \citenamefont {Stawarz}, \citenamefont
  {Steenkamp}, \citenamefont {Stegmann}, \citenamefont {Steppa}, \citenamefont
  {Sushch}, \citenamefont {Takahashi}, \citenamefont {Tavernet}, \citenamefont
  {Tavernier}, \citenamefont {Taylor}, \citenamefont {Terrier}, \citenamefont
  {Tibaldo}, \citenamefont {Tiziani}, \citenamefont {Tluczykont}, \citenamefont
  {Trichard}, \citenamefont {Tsirou}, \citenamefont {Tsuji}, \citenamefont
  {Tuffs}, \citenamefont {Uchiyama}, \citenamefont {van~der Walt},
  \citenamefont {van Eldik}, \citenamefont {van Rensburg}, \citenamefont {van
  Soelen}, \citenamefont {Vasileiadis}, \citenamefont {Veh}, \citenamefont
  {Venter}, \citenamefont {Viana}, \citenamefont {Vincent}, \citenamefont
  {Vink}, \citenamefont {Voisin}, \citenamefont {V\"olk}, \citenamefont
  {Vuillaume}, \citenamefont {Wadiasingh}, \citenamefont {Wagner},
  \citenamefont {Wagner}, \citenamefont {Wagner}, \citenamefont {White},
  \citenamefont {Wierzcholska}, \citenamefont {Willmann}, \citenamefont
  {W\"ornlein}, \citenamefont {Wouters}, \citenamefont {Yang}, \citenamefont
  {Zaborov}, \citenamefont {Zacharias}, \citenamefont {Zanin}, \citenamefont
  {Zdziarski}, \citenamefont {Zech}, \citenamefont {Zefi}, \citenamefont
  {Ziegler}, \citenamefont {Zorn},\ and\ \citenamefont {\ifmmode~\dot{Z}\else
  \.{Z}\fi{}ywucka}}]{dm_hessGCline}%
  \BibitemOpen
  \bibfield  {author} {\bibinfo {author} {\bibfnamefont {H.}~\bibnamefont
  {Abdallah}}, \bibinfo {author} {\bibfnamefont {A.}~\bibnamefont
  {Abramowski}}, \bibinfo {author} {\bibfnamefont {F.}~\bibnamefont
  {Aharonian}}, \bibinfo {author} {\bibfnamefont {F.}~\bibnamefont
  {Ait~Benkhali}}, \bibinfo {author} {\bibfnamefont {E.~O.}\ \bibnamefont
  {Ang\"uner}}, \bibinfo {author} {\bibfnamefont {M.}~\bibnamefont {Arakawa}},
  \bibinfo {author} {\bibfnamefont {M.}~\bibnamefont {Arrieta}}, \bibinfo
  {author} {\bibfnamefont {P.}~\bibnamefont {Aubert}}, \bibinfo {author}
  {\bibfnamefont {M.}~\bibnamefont {Backes}}, \bibinfo {author} {\bibfnamefont
  {A.}~\bibnamefont {Balzer}}, \bibinfo {author} {\bibfnamefont
  {M.}~\bibnamefont {Barnard}}, \bibinfo {author} {\bibfnamefont
  {Y.}~\bibnamefont {Becherini}}, \bibinfo {author} {\bibfnamefont
  {J.}~\bibnamefont {Becker~Tjus}}, \bibinfo {author} {\bibfnamefont
  {D.}~\bibnamefont {Berge}}, \bibinfo {author} {\bibfnamefont
  {S.}~\bibnamefont {Bernhard}}, \bibinfo {author} {\bibfnamefont
  {K.}~\bibnamefont {Bernl\"ohr}}, \bibinfo {author} {\bibfnamefont
  {R.}~\bibnamefont {Blackwell}}, \bibinfo {author} {\bibfnamefont
  {M.}~\bibnamefont {B\"ottcher}}, \bibinfo {author} {\bibfnamefont
  {C.}~\bibnamefont {Boisson}}, \bibinfo {author} {\bibfnamefont
  {J.}~\bibnamefont {Bolmont}}, \bibinfo {author} {\bibfnamefont
  {S.}~\bibnamefont {Bonnefoy}}, \bibinfo {author} {\bibfnamefont
  {P.}~\bibnamefont {Bordas}}, \bibinfo {author} {\bibfnamefont
  {J.}~\bibnamefont {Bregeon}}, \bibinfo {author} {\bibfnamefont
  {F.}~\bibnamefont {Brun}}, \bibinfo {author} {\bibfnamefont {P.}~\bibnamefont
  {Brun}}, \bibinfo {author} {\bibfnamefont {M.}~\bibnamefont {Bryan}},
  \bibinfo {author} {\bibfnamefont {M.}~\bibnamefont {B\"uchele}}, \bibinfo
  {author} {\bibfnamefont {T.}~\bibnamefont {Bulik}}, \bibinfo {author}
  {\bibfnamefont {M.}~\bibnamefont {Capasso}}, \bibinfo {author} {\bibfnamefont
  {S.}~\bibnamefont {Caroff}}, \bibinfo {author} {\bibfnamefont
  {A.}~\bibnamefont {Carosi}}, \bibinfo {author} {\bibfnamefont
  {J.}~\bibnamefont {Carr}}, \bibinfo {author} {\bibfnamefont {S.}~\bibnamefont
  {Casanova}}, \bibinfo {author} {\bibfnamefont {M.}~\bibnamefont {Cerruti}},
  \bibinfo {author} {\bibfnamefont {N.}~\bibnamefont {Chakraborty}}, \bibinfo
  {author} {\bibfnamefont {R.~C.~G.}\ \bibnamefont {Chaves}}, \bibinfo {author}
  {\bibfnamefont {A.}~\bibnamefont {Chen}}, \bibinfo {author} {\bibfnamefont
  {J.}~\bibnamefont {Chevalier}}, \bibinfo {author} {\bibfnamefont
  {S.}~\bibnamefont {Colafrancesco}}, \bibinfo {author} {\bibfnamefont
  {B.}~\bibnamefont {Condon}}, \bibinfo {author} {\bibfnamefont
  {J.}~\bibnamefont {Conrad}}, \bibinfo {author} {\bibfnamefont {I.~D.}\
  \bibnamefont {Davids}}, \bibinfo {author} {\bibfnamefont {J.}~\bibnamefont
  {Decock}}, \bibinfo {author} {\bibfnamefont {C.}~\bibnamefont {Deil}},
  \bibinfo {author} {\bibfnamefont {J.}~\bibnamefont {Devin}}, \bibinfo
  {author} {\bibfnamefont {P.}~\bibnamefont {deWilt}}, \bibinfo {author}
  {\bibfnamefont {L.}~\bibnamefont {Dirson}}, \bibinfo {author} {\bibfnamefont
  {A.}~\bibnamefont {Djannati-Ata\"{\i}}}, \bibinfo {author} {\bibfnamefont
  {W.}~\bibnamefont {Domainko}}, \bibinfo {author} {\bibfnamefont
  {A.}~\bibnamefont {Donath}}, \bibinfo {author} {\bibfnamefont {L.~O.}\
  \bibnamefont {Drury}}, \bibinfo {author} {\bibfnamefont {K.}~\bibnamefont
  {Dutson}}, \bibinfo {author} {\bibfnamefont {J.}~\bibnamefont {Dyks}},
  \bibinfo {author} {\bibfnamefont {T.}~\bibnamefont {Edwards}}, \bibinfo
  {author} {\bibfnamefont {K.}~\bibnamefont {Egberts}}, \bibinfo {author}
  {\bibfnamefont {P.}~\bibnamefont {Eger}}, \bibinfo {author} {\bibfnamefont
  {G.}~\bibnamefont {Emery}}, \bibinfo {author} {\bibfnamefont {J.-P.}\
  \bibnamefont {Ernenwein}}, \bibinfo {author} {\bibfnamefont {S.}~\bibnamefont
  {Eschbach}}, \bibinfo {author} {\bibfnamefont {C.}~\bibnamefont {Farnier}},
  \bibinfo {author} {\bibfnamefont {S.}~\bibnamefont {Fegan}}, \bibinfo
  {author} {\bibfnamefont {M.~V.}\ \bibnamefont {Fernandes}}, \bibinfo {author}
  {\bibfnamefont {A.}~\bibnamefont {Fiasson}}, \bibinfo {author} {\bibfnamefont
  {G.}~\bibnamefont {Fontaine}}, \bibinfo {author} {\bibfnamefont
  {A.}~\bibnamefont {F\"orster}}, \bibinfo {author} {\bibfnamefont
  {S.}~\bibnamefont {Funk}}, \bibinfo {author} {\bibfnamefont {M.}~\bibnamefont
  {F\"u\ss{}ling}}, \bibinfo {author} {\bibfnamefont {S.}~\bibnamefont
  {Gabici}}, \bibinfo {author} {\bibfnamefont {Y.~A.}\ \bibnamefont {Gallant}},
  \bibinfo {author} {\bibfnamefont {T.}~\bibnamefont {Garrigoux}}, \bibinfo
  {author} {\bibfnamefont {F.}~\bibnamefont {Gat\'e}}, \bibinfo {author}
  {\bibfnamefont {G.}~\bibnamefont {Giavitto}}, \bibinfo {author}
  {\bibfnamefont {B.}~\bibnamefont {Giebels}}, \bibinfo {author} {\bibfnamefont
  {D.}~\bibnamefont {Glawion}}, \bibinfo {author} {\bibfnamefont {J.~F.}\
  \bibnamefont {Glicenstein}}, \bibinfo {author} {\bibfnamefont
  {D.}~\bibnamefont {Gottschall}}, \bibinfo {author} {\bibfnamefont {M.-H.}\
  \bibnamefont {Grondin}}, \bibinfo {author} {\bibfnamefont {J.}~\bibnamefont
  {Hahn}}, \bibinfo {author} {\bibfnamefont {M.}~\bibnamefont {Haupt}},
  \bibinfo {author} {\bibfnamefont {J.}~\bibnamefont {Hawkes}}, \bibinfo
  {author} {\bibfnamefont {G.}~\bibnamefont {Heinzelmann}}, \bibinfo {author}
  {\bibfnamefont {G.}~\bibnamefont {Henri}}, \bibinfo {author} {\bibfnamefont
  {G.}~\bibnamefont {Hermann}}, \bibinfo {author} {\bibfnamefont {J.~A.}\
  \bibnamefont {Hinton}}, \bibinfo {author} {\bibfnamefont {W.}~\bibnamefont
  {Hofmann}}, \bibinfo {author} {\bibfnamefont {C.}~\bibnamefont {Hoischen}},
  \bibinfo {author} {\bibfnamefont {T.~L.}\ \bibnamefont {Holch}}, \bibinfo
  {author} {\bibfnamefont {M.}~\bibnamefont {Holler}}, \bibinfo {author}
  {\bibfnamefont {D.}~\bibnamefont {Horns}}, \bibinfo {author} {\bibfnamefont
  {A.}~\bibnamefont {Ivascenko}}, \bibinfo {author} {\bibfnamefont
  {H.}~\bibnamefont {Iwasaki}}, \bibinfo {author} {\bibfnamefont
  {A.}~\bibnamefont {Jacholkowska}}, \bibinfo {author} {\bibfnamefont
  {M.}~\bibnamefont {Jamrozy}}, \bibinfo {author} {\bibfnamefont
  {M.}~\bibnamefont {Janiak}}, \bibinfo {author} {\bibfnamefont
  {D.}~\bibnamefont {Jankowsky}}, \bibinfo {author} {\bibfnamefont
  {F.}~\bibnamefont {Jankowsky}}, \bibinfo {author} {\bibfnamefont
  {M.}~\bibnamefont {Jingo}}, \bibinfo {author} {\bibfnamefont
  {L.}~\bibnamefont {Jouvin}}, \bibinfo {author} {\bibfnamefont
  {I.}~\bibnamefont {Jung-Richardt}}, \bibinfo {author} {\bibfnamefont {M.~A.}\
  \bibnamefont {Kastendieck}}, \bibinfo {author} {\bibfnamefont
  {K.}~\bibnamefont {Katarzy\ifmmode~\acute{n}\else \'{n}\fi{}ski}}, \bibinfo
  {author} {\bibfnamefont {M.}~\bibnamefont {Katsuragawa}}, \bibinfo {author}
  {\bibfnamefont {U.}~\bibnamefont {Katz}}, \bibinfo {author} {\bibfnamefont
  {D.}~\bibnamefont {Kerszberg}}, \bibinfo {author} {\bibfnamefont
  {D.}~\bibnamefont {Khangulyan}}, \bibinfo {author} {\bibfnamefont
  {B.}~\bibnamefont {Kh\'elifi}}, \bibinfo {author} {\bibfnamefont
  {J.}~\bibnamefont {King}}, \bibinfo {author} {\bibfnamefont {S.}~\bibnamefont
  {Klepser}}, \bibinfo {author} {\bibfnamefont {D.}~\bibnamefont {Klochkov}},
  \bibinfo {author} {\bibfnamefont {W.}~\bibnamefont
  {Klu\ifmmode~\acute{z}\else \'{z}\fi{}niak}}, \bibinfo {author}
  {\bibfnamefont {N.}~\bibnamefont {Komin}}, \bibinfo {author} {\bibfnamefont
  {K.}~\bibnamefont {Kosack}}, \bibinfo {author} {\bibfnamefont
  {S.}~\bibnamefont {Krakau}}, \bibinfo {author} {\bibfnamefont
  {M.}~\bibnamefont {Kraus}}, \bibinfo {author} {\bibfnamefont {P.~P.}\
  \bibnamefont {Kr\"uger}}, \bibinfo {author} {\bibfnamefont {H.}~\bibnamefont
  {Laffon}}, \bibinfo {author} {\bibfnamefont {G.}~\bibnamefont {Lamanna}},
  \bibinfo {author} {\bibfnamefont {J.}~\bibnamefont {Lau}}, \bibinfo {author}
  {\bibfnamefont {J.-P.}\ \bibnamefont {Lees}}, \bibinfo {author}
  {\bibfnamefont {J.}~\bibnamefont {Lefaucheur}}, \bibinfo {author}
  {\bibfnamefont {A.}~\bibnamefont {Lemi\`ere}}, \bibinfo {author}
  {\bibfnamefont {M.}~\bibnamefont {Lemoine-Goumard}}, \bibinfo {author}
  {\bibfnamefont {J.-P.}\ \bibnamefont {Lenain}}, \bibinfo {author}
  {\bibfnamefont {E.}~\bibnamefont {Leser}}, \bibinfo {author} {\bibfnamefont
  {R.}~\bibnamefont {Liu}}, \bibinfo {author} {\bibfnamefont {T.}~\bibnamefont
  {Lohse}}, \bibinfo {author} {\bibfnamefont {M.}~\bibnamefont {Lorentz}},
  \bibinfo {author} {\bibfnamefont {R.}~\bibnamefont {L\'opez-Coto}}, \bibinfo
  {author} {\bibfnamefont {I.}~\bibnamefont {Lypova}}, \bibinfo {author}
  {\bibfnamefont {D.}~\bibnamefont {Malyshev}}, \bibinfo {author}
  {\bibfnamefont {V.}~\bibnamefont {Marandon}}, \bibinfo {author}
  {\bibfnamefont {A.}~\bibnamefont {Marcowith}}, \bibinfo {author}
  {\bibfnamefont {C.}~\bibnamefont {Mariaud}}, \bibinfo {author} {\bibfnamefont
  {R.}~\bibnamefont {Marx}}, \bibinfo {author} {\bibfnamefont {G.}~\bibnamefont
  {Maurin}}, \bibinfo {author} {\bibfnamefont {N.}~\bibnamefont {Maxted}},
  \bibinfo {author} {\bibfnamefont {M.}~\bibnamefont {Mayer}}, \bibinfo
  {author} {\bibfnamefont {P.~J.}\ \bibnamefont {Meintjes}}, \bibinfo {author}
  {\bibfnamefont {M.}~\bibnamefont {Meyer}}, \bibinfo {author} {\bibfnamefont
  {A.~M.~W.}\ \bibnamefont {Mitchell}}, \bibinfo {author} {\bibfnamefont
  {R.}~\bibnamefont {Moderski}}, \bibinfo {author} {\bibfnamefont
  {M.}~\bibnamefont {Mohamed}}, \bibinfo {author} {\bibfnamefont
  {L.}~\bibnamefont {Mohrmann}}, \bibinfo {author} {\bibfnamefont
  {K.}~\bibnamefont {Mor\aa{}}}, \bibinfo {author} {\bibfnamefont
  {E.}~\bibnamefont {Moulin}}, \bibinfo {author} {\bibfnamefont
  {T.}~\bibnamefont {Murach}}, \bibinfo {author} {\bibfnamefont
  {S.}~\bibnamefont {Nakashima}}, \bibinfo {author} {\bibfnamefont
  {M.}~\bibnamefont {de~Naurois}}, \bibinfo {author} {\bibfnamefont
  {H.}~\bibnamefont {Ndiyavala}}, \bibinfo {author} {\bibfnamefont
  {F.}~\bibnamefont {Niederwanger}}, \bibinfo {author} {\bibfnamefont
  {J.}~\bibnamefont {Niemiec}}, \bibinfo {author} {\bibfnamefont
  {L.}~\bibnamefont {Oakes}}, \bibinfo {author} {\bibfnamefont
  {P.}~\bibnamefont {O'Brien}}, \bibinfo {author} {\bibfnamefont
  {H.}~\bibnamefont {Odaka}}, \bibinfo {author} {\bibfnamefont
  {S.}~\bibnamefont {Ohm}}, \bibinfo {author} {\bibfnamefont {M.}~\bibnamefont
  {Ostrowski}}, \bibinfo {author} {\bibfnamefont {I.}~\bibnamefont {Oya}},
  \bibinfo {author} {\bibfnamefont {M.}~\bibnamefont {Padovani}}, \bibinfo
  {author} {\bibfnamefont {M.}~\bibnamefont {Panter}}, \bibinfo {author}
  {\bibfnamefont {R.~D.}\ \bibnamefont {Parsons}}, \bibinfo {author}
  {\bibfnamefont {N.~W.}\ \bibnamefont {Pekeur}}, \bibinfo {author}
  {\bibfnamefont {G.}~\bibnamefont {Pelletier}}, \bibinfo {author}
  {\bibfnamefont {C.}~\bibnamefont {Perennes}}, \bibinfo {author}
  {\bibfnamefont {P.-O.}\ \bibnamefont {Petrucci}}, \bibinfo {author}
  {\bibfnamefont {B.}~\bibnamefont {Peyaud}}, \bibinfo {author} {\bibfnamefont
  {Q.}~\bibnamefont {Piel}}, \bibinfo {author} {\bibfnamefont {S.}~\bibnamefont
  {Pita}}, \bibinfo {author} {\bibfnamefont {V.}~\bibnamefont {Poireau}},
  \bibinfo {author} {\bibfnamefont {H.}~\bibnamefont {Poon}}, \bibinfo {author}
  {\bibfnamefont {D.}~\bibnamefont {Prokhorov}}, \bibinfo {author}
  {\bibfnamefont {H.}~\bibnamefont {Prokoph}}, \bibinfo {author} {\bibfnamefont
  {G.}~\bibnamefont {P\"uhlhofer}}, \bibinfo {author} {\bibfnamefont
  {M.}~\bibnamefont {Punch}}, \bibinfo {author} {\bibfnamefont
  {A.}~\bibnamefont {Quirrenbach}}, \bibinfo {author} {\bibfnamefont
  {S.}~\bibnamefont {Raab}}, \bibinfo {author} {\bibfnamefont {R.}~\bibnamefont
  {Rauth}}, \bibinfo {author} {\bibfnamefont {A.}~\bibnamefont {Reimer}},
  \bibinfo {author} {\bibfnamefont {O.}~\bibnamefont {Reimer}}, \bibinfo
  {author} {\bibfnamefont {M.}~\bibnamefont {Renaud}}, \bibinfo {author}
  {\bibfnamefont {R.}~\bibnamefont {de~los Reyes}}, \bibinfo {author}
  {\bibfnamefont {F.}~\bibnamefont {Rieger}}, \bibinfo {author} {\bibfnamefont
  {L.}~\bibnamefont {Rinchiuso}}, \bibinfo {author} {\bibfnamefont
  {C.}~\bibnamefont {Romoli}}, \bibinfo {author} {\bibfnamefont
  {G.}~\bibnamefont {Rowell}}, \bibinfo {author} {\bibfnamefont
  {B.}~\bibnamefont {Rudak}}, \bibinfo {author} {\bibfnamefont {C.~B.}\
  \bibnamefont {Rulten}}, \bibinfo {author} {\bibfnamefont {V.}~\bibnamefont
  {Sahakian}}, \bibinfo {author} {\bibfnamefont {S.}~\bibnamefont {Saito}},
  \bibinfo {author} {\bibfnamefont {D.~A.}\ \bibnamefont {Sanchez}}, \bibinfo
  {author} {\bibfnamefont {A.}~\bibnamefont {Santangelo}}, \bibinfo {author}
  {\bibfnamefont {M.}~\bibnamefont {Sasaki}}, \bibinfo {author} {\bibfnamefont
  {M.}~\bibnamefont {Schandri}}, \bibinfo {author} {\bibfnamefont
  {R.}~\bibnamefont {Schlickeiser}}, \bibinfo {author} {\bibfnamefont
  {F.}~\bibnamefont {Sch\"ussler}}, \bibinfo {author} {\bibfnamefont
  {A.}~\bibnamefont {Schulz}}, \bibinfo {author} {\bibfnamefont
  {U.}~\bibnamefont {Schwanke}}, \bibinfo {author} {\bibfnamefont
  {S.}~\bibnamefont {Schwemmer}}, \bibinfo {author} {\bibfnamefont
  {M.}~\bibnamefont {Seglar-Arroyo}}, \bibinfo {author} {\bibfnamefont
  {M.}~\bibnamefont {Settimo}}, \bibinfo {author} {\bibfnamefont {A.~S.}\
  \bibnamefont {Seyffert}}, \bibinfo {author} {\bibfnamefont {N.}~\bibnamefont
  {Shafi}}, \bibinfo {author} {\bibfnamefont {I.}~\bibnamefont {Shilon}},
  \bibinfo {author} {\bibfnamefont {K.}~\bibnamefont {Shiningayamwe}}, \bibinfo
  {author} {\bibfnamefont {R.}~\bibnamefont {Simoni}}, \bibinfo {author}
  {\bibfnamefont {H.}~\bibnamefont {Sol}}, \bibinfo {author} {\bibfnamefont
  {F.}~\bibnamefont {Spanier}}, \bibinfo {author} {\bibfnamefont
  {M.}~\bibnamefont {Spir-Jacob}}, \bibinfo {author} {\bibfnamefont
  {L.}~\bibnamefont {Stawarz}}, \bibinfo {author} {\bibfnamefont
  {R.}~\bibnamefont {Steenkamp}}, \bibinfo {author} {\bibfnamefont
  {C.}~\bibnamefont {Stegmann}}, \bibinfo {author} {\bibfnamefont
  {C.}~\bibnamefont {Steppa}}, \bibinfo {author} {\bibfnamefont
  {I.}~\bibnamefont {Sushch}}, \bibinfo {author} {\bibfnamefont
  {T.}~\bibnamefont {Takahashi}}, \bibinfo {author} {\bibfnamefont {J.-P.}\
  \bibnamefont {Tavernet}}, \bibinfo {author} {\bibfnamefont {T.}~\bibnamefont
  {Tavernier}}, \bibinfo {author} {\bibfnamefont {A.~M.}\ \bibnamefont
  {Taylor}}, \bibinfo {author} {\bibfnamefont {R.}~\bibnamefont {Terrier}},
  \bibinfo {author} {\bibfnamefont {L.}~\bibnamefont {Tibaldo}}, \bibinfo
  {author} {\bibfnamefont {D.}~\bibnamefont {Tiziani}}, \bibinfo {author}
  {\bibfnamefont {M.}~\bibnamefont {Tluczykont}}, \bibinfo {author}
  {\bibfnamefont {C.}~\bibnamefont {Trichard}}, \bibinfo {author}
  {\bibfnamefont {M.}~\bibnamefont {Tsirou}}, \bibinfo {author} {\bibfnamefont
  {N.}~\bibnamefont {Tsuji}}, \bibinfo {author} {\bibfnamefont
  {R.}~\bibnamefont {Tuffs}}, \bibinfo {author} {\bibfnamefont
  {Y.}~\bibnamefont {Uchiyama}}, \bibinfo {author} {\bibfnamefont
  {J.}~\bibnamefont {van~der Walt}}, \bibinfo {author} {\bibfnamefont
  {C.}~\bibnamefont {van Eldik}}, \bibinfo {author} {\bibfnamefont
  {C.}~\bibnamefont {van Rensburg}}, \bibinfo {author} {\bibfnamefont
  {B.}~\bibnamefont {van Soelen}}, \bibinfo {author} {\bibfnamefont
  {G.}~\bibnamefont {Vasileiadis}}, \bibinfo {author} {\bibfnamefont
  {J.}~\bibnamefont {Veh}}, \bibinfo {author} {\bibfnamefont {C.}~\bibnamefont
  {Venter}}, \bibinfo {author} {\bibfnamefont {A.}~\bibnamefont {Viana}},
  \bibinfo {author} {\bibfnamefont {P.}~\bibnamefont {Vincent}}, \bibinfo
  {author} {\bibfnamefont {J.}~\bibnamefont {Vink}}, \bibinfo {author}
  {\bibfnamefont {F.}~\bibnamefont {Voisin}}, \bibinfo {author} {\bibfnamefont
  {H.~J.}\ \bibnamefont {V\"olk}}, \bibinfo {author} {\bibfnamefont
  {T.}~\bibnamefont {Vuillaume}}, \bibinfo {author} {\bibfnamefont
  {Z.}~\bibnamefont {Wadiasingh}}, \bibinfo {author} {\bibfnamefont {S.~J.}\
  \bibnamefont {Wagner}}, \bibinfo {author} {\bibfnamefont {P.}~\bibnamefont
  {Wagner}}, \bibinfo {author} {\bibfnamefont {R.~M.}\ \bibnamefont {Wagner}},
  \bibinfo {author} {\bibfnamefont {R.}~\bibnamefont {White}}, \bibinfo
  {author} {\bibfnamefont {A.}~\bibnamefont {Wierzcholska}}, \bibinfo {author}
  {\bibfnamefont {P.}~\bibnamefont {Willmann}}, \bibinfo {author}
  {\bibfnamefont {A.}~\bibnamefont {W\"ornlein}}, \bibinfo {author}
  {\bibfnamefont {D.}~\bibnamefont {Wouters}}, \bibinfo {author} {\bibfnamefont
  {R.}~\bibnamefont {Yang}}, \bibinfo {author} {\bibfnamefont {D.}~\bibnamefont
  {Zaborov}}, \bibinfo {author} {\bibfnamefont {M.}~\bibnamefont {Zacharias}},
  \bibinfo {author} {\bibfnamefont {R.}~\bibnamefont {Zanin}}, \bibinfo
  {author} {\bibfnamefont {A.~A.}\ \bibnamefont {Zdziarski}}, \bibinfo {author}
  {\bibfnamefont {A.}~\bibnamefont {Zech}}, \bibinfo {author} {\bibfnamefont
  {F.}~\bibnamefont {Zefi}}, \bibinfo {author} {\bibfnamefont {A.}~\bibnamefont
  {Ziegler}}, \bibinfo {author} {\bibfnamefont {J.}~\bibnamefont {Zorn}},\ and\
  \bibinfo {author} {\bibfnamefont {N.}~\bibnamefont {\ifmmode~\dot{Z}\else
  \.{Z}\fi{}ywucka}} (\bibinfo {collaboration} {H.E.S.S. Collaboration}),\
  }\href {https://doi.org/10.1103/PhysRevLett.120.201101} {\bibfield  {journal}
  {\bibinfo  {journal} {Phys. Rev. Lett.}\ }\textbf {\bibinfo {volume} {120}},\
  \bibinfo {pages} {201101} (\bibinfo {year} {2018})}\BibitemShut {NoStop}%
\bibitem [{\citenamefont {Abdallah}\ \emph {et~al.}(2020)\citenamefont
  {Abdallah} \emph {et~al.}}]{dm_hess2}%
  \BibitemOpen
  \bibfield  {author} {\bibinfo {author} {\bibfnamefont {H.}~\bibnamefont
  {Abdallah}} \emph {et~al.},\ }\href
  {https://doi.org/10.1103/PhysRevD.102.062001} {\bibfield  {journal} {\bibinfo
   {journal} {\prd}\ }\textbf {\bibinfo {volume} {102}},\ \bibinfo {eid}
  {062001} (\bibinfo {year} {2020})},\ \Eprint
  {https://arxiv.org/abs/2008.00688} {arXiv:2008.00688 [astro-ph.HE]}
  \BibitemShut {NoStop}%
\bibitem [{\citenamefont {Abdalla}\ \emph {et~al.}(2022)\citenamefont
  {Abdalla}, \citenamefont {Aharonian}, \citenamefont {Benkhali}, \citenamefont
  {Ang\"uner}, \citenamefont {Armand}, \citenamefont {Ashkar}, \citenamefont
  {Backes}, \citenamefont {Baghmanyan}, \citenamefont {Martins}, \citenamefont
  {Batzofin}, \citenamefont {Becherini}, \citenamefont {Berge}, \citenamefont
  {Bernl\"ohr}, \citenamefont {Bi}, \citenamefont {B\"ottcher}, \citenamefont
  {Bolmont}, \citenamefont {de~Lavergne}, \citenamefont {Brose}, \citenamefont
  {Brun}, \citenamefont {Cangemi}, \citenamefont {Caroff}, \citenamefont
  {Cerruti}, \citenamefont {Chand}, \citenamefont {Chen}, \citenamefont
  {Cotter}, \citenamefont {Mbarubucyeye}, \citenamefont {Devin}, \citenamefont
  {Djannati-Ata\"{\i}}, \citenamefont {Dmytriiev}, \citenamefont {Doroshenko},
  \citenamefont {Egberts}, \citenamefont {Fiasson}, \citenamefont
  {de~Clairfontaine}, \citenamefont {Fontaine}, \citenamefont {Funk},
  \citenamefont {Gabici}, \citenamefont {Giavitto}, \citenamefont {Glawion},
  \citenamefont {Glicenstein}, \citenamefont {Grondin}, \citenamefont {Hinton},
  \citenamefont {Hofmann}, \citenamefont {Holch}, \citenamefont {Holler},
  \citenamefont {Horns}, \citenamefont {Huang}, \citenamefont {Jamrozy},
  \citenamefont {Jankowsky}, \citenamefont {Kasai}, \citenamefont
  {Katarzy\ifmmode~\acute{n}\else \'{n}\fi{}ski}, \citenamefont {Katz},
  \citenamefont {Kh\'elifi}, \citenamefont {Klu\ifmmode~\acute{z}\else
  \'{z}\fi{}niak}, \citenamefont {Komin}, \citenamefont {Kosack}, \citenamefont
  {Kostunin}, \citenamefont {Lamanna}, \citenamefont {Lemoine-Goumard},
  \citenamefont {Lenain}, \citenamefont {Leuschner}, \citenamefont {Lohse},
  \citenamefont {Luashvili}, \citenamefont {Lypova}, \citenamefont {Mackey},
  \citenamefont {Malyshev}, \citenamefont {Malyshev}, \citenamefont {Marandon},
  \citenamefont {Marchegiani}, \citenamefont {Mart\'{\i}-Devesa}, \citenamefont
  {Marx}, \citenamefont {Maurin}, \citenamefont {Meyer}, \citenamefont
  {Mitchell}, \citenamefont {Moderski}, \citenamefont {Montanari},
  \citenamefont {Moulin}, \citenamefont {Muller}, \citenamefont {de~Naurois},
  \citenamefont {Niemiec}, \citenamefont {Noel}, \citenamefont {Ohm},
  \citenamefont {Olivera-Nieto}, \citenamefont {Wilhelmi}, \citenamefont
  {Ostrowski}, \citenamefont {Panny}, \citenamefont {Panter}, \citenamefont
  {Parsons}, \citenamefont {Peron}, \citenamefont {Poireau}, \citenamefont
  {Prokoph}, \citenamefont {P\"uhlhofer}, \citenamefont {Punch}, \citenamefont
  {Quirrenbach}, \citenamefont {Reichherzer}, \citenamefont {Reimer},
  \citenamefont {Reimer}, \citenamefont {Renaud}, \citenamefont {Rieger},
  \citenamefont {Rowell}, \citenamefont {Rudak}, \citenamefont {Ricarte},
  \citenamefont {Ruiz-Velasco}, \citenamefont {Sahakian}, \citenamefont
  {Salzmann}, \citenamefont {Santangelo}, \citenamefont {Sasaki}, \citenamefont
  {Sch\"ussler}, \citenamefont {Schutte}, \citenamefont {Schwanke},
  \citenamefont {Senniappan}, \citenamefont {Shapopi}, \citenamefont {Sol},
  \citenamefont {Specovius}, \citenamefont {Spencer}, \citenamefont {Stawarz},
  \citenamefont {Stegmann}, \citenamefont {Steinmassl}, \citenamefont {Steppa},
  \citenamefont {Takahashi}, \citenamefont {Tanaka}, \citenamefont {Terrier},
  \citenamefont {Thorpe-Morgan}, \citenamefont {Tluczykont}, \citenamefont
  {Tsirou}, \citenamefont {Tsuji}, \citenamefont {Uchiyama}, \citenamefont {van
  Eldik}, \citenamefont {Veh}, \citenamefont {Vink}, \citenamefont {Wagner},
  \citenamefont {White}, \citenamefont {Wierzcholska}, \citenamefont {Wong},
  \citenamefont {Zacharias}, \citenamefont {Zargaryan}, \citenamefont
  {Zdziarski}, \citenamefont {Zech}, \citenamefont {Zhu}, \citenamefont
  {Zouari},\ and\ \citenamefont {\ifmmode~\dot{Z}\else
  \.{Z}\fi{}ywucka}}]{dm_hessGC}%
  \BibitemOpen
  \bibfield  {author} {\bibinfo {author} {\bibfnamefont {H.}~\bibnamefont
  {Abdalla}}, \bibinfo {author} {\bibfnamefont {F.}~\bibnamefont {Aharonian}},
  \bibinfo {author} {\bibfnamefont {F.~A.}\ \bibnamefont {Benkhali}}, \bibinfo
  {author} {\bibfnamefont {E.~O.}\ \bibnamefont {Ang\"uner}}, \bibinfo {author}
  {\bibfnamefont {C.}~\bibnamefont {Armand}}, \bibinfo {author} {\bibfnamefont
  {H.}~\bibnamefont {Ashkar}}, \bibinfo {author} {\bibfnamefont
  {M.}~\bibnamefont {Backes}}, \bibinfo {author} {\bibfnamefont
  {V.}~\bibnamefont {Baghmanyan}}, \bibinfo {author} {\bibfnamefont {V.~B.}\
  \bibnamefont {Martins}}, \bibinfo {author} {\bibfnamefont {R.}~\bibnamefont
  {Batzofin}}, \bibinfo {author} {\bibfnamefont {Y.}~\bibnamefont {Becherini}},
  \bibinfo {author} {\bibfnamefont {D.}~\bibnamefont {Berge}}, \bibinfo
  {author} {\bibfnamefont {K.}~\bibnamefont {Bernl\"ohr}}, \bibinfo {author}
  {\bibfnamefont {B.}~\bibnamefont {Bi}}, \bibinfo {author} {\bibfnamefont
  {M.}~\bibnamefont {B\"ottcher}}, \bibinfo {author} {\bibfnamefont
  {J.}~\bibnamefont {Bolmont}}, \bibinfo {author} {\bibfnamefont {M.~d.~B.}\
  \bibnamefont {de~Lavergne}}, \bibinfo {author} {\bibfnamefont
  {R.}~\bibnamefont {Brose}}, \bibinfo {author} {\bibfnamefont
  {F.}~\bibnamefont {Brun}}, \bibinfo {author} {\bibfnamefont {F.}~\bibnamefont
  {Cangemi}}, \bibinfo {author} {\bibfnamefont {S.}~\bibnamefont {Caroff}},
  \bibinfo {author} {\bibfnamefont {M.}~\bibnamefont {Cerruti}}, \bibinfo
  {author} {\bibfnamefont {T.}~\bibnamefont {Chand}}, \bibinfo {author}
  {\bibfnamefont {A.}~\bibnamefont {Chen}}, \bibinfo {author} {\bibfnamefont
  {G.}~\bibnamefont {Cotter}}, \bibinfo {author} {\bibfnamefont {J.~D.}\
  \bibnamefont {Mbarubucyeye}}, \bibinfo {author} {\bibfnamefont
  {J.}~\bibnamefont {Devin}}, \bibinfo {author} {\bibfnamefont
  {A.}~\bibnamefont {Djannati-Ata\"{\i}}}, \bibinfo {author} {\bibfnamefont
  {A.}~\bibnamefont {Dmytriiev}}, \bibinfo {author} {\bibfnamefont
  {V.}~\bibnamefont {Doroshenko}}, \bibinfo {author} {\bibfnamefont
  {K.}~\bibnamefont {Egberts}}, \bibinfo {author} {\bibfnamefont
  {A.}~\bibnamefont {Fiasson}}, \bibinfo {author} {\bibfnamefont {G.~F.}\
  \bibnamefont {de~Clairfontaine}}, \bibinfo {author} {\bibfnamefont
  {G.}~\bibnamefont {Fontaine}}, \bibinfo {author} {\bibfnamefont
  {S.}~\bibnamefont {Funk}}, \bibinfo {author} {\bibfnamefont {S.}~\bibnamefont
  {Gabici}}, \bibinfo {author} {\bibfnamefont {G.}~\bibnamefont {Giavitto}},
  \bibinfo {author} {\bibfnamefont {D.}~\bibnamefont {Glawion}}, \bibinfo
  {author} {\bibfnamefont {J.~F.}\ \bibnamefont {Glicenstein}}, \bibinfo
  {author} {\bibfnamefont {M.-H.}\ \bibnamefont {Grondin}}, \bibinfo {author}
  {\bibfnamefont {J.~A.}\ \bibnamefont {Hinton}}, \bibinfo {author}
  {\bibfnamefont {W.}~\bibnamefont {Hofmann}}, \bibinfo {author} {\bibfnamefont
  {T.~L.}\ \bibnamefont {Holch}}, \bibinfo {author} {\bibfnamefont
  {M.}~\bibnamefont {Holler}}, \bibinfo {author} {\bibfnamefont
  {D.}~\bibnamefont {Horns}}, \bibinfo {author} {\bibfnamefont
  {Z.}~\bibnamefont {Huang}}, \bibinfo {author} {\bibfnamefont
  {M.}~\bibnamefont {Jamrozy}}, \bibinfo {author} {\bibfnamefont
  {F.}~\bibnamefont {Jankowsky}}, \bibinfo {author} {\bibfnamefont
  {E.}~\bibnamefont {Kasai}}, \bibinfo {author} {\bibfnamefont
  {K.}~\bibnamefont {Katarzy\ifmmode~\acute{n}\else \'{n}\fi{}ski}}, \bibinfo
  {author} {\bibfnamefont {U.}~\bibnamefont {Katz}}, \bibinfo {author}
  {\bibfnamefont {B.}~\bibnamefont {Kh\'elifi}}, \bibinfo {author}
  {\bibfnamefont {W.}~\bibnamefont {Klu\ifmmode~\acute{z}\else
  \'{z}\fi{}niak}}, \bibinfo {author} {\bibfnamefont {N.}~\bibnamefont
  {Komin}}, \bibinfo {author} {\bibfnamefont {K.}~\bibnamefont {Kosack}},
  \bibinfo {author} {\bibfnamefont {D.}~\bibnamefont {Kostunin}}, \bibinfo
  {author} {\bibfnamefont {G.}~\bibnamefont {Lamanna}}, \bibinfo {author}
  {\bibfnamefont {M.}~\bibnamefont {Lemoine-Goumard}}, \bibinfo {author}
  {\bibfnamefont {J.-P.}\ \bibnamefont {Lenain}}, \bibinfo {author}
  {\bibfnamefont {F.}~\bibnamefont {Leuschner}}, \bibinfo {author}
  {\bibfnamefont {T.}~\bibnamefont {Lohse}}, \bibinfo {author} {\bibfnamefont
  {A.}~\bibnamefont {Luashvili}}, \bibinfo {author} {\bibfnamefont
  {I.}~\bibnamefont {Lypova}}, \bibinfo {author} {\bibfnamefont
  {J.}~\bibnamefont {Mackey}}, \bibinfo {author} {\bibfnamefont
  {D.}~\bibnamefont {Malyshev}}, \bibinfo {author} {\bibfnamefont
  {D.}~\bibnamefont {Malyshev}}, \bibinfo {author} {\bibfnamefont
  {V.}~\bibnamefont {Marandon}}, \bibinfo {author} {\bibfnamefont
  {P.}~\bibnamefont {Marchegiani}}, \bibinfo {author} {\bibfnamefont
  {G.}~\bibnamefont {Mart\'{\i}-Devesa}}, \bibinfo {author} {\bibfnamefont
  {R.}~\bibnamefont {Marx}}, \bibinfo {author} {\bibfnamefont {G.}~\bibnamefont
  {Maurin}}, \bibinfo {author} {\bibfnamefont {M.}~\bibnamefont {Meyer}},
  \bibinfo {author} {\bibfnamefont {A.}~\bibnamefont {Mitchell}}, \bibinfo
  {author} {\bibfnamefont {R.}~\bibnamefont {Moderski}}, \bibinfo {author}
  {\bibfnamefont {A.}~\bibnamefont {Montanari}}, \bibinfo {author}
  {\bibfnamefont {E.}~\bibnamefont {Moulin}}, \bibinfo {author} {\bibfnamefont
  {J.}~\bibnamefont {Muller}}, \bibinfo {author} {\bibfnamefont
  {M.}~\bibnamefont {de~Naurois}}, \bibinfo {author} {\bibfnamefont
  {J.}~\bibnamefont {Niemiec}}, \bibinfo {author} {\bibfnamefont {A.~P.}\
  \bibnamefont {Noel}}, \bibinfo {author} {\bibfnamefont {S.}~\bibnamefont
  {Ohm}}, \bibinfo {author} {\bibfnamefont {L.}~\bibnamefont {Olivera-Nieto}},
  \bibinfo {author} {\bibfnamefont {E.~d.~O.}\ \bibnamefont {Wilhelmi}},
  \bibinfo {author} {\bibfnamefont {M.}~\bibnamefont {Ostrowski}}, \bibinfo
  {author} {\bibfnamefont {S.}~\bibnamefont {Panny}}, \bibinfo {author}
  {\bibfnamefont {M.}~\bibnamefont {Panter}}, \bibinfo {author} {\bibfnamefont
  {R.~D.}\ \bibnamefont {Parsons}}, \bibinfo {author} {\bibfnamefont
  {G.}~\bibnamefont {Peron}}, \bibinfo {author} {\bibfnamefont
  {V.}~\bibnamefont {Poireau}}, \bibinfo {author} {\bibfnamefont
  {H.}~\bibnamefont {Prokoph}}, \bibinfo {author} {\bibfnamefont
  {G.}~\bibnamefont {P\"uhlhofer}}, \bibinfo {author} {\bibfnamefont
  {M.}~\bibnamefont {Punch}}, \bibinfo {author} {\bibfnamefont
  {A.}~\bibnamefont {Quirrenbach}}, \bibinfo {author} {\bibfnamefont
  {P.}~\bibnamefont {Reichherzer}}, \bibinfo {author} {\bibfnamefont
  {A.}~\bibnamefont {Reimer}}, \bibinfo {author} {\bibfnamefont
  {O.}~\bibnamefont {Reimer}}, \bibinfo {author} {\bibfnamefont
  {M.}~\bibnamefont {Renaud}}, \bibinfo {author} {\bibfnamefont
  {F.}~\bibnamefont {Rieger}}, \bibinfo {author} {\bibfnamefont
  {G.}~\bibnamefont {Rowell}}, \bibinfo {author} {\bibfnamefont
  {B.}~\bibnamefont {Rudak}}, \bibinfo {author} {\bibfnamefont {H.~R.}\
  \bibnamefont {Ricarte}}, \bibinfo {author} {\bibfnamefont {E.}~\bibnamefont
  {Ruiz-Velasco}}, \bibinfo {author} {\bibfnamefont {V.}~\bibnamefont
  {Sahakian}}, \bibinfo {author} {\bibfnamefont {H.}~\bibnamefont {Salzmann}},
  \bibinfo {author} {\bibfnamefont {A.}~\bibnamefont {Santangelo}}, \bibinfo
  {author} {\bibfnamefont {M.}~\bibnamefont {Sasaki}}, \bibinfo {author}
  {\bibfnamefont {F.}~\bibnamefont {Sch\"ussler}}, \bibinfo {author}
  {\bibfnamefont {H.~M.}\ \bibnamefont {Schutte}}, \bibinfo {author}
  {\bibfnamefont {U.}~\bibnamefont {Schwanke}}, \bibinfo {author}
  {\bibfnamefont {M.}~\bibnamefont {Senniappan}}, \bibinfo {author}
  {\bibfnamefont {J.~N.~S.}\ \bibnamefont {Shapopi}}, \bibinfo {author}
  {\bibfnamefont {H.}~\bibnamefont {Sol}}, \bibinfo {author} {\bibfnamefont
  {A.}~\bibnamefont {Specovius}}, \bibinfo {author} {\bibfnamefont
  {S.}~\bibnamefont {Spencer}}, \bibinfo {author} {\bibfnamefont
  {L.}~\bibnamefont {Stawarz}}, \bibinfo {author} {\bibfnamefont
  {C.}~\bibnamefont {Stegmann}}, \bibinfo {author} {\bibfnamefont
  {S.}~\bibnamefont {Steinmassl}}, \bibinfo {author} {\bibfnamefont
  {C.}~\bibnamefont {Steppa}}, \bibinfo {author} {\bibfnamefont
  {T.}~\bibnamefont {Takahashi}}, \bibinfo {author} {\bibfnamefont
  {T.}~\bibnamefont {Tanaka}}, \bibinfo {author} {\bibfnamefont
  {R.}~\bibnamefont {Terrier}}, \bibinfo {author} {\bibfnamefont
  {C.}~\bibnamefont {Thorpe-Morgan}}, \bibinfo {author} {\bibfnamefont
  {M.}~\bibnamefont {Tluczykont}}, \bibinfo {author} {\bibfnamefont
  {M.}~\bibnamefont {Tsirou}}, \bibinfo {author} {\bibfnamefont
  {N.}~\bibnamefont {Tsuji}}, \bibinfo {author} {\bibfnamefont
  {Y.}~\bibnamefont {Uchiyama}}, \bibinfo {author} {\bibfnamefont
  {C.}~\bibnamefont {van Eldik}}, \bibinfo {author} {\bibfnamefont
  {J.}~\bibnamefont {Veh}}, \bibinfo {author} {\bibfnamefont {J.}~\bibnamefont
  {Vink}}, \bibinfo {author} {\bibfnamefont {S.~J.}\ \bibnamefont {Wagner}},
  \bibinfo {author} {\bibfnamefont {R.}~\bibnamefont {White}}, \bibinfo
  {author} {\bibfnamefont {A.}~\bibnamefont {Wierzcholska}}, \bibinfo {author}
  {\bibfnamefont {Y.~W.}\ \bibnamefont {Wong}}, \bibinfo {author}
  {\bibfnamefont {M.}~\bibnamefont {Zacharias}}, \bibinfo {author}
  {\bibfnamefont {D.}~\bibnamefont {Zargaryan}}, \bibinfo {author}
  {\bibfnamefont {A.~A.}\ \bibnamefont {Zdziarski}}, \bibinfo {author}
  {\bibfnamefont {A.}~\bibnamefont {Zech}}, \bibinfo {author} {\bibfnamefont
  {S.~J.}\ \bibnamefont {Zhu}}, \bibinfo {author} {\bibfnamefont
  {S.}~\bibnamefont {Zouari}},\ and\ \bibinfo {author} {\bibfnamefont
  {N.}~\bibnamefont {\ifmmode~\dot{Z}\else \.{Z}\fi{}ywucka}} (\bibinfo
  {collaboration} {H.E.S.S. Collaboration}),\ }\href
  {https://doi.org/10.1103/PhysRevLett.129.111101} {\bibfield  {journal}
  {\bibinfo  {journal} {Phys. Rev. Lett.}\ }\textbf {\bibinfo {volume} {129}},\
  \bibinfo {pages} {111101} (\bibinfo {year} {2022})}\BibitemShut {NoStop}%
\bibitem [{\citenamefont {Acciari}\ \emph {et~al.}(2022)\citenamefont {Acciari}
  \emph {et~al.}}]{dm_magic2}%
  \BibitemOpen
  \bibfield  {author} {\bibinfo {author} {\bibfnamefont {V.~A.}\ \bibnamefont
  {Acciari}} \emph {et~al.},\ }\href
  {https://doi.org/10.1016/j.dark.2021.100912} {\bibfield  {journal} {\bibinfo
  {journal} {Physics of the Dark Universe}\ }\textbf {\bibinfo {volume} {35}},\
  \bibinfo {eid} {100912} (\bibinfo {year} {2022})},\ \Eprint
  {https://arxiv.org/abs/2111.15009} {arXiv:2111.15009 [astro-ph.HE]}
  \BibitemShut {NoStop}%
\bibitem [{\citenamefont {Abe}\ \emph {et~al.}(2023)\citenamefont {Abe},
  \citenamefont {Abe}, \citenamefont {Acciari}, \citenamefont {Aniello},
  \citenamefont {Ansoldi}, \citenamefont {Antonelli}, \citenamefont
  {Arbet~Engels}, \citenamefont {Arcaro}, \citenamefont {Artero}, \citenamefont
  {Asano}, \citenamefont {Baack}, \citenamefont {Babi\ifmmode~\acute{c}\else
  \'{c}\fi{}}, \citenamefont {Baquero}, \citenamefont {Barres~de Almeida},
  \citenamefont {Barrio}, \citenamefont {Batkovi\ifmmode~\acute{c}\else
  \'{c}\fi{}}, \citenamefont {Baxter}, \citenamefont {Becerra~Gonz\'alez},
  \citenamefont {Bednarek}, \citenamefont {Bernardini}, \citenamefont
  {Bernardos}, \citenamefont {Berti}, \citenamefont {Besenrieder},
  \citenamefont {Bhattacharyya}, \citenamefont {Bigongiari}, \citenamefont
  {Biland}, \citenamefont {Blanch}, \citenamefont {Bonnoli}, \citenamefont
  {Bo\ifmmode~\check{s}\else \v{s}\fi{}njak}, \citenamefont {Burelli},
  \citenamefont {Busetto}, \citenamefont {Carosi}, \citenamefont
  {Carretero-Castrillo}, \citenamefont {Ceribella}, \citenamefont {Chai},
  \citenamefont {Chilingarian}, \citenamefont {Cikota}, \citenamefont
  {Colombo}, \citenamefont {Contreras}, \citenamefont {Cortina}, \citenamefont
  {Covino}, \citenamefont {D'Amico}, \citenamefont {D'Elia}, \citenamefont
  {Da~Vela}, \citenamefont {Dazzi}, \citenamefont {De~Angelis}, \citenamefont
  {De~Lotto}, \citenamefont {Del~Popolo}, \citenamefont {Delfino},
  \citenamefont {Delgado}, \citenamefont {Delgado~Mendez}, \citenamefont
  {Depaoli}, \citenamefont {Di~Pierro}, \citenamefont {Di~Venere},
  \citenamefont {Do~Souto Espi\~neira}, \citenamefont {Dominis~Prester},
  \citenamefont {Donini}, \citenamefont {Dorner}, \citenamefont {Doro},
  \citenamefont {Elsaesser}, \citenamefont {Emery}, \citenamefont
  {Fallah~Ramazani}, \citenamefont {Fari\~na}, \citenamefont {Fattorini},
  \citenamefont {Font}, \citenamefont {Fruck}, \citenamefont {Fukami},
  \citenamefont {Fukazawa}, \citenamefont {Garc\'{\i}a~L\'opez}, \citenamefont
  {Garczarczyk}, \citenamefont {Gasparyan}, \citenamefont {Gaug}, \citenamefont
  {Giesbrecht~Paiva}, \citenamefont {Giglietto}, \citenamefont {Giordano},
  \citenamefont {Gliwny}, \citenamefont {Godinovi\ifmmode~\acute{c}\else
  \'{c}\fi{}}, \citenamefont {Green}, \citenamefont {Green}, \citenamefont
  {Hadasch}, \citenamefont {Hahn}, \citenamefont {Hassan}, \citenamefont
  {Heckmann}, \citenamefont {Herrera}, \citenamefont {Hrupec}, \citenamefont
  {H\"utten}, \citenamefont {Imazawa}, \citenamefont {Inada}, \citenamefont
  {Iotov}, \citenamefont {Ishio}, \citenamefont {Jim\'enez~Mart\'{\i}nez},
  \citenamefont {Jormanainen}, \citenamefont {Kerszberg}, \citenamefont
  {Kobayashi}, \citenamefont {Kubo}, \citenamefont {Kushida}, \citenamefont
  {Lamastra}, \citenamefont {Lelas}, \citenamefont {Leone}, \citenamefont
  {Lindfors}, \citenamefont {Linhoff}, \citenamefont {Lombardi}, \citenamefont
  {Longo}, \citenamefont {L\'opez-Coto}, \citenamefont {L\'opez-Moya},
  \citenamefont {L\'opez-Oramas}, \citenamefont {Loporchio}, \citenamefont
  {Lorini}, \citenamefont {Lyard}, \citenamefont {Machado~de Oliveira~Fraga},
  \citenamefont {Majumdar}, \citenamefont {Makariev}, \citenamefont {Maneva},
  \citenamefont {Mang}, \citenamefont {Manganaro}, \citenamefont {Mangano},
  \citenamefont {Mannheim}, \citenamefont {Mariotti}, \citenamefont
  {Mart\'{\i}nez}, \citenamefont {Mas~Aguilar}, \citenamefont {Mazin},
  \citenamefont {Menchiari}, \citenamefont {Mender}, \citenamefont {Mi\ifmmode
  \acute{c}\else \'{c}\fi{}anovi\ifmmode~\acute{c}\else \'{c}\fi{}},
  \citenamefont {Miceli}, \citenamefont {Miener}, \citenamefont {Miranda},
  \citenamefont {Mirzoyan}, \citenamefont {Molina}, \citenamefont {Mondal},
  \citenamefont {Moralejo}, \citenamefont {Morcuende}, \citenamefont {Moreno},
  \citenamefont {Nakamori}, \citenamefont {Nanci}, \citenamefont {Nava},
  \citenamefont {Neustroev}, \citenamefont {Nievas~Rosillo}, \citenamefont
  {Nigro}, \citenamefont {Nilsson}, \citenamefont {Nishijima}, \citenamefont
  {Njoh~Ekoume}, \citenamefont {Noda}, \citenamefont {Nozaki}, \citenamefont
  {Ohtani}, \citenamefont {Oka}, \citenamefont {Otero-Santos}, \citenamefont
  {Paiano}, \citenamefont {Palatiello}, \citenamefont {Paneque}, \citenamefont
  {Paoletti}, \citenamefont {Paredes}, \citenamefont
  {Pavleti\ifmmode~\acute{c}\else \'{c}\fi{}}, \citenamefont {Persic},
  \citenamefont {Pihet}, \citenamefont {Podobnik}, \citenamefont
  {Prada~Moroni}, \citenamefont {Prandini}, \citenamefont {Principe},
  \citenamefont {Priyadarshi}, \citenamefont {Puljak}, \citenamefont {Rhode},
  \citenamefont {Rib\'o}, \citenamefont {Rico}, \citenamefont {Righi},
  \citenamefont {Rugliancich}, \citenamefont {Sahakyan}, \citenamefont {Saito},
  \citenamefont {Sakurai}, \citenamefont {Satalecka}, \citenamefont {Saturni},
  \citenamefont {Schleicher}, \citenamefont {Schmidt}, \citenamefont
  {Schmuckermaier}, \citenamefont {Schubert}, \citenamefont {Schweizer},
  \citenamefont {Sitarek}, \citenamefont {Sliusar}, \citenamefont {Sobczynska},
  \citenamefont {Spolon}, \citenamefont {Stamerra}, \citenamefont {Stri\ifmmode
  \check{s}\else \v{s}\fi{}kovi\ifmmode~\acute{c}\else \'{c}\fi{}},
  \citenamefont {Strom}, \citenamefont {Strzys}, \citenamefont {Suda},
  \citenamefont {Suri\ifmmode~\acute{c}\else \'{c}\fi{}}, \citenamefont
  {Takahashi}, \citenamefont {Takeishi}, \citenamefont {Tavecchio},
  \citenamefont {Temnikov}, \citenamefont {Terauchi}, \citenamefont
  {Terzi\ifmmode~\acute{c}\else \'{c}\fi{}}, \citenamefont {Teshima},
  \citenamefont {Tosti}, \citenamefont {Truzzi}, \citenamefont {Tutone},
  \citenamefont {Ubach}, \citenamefont {van Scherpenberg}, \citenamefont
  {Vazquez~Acosta}, \citenamefont {Ventura}, \citenamefont {Verguilov},
  \citenamefont {Viale}, \citenamefont {Vigorito}, \citenamefont {Vitale},
  \citenamefont {Vovk}, \citenamefont {Walter}, \citenamefont {Will},
  \citenamefont {Wunderlich}, \citenamefont {Yamamoto}, \citenamefont
  {Zari\ifmmode~\acute{c}\else \'{c}\fi{}}, \citenamefont {Hiroshima},\ and\
  \citenamefont {Kohri}}]{dm_magicGCline}%
  \BibitemOpen
  \bibfield  {author} {\bibinfo {author} {\bibfnamefont {H.}~\bibnamefont
  {Abe}}, \bibinfo {author} {\bibfnamefont {S.}~\bibnamefont {Abe}}, \bibinfo
  {author} {\bibfnamefont {V.~A.}\ \bibnamefont {Acciari}}, \bibinfo {author}
  {\bibfnamefont {T.}~\bibnamefont {Aniello}}, \bibinfo {author} {\bibfnamefont
  {S.}~\bibnamefont {Ansoldi}}, \bibinfo {author} {\bibfnamefont {L.~A.}\
  \bibnamefont {Antonelli}}, \bibinfo {author} {\bibfnamefont {A.}~\bibnamefont
  {Arbet~Engels}}, \bibinfo {author} {\bibfnamefont {C.}~\bibnamefont
  {Arcaro}}, \bibinfo {author} {\bibfnamefont {M.}~\bibnamefont {Artero}},
  \bibinfo {author} {\bibfnamefont {K.}~\bibnamefont {Asano}}, \bibinfo
  {author} {\bibfnamefont {D.}~\bibnamefont {Baack}}, \bibinfo {author}
  {\bibfnamefont {A.}~\bibnamefont {Babi\ifmmode~\acute{c}\else \'{c}\fi{}}},
  \bibinfo {author} {\bibfnamefont {A.}~\bibnamefont {Baquero}}, \bibinfo
  {author} {\bibfnamefont {U.}~\bibnamefont {Barres~de Almeida}}, \bibinfo
  {author} {\bibfnamefont {J.~A.}\ \bibnamefont {Barrio}}, \bibinfo {author}
  {\bibfnamefont {I.}~\bibnamefont {Batkovi\ifmmode~\acute{c}\else
  \'{c}\fi{}}}, \bibinfo {author} {\bibfnamefont {J.}~\bibnamefont {Baxter}},
  \bibinfo {author} {\bibfnamefont {J.}~\bibnamefont {Becerra~Gonz\'alez}},
  \bibinfo {author} {\bibfnamefont {W.}~\bibnamefont {Bednarek}}, \bibinfo
  {author} {\bibfnamefont {E.}~\bibnamefont {Bernardini}}, \bibinfo {author}
  {\bibfnamefont {M.}~\bibnamefont {Bernardos}}, \bibinfo {author}
  {\bibfnamefont {A.}~\bibnamefont {Berti}}, \bibinfo {author} {\bibfnamefont
  {J.}~\bibnamefont {Besenrieder}}, \bibinfo {author} {\bibfnamefont
  {W.}~\bibnamefont {Bhattacharyya}}, \bibinfo {author} {\bibfnamefont
  {C.}~\bibnamefont {Bigongiari}}, \bibinfo {author} {\bibfnamefont
  {A.}~\bibnamefont {Biland}}, \bibinfo {author} {\bibfnamefont
  {O.}~\bibnamefont {Blanch}}, \bibinfo {author} {\bibfnamefont
  {G.}~\bibnamefont {Bonnoli}}, \bibinfo {author} {\bibfnamefont {i.~c.~v.}\
  \bibnamefont {Bo\ifmmode~\check{s}\else \v{s}\fi{}njak}}, \bibinfo {author}
  {\bibfnamefont {I.}~\bibnamefont {Burelli}}, \bibinfo {author} {\bibfnamefont
  {G.}~\bibnamefont {Busetto}}, \bibinfo {author} {\bibfnamefont
  {R.}~\bibnamefont {Carosi}}, \bibinfo {author} {\bibfnamefont
  {M.}~\bibnamefont {Carretero-Castrillo}}, \bibinfo {author} {\bibfnamefont
  {G.}~\bibnamefont {Ceribella}}, \bibinfo {author} {\bibfnamefont
  {Y.}~\bibnamefont {Chai}}, \bibinfo {author} {\bibfnamefont {A.}~\bibnamefont
  {Chilingarian}}, \bibinfo {author} {\bibfnamefont {S.}~\bibnamefont
  {Cikota}}, \bibinfo {author} {\bibfnamefont {E.}~\bibnamefont {Colombo}},
  \bibinfo {author} {\bibfnamefont {J.~L.}\ \bibnamefont {Contreras}}, \bibinfo
  {author} {\bibfnamefont {J.}~\bibnamefont {Cortina}}, \bibinfo {author}
  {\bibfnamefont {S.}~\bibnamefont {Covino}}, \bibinfo {author} {\bibfnamefont
  {G.}~\bibnamefont {D'Amico}}, \bibinfo {author} {\bibfnamefont
  {V.}~\bibnamefont {D'Elia}}, \bibinfo {author} {\bibfnamefont
  {P.}~\bibnamefont {Da~Vela}}, \bibinfo {author} {\bibfnamefont
  {F.}~\bibnamefont {Dazzi}}, \bibinfo {author} {\bibfnamefont
  {A.}~\bibnamefont {De~Angelis}}, \bibinfo {author} {\bibfnamefont
  {B.}~\bibnamefont {De~Lotto}}, \bibinfo {author} {\bibfnamefont
  {A.}~\bibnamefont {Del~Popolo}}, \bibinfo {author} {\bibfnamefont
  {M.}~\bibnamefont {Delfino}}, \bibinfo {author} {\bibfnamefont
  {J.}~\bibnamefont {Delgado}}, \bibinfo {author} {\bibfnamefont
  {C.}~\bibnamefont {Delgado~Mendez}}, \bibinfo {author} {\bibfnamefont
  {D.}~\bibnamefont {Depaoli}}, \bibinfo {author} {\bibfnamefont
  {F.}~\bibnamefont {Di~Pierro}}, \bibinfo {author} {\bibfnamefont
  {L.}~\bibnamefont {Di~Venere}}, \bibinfo {author} {\bibfnamefont
  {E.}~\bibnamefont {Do~Souto Espi\~neira}}, \bibinfo {author} {\bibfnamefont
  {D.}~\bibnamefont {Dominis~Prester}}, \bibinfo {author} {\bibfnamefont
  {A.}~\bibnamefont {Donini}}, \bibinfo {author} {\bibfnamefont
  {D.}~\bibnamefont {Dorner}}, \bibinfo {author} {\bibfnamefont
  {M.}~\bibnamefont {Doro}}, \bibinfo {author} {\bibfnamefont {D.}~\bibnamefont
  {Elsaesser}}, \bibinfo {author} {\bibfnamefont {G.}~\bibnamefont {Emery}},
  \bibinfo {author} {\bibfnamefont {V.}~\bibnamefont {Fallah~Ramazani}},
  \bibinfo {author} {\bibfnamefont {L.}~\bibnamefont {Fari\~na}}, \bibinfo
  {author} {\bibfnamefont {A.}~\bibnamefont {Fattorini}}, \bibinfo {author}
  {\bibfnamefont {L.}~\bibnamefont {Font}}, \bibinfo {author} {\bibfnamefont
  {C.}~\bibnamefont {Fruck}}, \bibinfo {author} {\bibfnamefont
  {S.}~\bibnamefont {Fukami}}, \bibinfo {author} {\bibfnamefont
  {Y.}~\bibnamefont {Fukazawa}}, \bibinfo {author} {\bibfnamefont {R.~J.}\
  \bibnamefont {Garc\'{\i}a~L\'opez}}, \bibinfo {author} {\bibfnamefont
  {M.}~\bibnamefont {Garczarczyk}}, \bibinfo {author} {\bibfnamefont
  {S.}~\bibnamefont {Gasparyan}}, \bibinfo {author} {\bibfnamefont
  {M.}~\bibnamefont {Gaug}}, \bibinfo {author} {\bibfnamefont {J.~G.}\
  \bibnamefont {Giesbrecht~Paiva}}, \bibinfo {author} {\bibfnamefont
  {N.}~\bibnamefont {Giglietto}}, \bibinfo {author} {\bibfnamefont
  {F.}~\bibnamefont {Giordano}}, \bibinfo {author} {\bibfnamefont
  {P.}~\bibnamefont {Gliwny}}, \bibinfo {author} {\bibfnamefont
  {N.}~\bibnamefont {Godinovi\ifmmode~\acute{c}\else \'{c}\fi{}}}, \bibinfo
  {author} {\bibfnamefont {J.~G.}\ \bibnamefont {Green}}, \bibinfo {author}
  {\bibfnamefont {D.}~\bibnamefont {Green}}, \bibinfo {author} {\bibfnamefont
  {D.}~\bibnamefont {Hadasch}}, \bibinfo {author} {\bibfnamefont
  {A.}~\bibnamefont {Hahn}}, \bibinfo {author} {\bibfnamefont {T.}~\bibnamefont
  {Hassan}}, \bibinfo {author} {\bibfnamefont {L.}~\bibnamefont {Heckmann}},
  \bibinfo {author} {\bibfnamefont {J.}~\bibnamefont {Herrera}}, \bibinfo
  {author} {\bibfnamefont {D.}~\bibnamefont {Hrupec}}, \bibinfo {author}
  {\bibfnamefont {M.}~\bibnamefont {H\"utten}}, \bibinfo {author}
  {\bibfnamefont {R.}~\bibnamefont {Imazawa}}, \bibinfo {author} {\bibfnamefont
  {T.}~\bibnamefont {Inada}}, \bibinfo {author} {\bibfnamefont
  {R.}~\bibnamefont {Iotov}}, \bibinfo {author} {\bibfnamefont
  {K.}~\bibnamefont {Ishio}}, \bibinfo {author} {\bibfnamefont
  {I.}~\bibnamefont {Jim\'enez~Mart\'{\i}nez}}, \bibinfo {author}
  {\bibfnamefont {J.}~\bibnamefont {Jormanainen}}, \bibinfo {author}
  {\bibfnamefont {D.}~\bibnamefont {Kerszberg}}, \bibinfo {author}
  {\bibfnamefont {Y.}~\bibnamefont {Kobayashi}}, \bibinfo {author}
  {\bibfnamefont {H.}~\bibnamefont {Kubo}}, \bibinfo {author} {\bibfnamefont
  {J.}~\bibnamefont {Kushida}}, \bibinfo {author} {\bibfnamefont
  {A.}~\bibnamefont {Lamastra}}, \bibinfo {author} {\bibfnamefont
  {D.}~\bibnamefont {Lelas}}, \bibinfo {author} {\bibfnamefont
  {F.}~\bibnamefont {Leone}}, \bibinfo {author} {\bibfnamefont
  {E.}~\bibnamefont {Lindfors}}, \bibinfo {author} {\bibfnamefont
  {L.}~\bibnamefont {Linhoff}}, \bibinfo {author} {\bibfnamefont
  {S.}~\bibnamefont {Lombardi}}, \bibinfo {author} {\bibfnamefont
  {F.}~\bibnamefont {Longo}}, \bibinfo {author} {\bibfnamefont
  {R.}~\bibnamefont {L\'opez-Coto}}, \bibinfo {author} {\bibfnamefont
  {M.}~\bibnamefont {L\'opez-Moya}}, \bibinfo {author} {\bibfnamefont
  {A.}~\bibnamefont {L\'opez-Oramas}}, \bibinfo {author} {\bibfnamefont
  {S.}~\bibnamefont {Loporchio}}, \bibinfo {author} {\bibfnamefont
  {A.}~\bibnamefont {Lorini}}, \bibinfo {author} {\bibfnamefont
  {E.}~\bibnamefont {Lyard}}, \bibinfo {author} {\bibfnamefont
  {B.}~\bibnamefont {Machado~de Oliveira~Fraga}}, \bibinfo {author}
  {\bibfnamefont {P.}~\bibnamefont {Majumdar}}, \bibinfo {author}
  {\bibfnamefont {M.}~\bibnamefont {Makariev}}, \bibinfo {author}
  {\bibfnamefont {G.}~\bibnamefont {Maneva}}, \bibinfo {author} {\bibfnamefont
  {N.}~\bibnamefont {Mang}}, \bibinfo {author} {\bibfnamefont {M.}~\bibnamefont
  {Manganaro}}, \bibinfo {author} {\bibfnamefont {S.}~\bibnamefont {Mangano}},
  \bibinfo {author} {\bibfnamefont {K.}~\bibnamefont {Mannheim}}, \bibinfo
  {author} {\bibfnamefont {M.}~\bibnamefont {Mariotti}}, \bibinfo {author}
  {\bibfnamefont {M.}~\bibnamefont {Mart\'{\i}nez}}, \bibinfo {author}
  {\bibfnamefont {A.}~\bibnamefont {Mas~Aguilar}}, \bibinfo {author}
  {\bibfnamefont {D.}~\bibnamefont {Mazin}}, \bibinfo {author} {\bibfnamefont
  {S.}~\bibnamefont {Menchiari}}, \bibinfo {author} {\bibfnamefont
  {S.}~\bibnamefont {Mender}}, \bibinfo {author} {\bibfnamefont
  {S.}~\bibnamefont {Mi\ifmmode \acute{c}\else
  \'{c}\fi{}anovi\ifmmode~\acute{c}\else \'{c}\fi{}}}, \bibinfo {author}
  {\bibfnamefont {D.}~\bibnamefont {Miceli}}, \bibinfo {author} {\bibfnamefont
  {T.}~\bibnamefont {Miener}}, \bibinfo {author} {\bibfnamefont {J.~M.}\
  \bibnamefont {Miranda}}, \bibinfo {author} {\bibfnamefont {R.}~\bibnamefont
  {Mirzoyan}}, \bibinfo {author} {\bibfnamefont {E.}~\bibnamefont {Molina}},
  \bibinfo {author} {\bibfnamefont {H.~A.}\ \bibnamefont {Mondal}}, \bibinfo
  {author} {\bibfnamefont {A.}~\bibnamefont {Moralejo}}, \bibinfo {author}
  {\bibfnamefont {D.}~\bibnamefont {Morcuende}}, \bibinfo {author}
  {\bibfnamefont {V.}~\bibnamefont {Moreno}}, \bibinfo {author} {\bibfnamefont
  {T.}~\bibnamefont {Nakamori}}, \bibinfo {author} {\bibfnamefont
  {C.}~\bibnamefont {Nanci}}, \bibinfo {author} {\bibfnamefont
  {L.}~\bibnamefont {Nava}}, \bibinfo {author} {\bibfnamefont {V.}~\bibnamefont
  {Neustroev}}, \bibinfo {author} {\bibfnamefont {M.}~\bibnamefont
  {Nievas~Rosillo}}, \bibinfo {author} {\bibfnamefont {C.}~\bibnamefont
  {Nigro}}, \bibinfo {author} {\bibfnamefont {K.}~\bibnamefont {Nilsson}},
  \bibinfo {author} {\bibfnamefont {K.}~\bibnamefont {Nishijima}}, \bibinfo
  {author} {\bibfnamefont {T.}~\bibnamefont {Njoh~Ekoume}}, \bibinfo {author}
  {\bibfnamefont {K.}~\bibnamefont {Noda}}, \bibinfo {author} {\bibfnamefont
  {S.}~\bibnamefont {Nozaki}}, \bibinfo {author} {\bibfnamefont
  {Y.}~\bibnamefont {Ohtani}}, \bibinfo {author} {\bibfnamefont
  {T.}~\bibnamefont {Oka}}, \bibinfo {author} {\bibfnamefont {J.}~\bibnamefont
  {Otero-Santos}}, \bibinfo {author} {\bibfnamefont {S.}~\bibnamefont
  {Paiano}}, \bibinfo {author} {\bibfnamefont {M.}~\bibnamefont {Palatiello}},
  \bibinfo {author} {\bibfnamefont {D.}~\bibnamefont {Paneque}}, \bibinfo
  {author} {\bibfnamefont {R.}~\bibnamefont {Paoletti}}, \bibinfo {author}
  {\bibfnamefont {J.~M.}\ \bibnamefont {Paredes}}, \bibinfo {author}
  {\bibfnamefont {L.}~\bibnamefont {Pavleti\ifmmode~\acute{c}\else
  \'{c}\fi{}}}, \bibinfo {author} {\bibfnamefont {M.}~\bibnamefont {Persic}},
  \bibinfo {author} {\bibfnamefont {M.}~\bibnamefont {Pihet}}, \bibinfo
  {author} {\bibfnamefont {F.}~\bibnamefont {Podobnik}}, \bibinfo {author}
  {\bibfnamefont {P.~G.}\ \bibnamefont {Prada~Moroni}}, \bibinfo {author}
  {\bibfnamefont {E.}~\bibnamefont {Prandini}}, \bibinfo {author}
  {\bibfnamefont {G.}~\bibnamefont {Principe}}, \bibinfo {author}
  {\bibfnamefont {C.}~\bibnamefont {Priyadarshi}}, \bibinfo {author}
  {\bibfnamefont {I.}~\bibnamefont {Puljak}}, \bibinfo {author} {\bibfnamefont
  {W.}~\bibnamefont {Rhode}}, \bibinfo {author} {\bibfnamefont
  {M.}~\bibnamefont {Rib\'o}}, \bibinfo {author} {\bibfnamefont
  {J.}~\bibnamefont {Rico}}, \bibinfo {author} {\bibfnamefont {C.}~\bibnamefont
  {Righi}}, \bibinfo {author} {\bibfnamefont {A.}~\bibnamefont {Rugliancich}},
  \bibinfo {author} {\bibfnamefont {N.}~\bibnamefont {Sahakyan}}, \bibinfo
  {author} {\bibfnamefont {T.}~\bibnamefont {Saito}}, \bibinfo {author}
  {\bibfnamefont {S.}~\bibnamefont {Sakurai}}, \bibinfo {author} {\bibfnamefont
  {K.}~\bibnamefont {Satalecka}}, \bibinfo {author} {\bibfnamefont {F.~G.}\
  \bibnamefont {Saturni}}, \bibinfo {author} {\bibfnamefont {B.}~\bibnamefont
  {Schleicher}}, \bibinfo {author} {\bibfnamefont {K.}~\bibnamefont {Schmidt}},
  \bibinfo {author} {\bibfnamefont {F.}~\bibnamefont {Schmuckermaier}},
  \bibinfo {author} {\bibfnamefont {J.~L.}\ \bibnamefont {Schubert}}, \bibinfo
  {author} {\bibfnamefont {T.}~\bibnamefont {Schweizer}}, \bibinfo {author}
  {\bibfnamefont {J.}~\bibnamefont {Sitarek}}, \bibinfo {author} {\bibfnamefont
  {V.}~\bibnamefont {Sliusar}}, \bibinfo {author} {\bibfnamefont
  {D.}~\bibnamefont {Sobczynska}}, \bibinfo {author} {\bibfnamefont
  {A.}~\bibnamefont {Spolon}}, \bibinfo {author} {\bibfnamefont
  {A.}~\bibnamefont {Stamerra}}, \bibinfo {author} {\bibfnamefont
  {J.}~\bibnamefont {Stri\ifmmode \check{s}\else
  \v{s}\fi{}kovi\ifmmode~\acute{c}\else \'{c}\fi{}}}, \bibinfo {author}
  {\bibfnamefont {D.}~\bibnamefont {Strom}}, \bibinfo {author} {\bibfnamefont
  {M.}~\bibnamefont {Strzys}}, \bibinfo {author} {\bibfnamefont
  {Y.}~\bibnamefont {Suda}}, \bibinfo {author} {\bibfnamefont {T.}~\bibnamefont
  {Suri\ifmmode~\acute{c}\else \'{c}\fi{}}}, \bibinfo {author} {\bibfnamefont
  {M.}~\bibnamefont {Takahashi}}, \bibinfo {author} {\bibfnamefont
  {R.}~\bibnamefont {Takeishi}}, \bibinfo {author} {\bibfnamefont
  {F.}~\bibnamefont {Tavecchio}}, \bibinfo {author} {\bibfnamefont
  {P.}~\bibnamefont {Temnikov}}, \bibinfo {author} {\bibfnamefont
  {K.}~\bibnamefont {Terauchi}}, \bibinfo {author} {\bibfnamefont
  {T.}~\bibnamefont {Terzi\ifmmode~\acute{c}\else \'{c}\fi{}}}, \bibinfo
  {author} {\bibfnamefont {M.}~\bibnamefont {Teshima}}, \bibinfo {author}
  {\bibfnamefont {L.}~\bibnamefont {Tosti}}, \bibinfo {author} {\bibfnamefont
  {S.}~\bibnamefont {Truzzi}}, \bibinfo {author} {\bibfnamefont
  {A.}~\bibnamefont {Tutone}}, \bibinfo {author} {\bibfnamefont
  {S.}~\bibnamefont {Ubach}}, \bibinfo {author} {\bibfnamefont
  {J.}~\bibnamefont {van Scherpenberg}}, \bibinfo {author} {\bibfnamefont
  {M.}~\bibnamefont {Vazquez~Acosta}}, \bibinfo {author} {\bibfnamefont
  {S.}~\bibnamefont {Ventura}}, \bibinfo {author} {\bibfnamefont
  {V.}~\bibnamefont {Verguilov}}, \bibinfo {author} {\bibfnamefont
  {I.}~\bibnamefont {Viale}}, \bibinfo {author} {\bibfnamefont {C.~F.}\
  \bibnamefont {Vigorito}}, \bibinfo {author} {\bibfnamefont {V.}~\bibnamefont
  {Vitale}}, \bibinfo {author} {\bibfnamefont {I.}~\bibnamefont {Vovk}},
  \bibinfo {author} {\bibfnamefont {R.}~\bibnamefont {Walter}}, \bibinfo
  {author} {\bibfnamefont {M.}~\bibnamefont {Will}}, \bibinfo {author}
  {\bibfnamefont {C.}~\bibnamefont {Wunderlich}}, \bibinfo {author}
  {\bibfnamefont {T.}~\bibnamefont {Yamamoto}}, \bibinfo {author}
  {\bibfnamefont {D.}~\bibnamefont {Zari\ifmmode~\acute{c}\else \'{c}\fi{}}},
  \bibinfo {author} {\bibfnamefont {N.}~\bibnamefont {Hiroshima}},\ and\
  \bibinfo {author} {\bibfnamefont {K.}~\bibnamefont {Kohri}} (\bibinfo
  {collaboration} {MAGIC Collaboration}),\ }\href
  {https://doi.org/10.1103/PhysRevLett.130.061002} {\bibfield  {journal}
  {\bibinfo  {journal} {Phys. Rev. Lett.}\ }\textbf {\bibinfo {volume} {130}},\
  \bibinfo {pages} {061002} (\bibinfo {year} {2023})}\BibitemShut {NoStop}%
\bibitem [{\citenamefont {Archambault}\ \emph {et~al.}(2017)\citenamefont
  {Archambault} \emph {et~al.}}]{dm_veritas}%
  \BibitemOpen
  \bibfield  {author} {\bibinfo {author} {\bibfnamefont {S.}~\bibnamefont
  {Archambault}} \emph {et~al.},\ }\href
  {https://doi.org/10.1103/PhysRevD.95.082001} {\bibfield  {journal} {\bibinfo
  {journal} {\prd}\ }\textbf {\bibinfo {volume} {95}},\ \bibinfo {eid} {082001}
  (\bibinfo {year} {2017})},\ \Eprint {https://arxiv.org/abs/1703.04937}
  {arXiv:1703.04937 [astro-ph.HE]} \BibitemShut {NoStop}%
\bibitem [{\citenamefont {Albert}\ \emph {et~al.}(2018)\citenamefont {Albert}
  \emph {et~al.}}]{dm_hawc}%
  \BibitemOpen
  \bibfield  {author} {\bibinfo {author} {\bibfnamefont {A.}~\bibnamefont
  {Albert}} \emph {et~al.},\ }\href {https://doi.org/10.3847/1538-4357/aaa6d8}
  {\bibfield  {journal} {\bibinfo  {journal} {\apj}\ }\textbf {\bibinfo
  {volume} {853}},\ \bibinfo {eid} {154} (\bibinfo {year} {2018})},\ \Eprint
  {https://arxiv.org/abs/1706.01277} {arXiv:1706.01277 [astro-ph.HE]}
  \BibitemShut {NoStop}%
\bibitem [{\citenamefont {{Abeysekara}}\ \emph {et~al.}(2018)\citenamefont
  {{Abeysekara}}, \citenamefont {{Albert}}, \citenamefont {{Alfaro}},
  \citenamefont {{Alvarez}}, \citenamefont {{Arceo}}, \citenamefont
  {{Arteaga-Vel{\'a}zquez}}, \citenamefont {{Avila Rojas}}, \citenamefont
  {{Ayala Solares}}, \citenamefont {{Becerril}}, \citenamefont
  {{Belmont-Moreno}}, \citenamefont {{BenZvi}}, \citenamefont {{Bernal}},
  \citenamefont {{Brisbois}}, \citenamefont {{Caballero-Mora}}, \citenamefont
  {{Capistr{\'a}n}}, \citenamefont {{Carrami{\~n}ana}}, \citenamefont
  {{Casanova}}, \citenamefont {{Castillo}}, \citenamefont {{Cotti}},
  \citenamefont {{Cotzomi}}, \citenamefont {{De Le{\'o}n}}, \citenamefont {{De
  la Fuente}}, \citenamefont {{Diaz Hernandez}}, \citenamefont {{Dingus}},
  \citenamefont {{DuVernois}}, \citenamefont {{D{\'\i}az-V{\'e}lez}},
  \citenamefont {{Engel}}, \citenamefont {{Enr{\'\i}quez-Rivera}},
  \citenamefont {{Fiorino}}, \citenamefont {{Fleischhack}}, \citenamefont
  {{Fraija}}, \citenamefont {{Garc{\'\i}a-Gonz{\'a}lez}}, \citenamefont
  {{Garfias}}, \citenamefont {{Gonz{\'a}lez Mu{\~n}oz}}, \citenamefont
  {{Gonz{\'a}lez}}, \citenamefont {{Goodman}}, \citenamefont {{Hampel-Arias}},
  \citenamefont {{Harding}}, \citenamefont {{Hernandez}}, \citenamefont
  {{Hernandez-Almada}}, \citenamefont {{Hueyotl-Zahuantitla}}, \citenamefont
  {{H{\"u}ntemeyer}}, \citenamefont {{Iriarte}}, \citenamefont
  {{Jardin-Blicq}}, \citenamefont {{Joshi}}, \citenamefont {{Kaufmann}},
  \citenamefont {{Lauer}}, \citenamefont {{Lee}}, \citenamefont {{Lennarz}},
  \citenamefont {{Le{\'o}n Vargas}}, \citenamefont {{Linnemann}}, \citenamefont
  {{Longinotti}}, \citenamefont {{Luis-Raya}}, \citenamefont
  {{Luna-Garc{\'\i}a}}, \citenamefont {{L{\'o}pez-Coto}}, \citenamefont
  {{Malone}}, \citenamefont {{Marinelli}}, \citenamefont {{Martinez}},
  \citenamefont {{Martinez-Castellanos}}, \citenamefont
  {{Mart{\'\i}nez-Castro}}, \citenamefont {{Matthews}}, \citenamefont
  {{Miranda-Romagnoli}}, \citenamefont {{Moreno}}, \citenamefont
  {{Mostaf{\'a}}}, \citenamefont {{Nellen}}, \citenamefont {{Newbold}},
  \citenamefont {{Nisa}}, \citenamefont {{Noriega-Papaqui}}, \citenamefont
  {{Pelayo}}, \citenamefont {{Pretz}}, \citenamefont {{P{\'e}rez-P{\'e}rez}},
  \citenamefont {{Ren}}, \citenamefont {{Rho}}, \citenamefont {{Rodd}},
  \citenamefont {{Rosa-Gonz{\'a}lez}}, \citenamefont {{Rosenberg}},
  \citenamefont {{Ruiz-Velasco}}, \citenamefont {{Safdi}}, \citenamefont
  {{Salazar}}, \citenamefont {{Salesa Greus}}, \citenamefont {{Sandoval}},
  \citenamefont {{Schneider}}, \citenamefont {{Sinnis}}, \citenamefont
  {{Smith}}, \citenamefont {{Springer}}, \citenamefont {{Surajbali}},
  \citenamefont {{Taboada}}, \citenamefont {{Tibolla}}, \citenamefont
  {{Tollefson}}, \citenamefont {{Torres}}, \citenamefont {{Ukwatta}},
  \citenamefont {{Vianello}}, \citenamefont {{Villase{\~n}or}}, \citenamefont
  {{Weisgarber}}, \citenamefont {{Westerhoff}}, \citenamefont {{Wisher}},
  \citenamefont {{Wood}}, \citenamefont {{Yapici}}, \citenamefont {{Yodh}},
  \citenamefont {{Younk}}, \citenamefont {{Zepeda}}, \citenamefont {{Zhou}},\
  and\ \citenamefont {{{\'A}lvarez}}}]{dm_hawcGH}%
  \BibitemOpen
  \bibfield  {author} {\bibinfo {author} {\bibfnamefont {A.~U.}\ \bibnamefont
  {{Abeysekara}}}, \bibinfo {author} {\bibfnamefont {A.}~\bibnamefont
  {{Albert}}}, \bibinfo {author} {\bibfnamefont {R.}~\bibnamefont {{Alfaro}}},
  \bibinfo {author} {\bibfnamefont {C.}~\bibnamefont {{Alvarez}}}, \bibinfo
  {author} {\bibfnamefont {R.}~\bibnamefont {{Arceo}}}, \bibinfo {author}
  {\bibfnamefont {J.~C.}\ \bibnamefont {{Arteaga-Vel{\'a}zquez}}}, \bibinfo
  {author} {\bibfnamefont {D.}~\bibnamefont {{Avila Rojas}}}, \bibinfo {author}
  {\bibfnamefont {H.~A.}\ \bibnamefont {{Ayala Solares}}}, \bibinfo {author}
  {\bibfnamefont {A.}~\bibnamefont {{Becerril}}}, \bibinfo {author}
  {\bibfnamefont {E.}~\bibnamefont {{Belmont-Moreno}}}, \bibinfo {author}
  {\bibfnamefont {S.~Y.}\ \bibnamefont {{BenZvi}}}, \bibinfo {author}
  {\bibfnamefont {A.}~\bibnamefont {{Bernal}}}, \bibinfo {author}
  {\bibfnamefont {C.}~\bibnamefont {{Brisbois}}}, \bibinfo {author}
  {\bibfnamefont {K.~S.}\ \bibnamefont {{Caballero-Mora}}}, \bibinfo {author}
  {\bibfnamefont {T.}~\bibnamefont {{Capistr{\'a}n}}}, \bibinfo {author}
  {\bibfnamefont {A.}~\bibnamefont {{Carrami{\~n}ana}}}, \bibinfo {author}
  {\bibfnamefont {S.}~\bibnamefont {{Casanova}}}, \bibinfo {author}
  {\bibfnamefont {M.}~\bibnamefont {{Castillo}}}, \bibinfo {author}
  {\bibfnamefont {U.}~\bibnamefont {{Cotti}}}, \bibinfo {author} {\bibfnamefont
  {J.}~\bibnamefont {{Cotzomi}}}, \bibinfo {author} {\bibfnamefont
  {C.}~\bibnamefont {{De Le{\'o}n}}}, \bibinfo {author} {\bibfnamefont
  {E.}~\bibnamefont {{De la Fuente}}}, \bibinfo {author} {\bibfnamefont
  {R.}~\bibnamefont {{Diaz Hernandez}}}, \bibinfo {author} {\bibfnamefont
  {B.~L.}\ \bibnamefont {{Dingus}}}, \bibinfo {author} {\bibfnamefont {M.~A.}\
  \bibnamefont {{DuVernois}}}, \bibinfo {author} {\bibfnamefont {J.~C.}\
  \bibnamefont {{D{\'\i}az-V{\'e}lez}}}, \bibinfo {author} {\bibfnamefont
  {K.}~\bibnamefont {{Engel}}}, \bibinfo {author} {\bibfnamefont
  {O.}~\bibnamefont {{Enr{\'\i}quez-Rivera}}}, \bibinfo {author} {\bibfnamefont
  {D.~W.}\ \bibnamefont {{Fiorino}}}, \bibinfo {author} {\bibfnamefont
  {H.}~\bibnamefont {{Fleischhack}}}, \bibinfo {author} {\bibfnamefont
  {N.}~\bibnamefont {{Fraija}}}, \bibinfo {author} {\bibfnamefont {J.~A.}\
  \bibnamefont {{Garc{\'\i}a-Gonz{\'a}lez}}}, \bibinfo {author} {\bibfnamefont
  {F.}~\bibnamefont {{Garfias}}}, \bibinfo {author} {\bibfnamefont
  {A.}~\bibnamefont {{Gonz{\'a}lez Mu{\~n}oz}}}, \bibinfo {author}
  {\bibfnamefont {M.~M.}\ \bibnamefont {{Gonz{\'a}lez}}}, \bibinfo {author}
  {\bibfnamefont {J.~A.}\ \bibnamefont {{Goodman}}}, \bibinfo {author}
  {\bibfnamefont {Z.}~\bibnamefont {{Hampel-Arias}}}, \bibinfo {author}
  {\bibfnamefont {J.~P.}\ \bibnamefont {{Harding}}}, \bibinfo {author}
  {\bibfnamefont {S.}~\bibnamefont {{Hernandez}}}, \bibinfo {author}
  {\bibfnamefont {A.}~\bibnamefont {{Hernandez-Almada}}}, \bibinfo {author}
  {\bibfnamefont {F.}~\bibnamefont {{Hueyotl-Zahuantitla}}}, \bibinfo {author}
  {\bibfnamefont {P.}~\bibnamefont {{H{\"u}ntemeyer}}}, \bibinfo {author}
  {\bibfnamefont {A.}~\bibnamefont {{Iriarte}}}, \bibinfo {author}
  {\bibfnamefont {A.}~\bibnamefont {{Jardin-Blicq}}}, \bibinfo {author}
  {\bibfnamefont {V.}~\bibnamefont {{Joshi}}}, \bibinfo {author} {\bibfnamefont
  {S.}~\bibnamefont {{Kaufmann}}}, \bibinfo {author} {\bibfnamefont {R.~J.}\
  \bibnamefont {{Lauer}}}, \bibinfo {author} {\bibfnamefont {W.~H.}\
  \bibnamefont {{Lee}}}, \bibinfo {author} {\bibfnamefont {D.}~\bibnamefont
  {{Lennarz}}}, \bibinfo {author} {\bibfnamefont {H.}~\bibnamefont {{Le{\'o}n
  Vargas}}}, \bibinfo {author} {\bibfnamefont {J.~T.}\ \bibnamefont
  {{Linnemann}}}, \bibinfo {author} {\bibfnamefont {A.~L.}\ \bibnamefont
  {{Longinotti}}}, \bibinfo {author} {\bibfnamefont {G.}~\bibnamefont
  {{Luis-Raya}}}, \bibinfo {author} {\bibfnamefont {R.}~\bibnamefont
  {{Luna-Garc{\'\i}a}}}, \bibinfo {author} {\bibfnamefont {R.}~\bibnamefont
  {{L{\'o}pez-Coto}}}, \bibinfo {author} {\bibfnamefont {K.}~\bibnamefont
  {{Malone}}}, \bibinfo {author} {\bibfnamefont {S.~S.}\ \bibnamefont
  {{Marinelli}}}, \bibinfo {author} {\bibfnamefont {O.}~\bibnamefont
  {{Martinez}}}, \bibinfo {author} {\bibfnamefont {I.}~\bibnamefont
  {{Martinez-Castellanos}}}, \bibinfo {author} {\bibfnamefont {J.}~\bibnamefont
  {{Mart{\'\i}nez-Castro}}}, \bibinfo {author} {\bibfnamefont {J.~A.}\
  \bibnamefont {{Matthews}}}, \bibinfo {author} {\bibfnamefont
  {P.}~\bibnamefont {{Miranda-Romagnoli}}}, \bibinfo {author} {\bibfnamefont
  {E.}~\bibnamefont {{Moreno}}}, \bibinfo {author} {\bibfnamefont
  {M.}~\bibnamefont {{Mostaf{\'a}}}}, \bibinfo {author} {\bibfnamefont
  {L.}~\bibnamefont {{Nellen}}}, \bibinfo {author} {\bibfnamefont
  {M.}~\bibnamefont {{Newbold}}}, \bibinfo {author} {\bibfnamefont {M.~U.}\
  \bibnamefont {{Nisa}}}, \bibinfo {author} {\bibfnamefont {R.}~\bibnamefont
  {{Noriega-Papaqui}}}, \bibinfo {author} {\bibfnamefont {R.}~\bibnamefont
  {{Pelayo}}}, \bibinfo {author} {\bibfnamefont {J.}~\bibnamefont {{Pretz}}},
  \bibinfo {author} {\bibfnamefont {E.~G.}\ \bibnamefont
  {{P{\'e}rez-P{\'e}rez}}}, \bibinfo {author} {\bibfnamefont {Z.}~\bibnamefont
  {{Ren}}}, \bibinfo {author} {\bibfnamefont {C.~D.}\ \bibnamefont {{Rho}}},
  \bibinfo {author} {\bibfnamefont {N.~L.}\ \bibnamefont {{Rodd}}}, \bibinfo
  {author} {\bibfnamefont {D.}~\bibnamefont {{Rosa-Gonz{\'a}lez}}}, \bibinfo
  {author} {\bibfnamefont {M.}~\bibnamefont {{Rosenberg}}}, \bibinfo {author}
  {\bibfnamefont {E.}~\bibnamefont {{Ruiz-Velasco}}}, \bibinfo {author}
  {\bibfnamefont {B.~R.}\ \bibnamefont {{Safdi}}}, \bibinfo {author}
  {\bibfnamefont {H.}~\bibnamefont {{Salazar}}}, \bibinfo {author}
  {\bibfnamefont {F.}~\bibnamefont {{Salesa Greus}}}, \bibinfo {author}
  {\bibfnamefont {A.}~\bibnamefont {{Sandoval}}}, \bibinfo {author}
  {\bibfnamefont {M.}~\bibnamefont {{Schneider}}}, \bibinfo {author}
  {\bibfnamefont {G.}~\bibnamefont {{Sinnis}}}, \bibinfo {author}
  {\bibfnamefont {A.~J.}\ \bibnamefont {{Smith}}}, \bibinfo {author}
  {\bibfnamefont {R.~W.}\ \bibnamefont {{Springer}}}, \bibinfo {author}
  {\bibfnamefont {P.}~\bibnamefont {{Surajbali}}}, \bibinfo {author}
  {\bibfnamefont {I.}~\bibnamefont {{Taboada}}}, \bibinfo {author}
  {\bibfnamefont {O.}~\bibnamefont {{Tibolla}}}, \bibinfo {author}
  {\bibfnamefont {K.}~\bibnamefont {{Tollefson}}}, \bibinfo {author}
  {\bibfnamefont {I.}~\bibnamefont {{Torres}}}, \bibinfo {author}
  {\bibfnamefont {T.~N.}\ \bibnamefont {{Ukwatta}}}, \bibinfo {author}
  {\bibfnamefont {G.}~\bibnamefont {{Vianello}}}, \bibinfo {author}
  {\bibfnamefont {L.}~\bibnamefont {{Villase{\~n}or}}}, \bibinfo {author}
  {\bibfnamefont {T.}~\bibnamefont {{Weisgarber}}}, \bibinfo {author}
  {\bibfnamefont {S.}~\bibnamefont {{Westerhoff}}}, \bibinfo {author}
  {\bibfnamefont {I.~G.}\ \bibnamefont {{Wisher}}}, \bibinfo {author}
  {\bibfnamefont {J.}~\bibnamefont {{Wood}}}, \bibinfo {author} {\bibfnamefont
  {T.}~\bibnamefont {{Yapici}}}, \bibinfo {author} {\bibfnamefont {G.~B.}\
  \bibnamefont {{Yodh}}}, \bibinfo {author} {\bibfnamefont {P.~W.}\
  \bibnamefont {{Younk}}}, \bibinfo {author} {\bibfnamefont {A.}~\bibnamefont
  {{Zepeda}}}, \bibinfo {author} {\bibfnamefont {H.}~\bibnamefont {{Zhou}}},\
  and\ \bibinfo {author} {\bibfnamefont {J.~D.}\ \bibnamefont
  {{{\'A}lvarez}}},\ }\href {https://doi.org/10.1088/1475-7516/2018/02/049}
  {\bibfield  {journal} {\bibinfo  {journal} {\jcap}\ }\textbf {\bibinfo
  {volume} {2018}},\ \bibinfo {eid} {049} (\bibinfo {year} {2018})},\ \Eprint
  {https://arxiv.org/abs/1710.10288} {arXiv:1710.10288 [astro-ph.HE]}
  \BibitemShut {NoStop}%
\bibitem [{\citenamefont {{Albert}}\ \emph {et~al.}(2020)\citenamefont
  {{Albert}}, \citenamefont {{Alfaro}}, \citenamefont {{Alvarez}},
  \citenamefont {{Arteaga-Vel{\'a}zquez}} \emph {et~al.}}]{dm_hawc2}%
  \BibitemOpen
  \bibfield  {author} {\bibinfo {author} {\bibfnamefont {A.}~\bibnamefont
  {{Albert}}}, \bibinfo {author} {\bibfnamefont {R.}~\bibnamefont {{Alfaro}}},
  \bibinfo {author} {\bibfnamefont {C.}~\bibnamefont {{Alvarez}}}, \bibinfo
  {author} {\bibfnamefont {J.~C.}\ \bibnamefont {{Arteaga-Vel{\'a}zquez}}},
  \emph {et~al.},\ }\href {https://doi.org/10.1103/PhysRevD.101.103001}
  {\bibfield  {journal} {\bibinfo  {journal} {\prd}\ }\textbf {\bibinfo
  {volume} {101}},\ \bibinfo {eid} {103001} (\bibinfo {year}
  {2020})}\BibitemShut {NoStop}%
\bibitem [{\citenamefont {Tak}\ \emph {et~al.}(2022)\citenamefont {Tak},
  \citenamefont {Baumgart}, \citenamefont {Rodd},\ and\ \citenamefont
  {Pueschel}}]{Tak_2022}%
  \BibitemOpen
  \bibfield  {author} {\bibinfo {author} {\bibfnamefont {D.}~\bibnamefont
  {Tak}}, \bibinfo {author} {\bibfnamefont {M.}~\bibnamefont {Baumgart}},
  \bibinfo {author} {\bibfnamefont {N.~L.}\ \bibnamefont {Rodd}},\ and\
  \bibinfo {author} {\bibfnamefont {E.}~\bibnamefont {Pueschel}},\ }\href
  {https://doi.org/10.3847/2041-8213/ac9387} {\bibfield  {journal} {\bibinfo
  {journal} {The Astrophysical Journal Letters}\ }\textbf {\bibinfo {volume}
  {938}},\ \bibinfo {pages} {L4} (\bibinfo {year} {2022})}\BibitemShut
  {NoStop}%
\bibitem [{\citenamefont {{Griest}}\ and\ \citenamefont
  {{Kamionkowski}}(1990)}]{Griest1990}%
  \BibitemOpen
  \bibfield  {author} {\bibinfo {author} {\bibfnamefont {K.}~\bibnamefont
  {{Griest}}}\ and\ \bibinfo {author} {\bibfnamefont {M.}~\bibnamefont
  {{Kamionkowski}}},\ }\href {https://doi.org/10.1103/PhysRevLett.64.615}
  {\bibfield  {journal} {\bibinfo  {journal} {\prl}\ }\textbf {\bibinfo
  {volume} {64}},\ \bibinfo {pages} {615} (\bibinfo {year} {1990})}\BibitemShut
  {NoStop}%
\bibitem [{\citenamefont {Smirnov}\ and\ \citenamefont
  {Beacom}(2019)}]{Smirnov:2019ngs}%
  \BibitemOpen
  \bibfield  {author} {\bibinfo {author} {\bibfnamefont {J.}~\bibnamefont
  {Smirnov}}\ and\ \bibinfo {author} {\bibfnamefont {J.~F.}\ \bibnamefont
  {Beacom}},\ }\href {https://doi.org/10.1103/PhysRevD.100.043029} {\bibfield
  {journal} {\bibinfo  {journal} {Phys. Rev. D}\ }\textbf {\bibinfo {volume}
  {100}},\ \bibinfo {pages} {043029} (\bibinfo {year} {2019})},\ \Eprint
  {https://arxiv.org/abs/1904.11503} {arXiv:1904.11503 [hep-ph]} \BibitemShut
  {NoStop}%
\bibitem [{\citenamefont {{Harigaya}}\ \emph {et~al.}(2016)\citenamefont
  {{Harigaya}}, \citenamefont {{Ibe}}, \citenamefont {{Kaneta}}, \citenamefont
  {{Nakano}},\ and\ \citenamefont {{Suzuki}}}]{Harigaya2016}%
  \BibitemOpen
  \bibfield  {author} {\bibinfo {author} {\bibfnamefont {K.}~\bibnamefont
  {{Harigaya}}}, \bibinfo {author} {\bibfnamefont {M.}~\bibnamefont {{Ibe}}},
  \bibinfo {author} {\bibfnamefont {K.}~\bibnamefont {{Kaneta}}}, \bibinfo
  {author} {\bibfnamefont {W.}~\bibnamefont {{Nakano}}},\ and\ \bibinfo
  {author} {\bibfnamefont {M.}~\bibnamefont {{Suzuki}}},\ }\href
  {https://doi.org/10.1007/JHEP08(2016)151} {\bibfield  {journal} {\bibinfo
  {journal} {Journal of High Energy Physics}\ }\textbf {\bibinfo {volume}
  {2016}},\ \bibinfo {eid} {151} (\bibinfo {year} {2016})},\ \Eprint
  {https://arxiv.org/abs/1606.00159} {arXiv:1606.00159 [hep-ph]} \BibitemShut
  {NoStop}%
\bibitem [{\citenamefont {Geller}\ \emph {et~al.}(2018)\citenamefont {Geller},
  \citenamefont {Iwamoto}, \citenamefont {Lee}, \citenamefont {Shadmi},\ and\
  \citenamefont {Telem}}]{Geller2018}%
  \BibitemOpen
  \bibfield  {author} {\bibinfo {author} {\bibfnamefont {M.}~\bibnamefont
  {Geller}}, \bibinfo {author} {\bibfnamefont {S.}~\bibnamefont {Iwamoto}},
  \bibinfo {author} {\bibfnamefont {G.}~\bibnamefont {Lee}}, \bibinfo {author}
  {\bibfnamefont {Y.}~\bibnamefont {Shadmi}},\ and\ \bibinfo {author}
  {\bibfnamefont {O.}~\bibnamefont {Telem}},\ }\href
  {https://doi.org/10.1007/JHEP06(2018)135} {\bibfield  {journal} {\bibinfo
  {journal} {JHEP}\ }\textbf {\bibinfo {volume} {06}},\ \bibinfo {pages}
  {135}},\ \Eprint {https://arxiv.org/abs/1802.07720} {arXiv:1802.07720
  [hep-ph]} \BibitemShut {NoStop}%
\bibitem [{\citenamefont {{Acharyya}}\ \emph {et~al.}(2023)\citenamefont
  {{Acharyya}}, \citenamefont {{Archer}}, \citenamefont {{Bangale}},
  \citenamefont {{Bartkoske}}, \citenamefont {{Batista}}, \citenamefont
  {{Baumgart}}, \citenamefont {{Benbow}}, \citenamefont {{Buckley}},
  \citenamefont {{Falcone}}, \citenamefont {{Feng}}, \citenamefont {{Finley}},
  \citenamefont {{Foote}}, \citenamefont {{Fortson}}, \citenamefont
  {{Furniss}}, \citenamefont {{Gallagher}}, \citenamefont {{Hanlon}},
  \citenamefont {{Hervet}}, \citenamefont {{Hoang}}, \citenamefont {{Holder}},
  \citenamefont {{Humensky}}, \citenamefont {{Jin}}, \citenamefont {{Kaaret}},
  \citenamefont {{Kertzman}}, \citenamefont {{Kherlakian}}, \citenamefont
  {{Kieda}}, \citenamefont {{Kleiner}}, \citenamefont {{Korzoun}},
  \citenamefont {{Krennrich}}, \citenamefont {{Lang}}, \citenamefont {{Lundy}},
  \citenamefont {{Maier}}, \citenamefont {{McGrath}}, \citenamefont
  {{Moriarty}}, \citenamefont {{O'Brien}}, \citenamefont {{Ong}}, \citenamefont
  {{Pfrang}}, \citenamefont {{Pohl}}, \citenamefont {{Pueschel}}, \citenamefont
  {{Quinn}}, \citenamefont {{Ragan}}, \citenamefont {{Reynolds}}, \citenamefont
  {{Roache}}, \citenamefont {{Rodd}}, \citenamefont {{Ryan}}, \citenamefont
  {{Sadeh}}, \citenamefont {{Saha}}, \citenamefont {{Santander}}, \citenamefont
  {{Sembroski}}, \citenamefont {{Shang}}, \citenamefont {{Splettstoesser}},
  \citenamefont {{Tak}}, \citenamefont {{Tucci}}, \citenamefont {{Vassiliev}},\
  and\ \citenamefont {{Williams}}}]{Tak_2023}%
  \BibitemOpen
  \bibfield  {author} {\bibinfo {author} {\bibfnamefont {A.}~\bibnamefont
  {{Acharyya}}}, \bibinfo {author} {\bibfnamefont {A.}~\bibnamefont
  {{Archer}}}, \bibinfo {author} {\bibfnamefont {P.}~\bibnamefont {{Bangale}}},
  \bibinfo {author} {\bibfnamefont {J.~T.}\ \bibnamefont {{Bartkoske}}},
  \bibinfo {author} {\bibfnamefont {P.}~\bibnamefont {{Batista}}}, \bibinfo
  {author} {\bibfnamefont {M.}~\bibnamefont {{Baumgart}}}, \bibinfo {author}
  {\bibfnamefont {W.}~\bibnamefont {{Benbow}}}, \bibinfo {author}
  {\bibfnamefont {J.~H.}\ \bibnamefont {{Buckley}}}, \bibinfo {author}
  {\bibfnamefont {A.}~\bibnamefont {{Falcone}}}, \bibinfo {author}
  {\bibfnamefont {Q.}~\bibnamefont {{Feng}}}, \bibinfo {author} {\bibfnamefont
  {J.~P.}\ \bibnamefont {{Finley}}}, \bibinfo {author} {\bibfnamefont {G.~M.}\
  \bibnamefont {{Foote}}}, \bibinfo {author} {\bibfnamefont {L.}~\bibnamefont
  {{Fortson}}}, \bibinfo {author} {\bibfnamefont {A.}~\bibnamefont
  {{Furniss}}}, \bibinfo {author} {\bibfnamefont {G.}~\bibnamefont
  {{Gallagher}}}, \bibinfo {author} {\bibfnamefont {W.~F.}\ \bibnamefont
  {{Hanlon}}}, \bibinfo {author} {\bibfnamefont {O.}~\bibnamefont {{Hervet}}},
  \bibinfo {author} {\bibfnamefont {J.}~\bibnamefont {{Hoang}}}, \bibinfo
  {author} {\bibfnamefont {J.}~\bibnamefont {{Holder}}}, \bibinfo {author}
  {\bibfnamefont {T.~B.}\ \bibnamefont {{Humensky}}}, \bibinfo {author}
  {\bibfnamefont {W.}~\bibnamefont {{Jin}}}, \bibinfo {author} {\bibfnamefont
  {P.}~\bibnamefont {{Kaaret}}}, \bibinfo {author} {\bibfnamefont
  {M.}~\bibnamefont {{Kertzman}}}, \bibinfo {author} {\bibfnamefont
  {M.}~\bibnamefont {{Kherlakian}}}, \bibinfo {author} {\bibfnamefont
  {D.}~\bibnamefont {{Kieda}}}, \bibinfo {author} {\bibfnamefont {T.~K.}\
  \bibnamefont {{Kleiner}}}, \bibinfo {author} {\bibfnamefont {N.}~\bibnamefont
  {{Korzoun}}}, \bibinfo {author} {\bibfnamefont {F.}~\bibnamefont
  {{Krennrich}}}, \bibinfo {author} {\bibfnamefont {M.~J.}\ \bibnamefont
  {{Lang}}}, \bibinfo {author} {\bibfnamefont {M.}~\bibnamefont {{Lundy}}},
  \bibinfo {author} {\bibfnamefont {G.}~\bibnamefont {{Maier}}}, \bibinfo
  {author} {\bibfnamefont {C.~E.}\ \bibnamefont {{McGrath}}}, \bibinfo {author}
  {\bibfnamefont {P.}~\bibnamefont {{Moriarty}}}, \bibinfo {author}
  {\bibfnamefont {S.}~\bibnamefont {{O'Brien}}}, \bibinfo {author}
  {\bibfnamefont {R.~A.}\ \bibnamefont {{Ong}}}, \bibinfo {author}
  {\bibfnamefont {K.}~\bibnamefont {{Pfrang}}}, \bibinfo {author}
  {\bibfnamefont {M.}~\bibnamefont {{Pohl}}}, \bibinfo {author} {\bibfnamefont
  {E.}~\bibnamefont {{Pueschel}}}, \bibinfo {author} {\bibfnamefont
  {J.}~\bibnamefont {{Quinn}}}, \bibinfo {author} {\bibfnamefont
  {K.}~\bibnamefont {{Ragan}}}, \bibinfo {author} {\bibfnamefont {P.~T.}\
  \bibnamefont {{Reynolds}}}, \bibinfo {author} {\bibfnamefont
  {E.}~\bibnamefont {{Roache}}}, \bibinfo {author} {\bibfnamefont {N.~L.}\
  \bibnamefont {{Rodd}}}, \bibinfo {author} {\bibfnamefont {J.~L.}\
  \bibnamefont {{Ryan}}}, \bibinfo {author} {\bibfnamefont {I.}~\bibnamefont
  {{Sadeh}}}, \bibinfo {author} {\bibfnamefont {L.}~\bibnamefont {{Saha}}},
  \bibinfo {author} {\bibfnamefont {M.}~\bibnamefont {{Santander}}}, \bibinfo
  {author} {\bibfnamefont {G.~H.}\ \bibnamefont {{Sembroski}}}, \bibinfo
  {author} {\bibfnamefont {R.}~\bibnamefont {{Shang}}}, \bibinfo {author}
  {\bibfnamefont {M.}~\bibnamefont {{Splettstoesser}}}, \bibinfo {author}
  {\bibfnamefont {D.}~\bibnamefont {{Tak}}}, \bibinfo {author} {\bibfnamefont
  {J.~V.}\ \bibnamefont {{Tucci}}}, \bibinfo {author} {\bibfnamefont {V.~V.}\
  \bibnamefont {{Vassiliev}}},\ and\ \bibinfo {author} {\bibfnamefont {D.~A.}\
  \bibnamefont {{Williams}}},\ }\href
  {https://doi.org/10.3847/1538-4357/acbc7b} {\bibfield  {journal} {\bibinfo
  {journal} {\apj}\ }\textbf {\bibinfo {volume} {945}},\ \bibinfo {eid} {101}
  (\bibinfo {year} {2023})},\ \Eprint {https://arxiv.org/abs/2302.08784}
  {arXiv:2302.08784 [astro-ph.HE]} \BibitemShut {NoStop}%
\bibitem [{\citenamefont {{Albert}}\ \emph {et~al.}(2022)\citenamefont
  {{Albert}}, \citenamefont {{Alves}}, \citenamefont {{Andr{\'e}}},
  \citenamefont {{Anghinolfi}}, \citenamefont {{Anton}}, \citenamefont
  {{Ardid}}, \citenamefont {{Ardid}}, \citenamefont {{Aubert}}, \citenamefont
  {{Aublin}}, \citenamefont {{Baret}}, \citenamefont {{Basa}}, \citenamefont
  {{Belhorma}}, \citenamefont {{Bendahman}}, \citenamefont {{Benfenati}},
  \citenamefont {{Bertin}}, \citenamefont {{Biagi}}, \citenamefont
  {{Bissinger}}, \citenamefont {{Boumaaza}}, \citenamefont {{Bouta}},
  \citenamefont {{Bouwhuis}}, \citenamefont {{Br{\^a}nza{\c{s}}}},
  \citenamefont {{Bruijn}}, \citenamefont {{Brunner}}, \citenamefont {{Busto}},
  \citenamefont {{Caiffi}}, \citenamefont {{Calvo}}, \citenamefont {{Capone}},
  \citenamefont {{Caramete}}, \citenamefont {{Carr}}, \citenamefont
  {{Carretero}}, \citenamefont {{Celli}}, \citenamefont {{Chabab}},
  \citenamefont {{Chau}}, \citenamefont {{Cherkaoui El Moursli}}, \citenamefont
  {{Chiarusi}}, \citenamefont {{Circella}}, \citenamefont {{Coleiro}},
  \citenamefont {{Coniglione}}, \citenamefont {{Coyle}}, \citenamefont
  {{Creusot}}, \citenamefont {{D{\'\i}az}}, \citenamefont {{de Wasseige}},
  \citenamefont {{Distefano}}, \citenamefont {{di Palma}}, \citenamefont
  {{Domi}}, \citenamefont {{Donzaud}}, \citenamefont {{Dornic}}, \citenamefont
  {{Drouhin}}, \citenamefont {{Eberl}}, \citenamefont {{van Eeden}},
  \citenamefont {{van Eijk}}, \citenamefont {{El Khayati}}, \citenamefont
  {{Enzenh{\"o}fer}}, \citenamefont {{Fermani}}, \citenamefont {{Ferrara}},
  \citenamefont {{Filippini}}, \citenamefont {{Fusco}}, \citenamefont
  {{Gatelet}}, \citenamefont {{Gay}}, \citenamefont {{Glotin}}, \citenamefont
  {{Gozzini}}, \citenamefont {{Gracia Ruiz}}, \citenamefont {{Graf}},
  \citenamefont {{Guidi}}, \citenamefont {{Hallmann}}, \citenamefont {{van
  Haren}}, \citenamefont {{Heijboer}}, \citenamefont {{Hello}}, \citenamefont
  {{Hern{\'a}ndez-Rey}}, \citenamefont {{H{\"o}{\ss}l}}, \citenamefont
  {{Hofest{\"a}dt}}, \citenamefont {{Huang}}, \citenamefont {{Illuminati}},
  \citenamefont {{James}}, \citenamefont {{Jisse-Jung}}, \citenamefont {{de
  Jong}}, \citenamefont {{de Jong}}, \citenamefont {{Kadler}}, \citenamefont
  {{Kalekin}}, \citenamefont {{Katz}}, \citenamefont {{Khan-Chowdhury}},
  \citenamefont {{Kouchner}}, \citenamefont {{Kreykenbohm}}, \citenamefont
  {{Kulikovskiy}}, \citenamefont {{Lagunas Gualda}}, \citenamefont {{Lahmann}},
  \citenamefont {{Le Breton}}, \citenamefont {{Lestum}}, \citenamefont
  {{Lef{\`e}vre}}, \citenamefont {{Leonora}}, \citenamefont {{Levi}},
  \citenamefont {{Lincetto}}, \citenamefont {{Lopez-Coto}}, \citenamefont
  {{Loucatos}}, \citenamefont {{Maderer}}, \citenamefont {{Manczak}},
  \citenamefont {{Marcelin}}, \citenamefont {{Margiotta}}, \citenamefont
  {{Marinelli}}, \citenamefont {{Mart{\'\i}nez-Mora}}, \citenamefont
  {{Martino}}, \citenamefont {{Melis}}, \citenamefont {{Migliozzi}},
  \citenamefont {{Moussa}}, \citenamefont {{Muller}}, \citenamefont {{Nauta}},
  \citenamefont {{Navas}}, \citenamefont {{Nezri}}, \citenamefont {{{\'O}
  Fearraigh}}, \citenamefont {{P{\u{a}}un}}, \citenamefont
  {{P{\u{a}}v{\u{a}}la{\c{s}}}}, \citenamefont {{Pellegrino}}, \citenamefont
  {{Perrin-Terrin}}, \citenamefont {{Pestel}}, \citenamefont {{Piattelli}},
  \citenamefont {{Pieterse}}, \citenamefont {{Poir{\`e}}}, \citenamefont
  {{Popa}}, \citenamefont {{Pradier}}, \citenamefont {{Randazzo}},
  \citenamefont {{Real}}, \citenamefont {{Reck}}, \citenamefont {{Riccobene}},
  \citenamefont {{Romanov}}, \citenamefont {{Sala}}, \citenamefont
  {{S{\'a}nchez-Losa}}, \citenamefont {{Salesa Greus}}, \citenamefont
  {{Samtleben}}, \citenamefont {{Sanguineti}}, \citenamefont {{Sapienza}},
  \citenamefont {{Schnabel}}, \citenamefont {{Schumann}}, \citenamefont
  {{Sch{\"u}ssler}}, \citenamefont {{Seneca}}, \citenamefont {{Spurio}},
  \citenamefont {{Stolarczyk}}, \citenamefont {{Taiuti}}, \citenamefont
  {{Tayalati}}, \citenamefont {{Tingay}}, \citenamefont {{Vallage}},
  \citenamefont {{van Elewyck}}, \citenamefont {{Versari}}, \citenamefont
  {{Viola}}, \citenamefont {{Vivolo}}, \citenamefont {{Wilms}}, \citenamefont
  {{Zavatarelli}}, \citenamefont {{Zegarelli}}, \citenamefont {{Zornoza}},
  \citenamefont {{Z{\'u}{\~n}iga}},\ and\ \citenamefont {{ANTARES
  Collaboration}}}]{ANTARES2022}%
  \BibitemOpen
  \bibfield  {author} {\bibinfo {author} {\bibfnamefont {A.}~\bibnamefont
  {{Albert}}}, \bibinfo {author} {\bibfnamefont {S.}~\bibnamefont {{Alves}}},
  \bibinfo {author} {\bibfnamefont {M.}~\bibnamefont {{Andr{\'e}}}}, \bibinfo
  {author} {\bibfnamefont {M.}~\bibnamefont {{Anghinolfi}}}, \bibinfo {author}
  {\bibfnamefont {G.}~\bibnamefont {{Anton}}}, \bibinfo {author} {\bibfnamefont
  {M.}~\bibnamefont {{Ardid}}}, \bibinfo {author} {\bibfnamefont
  {S.}~\bibnamefont {{Ardid}}}, \bibinfo {author} {\bibfnamefont {J.~J.}\
  \bibnamefont {{Aubert}}}, \bibinfo {author} {\bibfnamefont {J.}~\bibnamefont
  {{Aublin}}}, \bibinfo {author} {\bibfnamefont {B.}~\bibnamefont {{Baret}}},
  \bibinfo {author} {\bibfnamefont {S.}~\bibnamefont {{Basa}}}, \bibinfo
  {author} {\bibfnamefont {B.}~\bibnamefont {{Belhorma}}}, \bibinfo {author}
  {\bibfnamefont {M.}~\bibnamefont {{Bendahman}}}, \bibinfo {author}
  {\bibfnamefont {F.}~\bibnamefont {{Benfenati}}}, \bibinfo {author}
  {\bibfnamefont {V.}~\bibnamefont {{Bertin}}}, \bibinfo {author}
  {\bibfnamefont {S.}~\bibnamefont {{Biagi}}}, \bibinfo {author} {\bibfnamefont
  {M.}~\bibnamefont {{Bissinger}}}, \bibinfo {author} {\bibfnamefont
  {J.}~\bibnamefont {{Boumaaza}}}, \bibinfo {author} {\bibfnamefont
  {M.}~\bibnamefont {{Bouta}}}, \bibinfo {author} {\bibfnamefont {M.~C.}\
  \bibnamefont {{Bouwhuis}}}, \bibinfo {author} {\bibfnamefont
  {H.}~\bibnamefont {{Br{\^a}nza{\c{s}}}}}, \bibinfo {author} {\bibfnamefont
  {R.}~\bibnamefont {{Bruijn}}}, \bibinfo {author} {\bibfnamefont
  {J.}~\bibnamefont {{Brunner}}}, \bibinfo {author} {\bibfnamefont
  {J.}~\bibnamefont {{Busto}}}, \bibinfo {author} {\bibfnamefont
  {B.}~\bibnamefont {{Caiffi}}}, \bibinfo {author} {\bibfnamefont
  {D.}~\bibnamefont {{Calvo}}}, \bibinfo {author} {\bibfnamefont
  {A.}~\bibnamefont {{Capone}}}, \bibinfo {author} {\bibfnamefont
  {L.}~\bibnamefont {{Caramete}}}, \bibinfo {author} {\bibfnamefont
  {J.}~\bibnamefont {{Carr}}}, \bibinfo {author} {\bibfnamefont
  {V.}~\bibnamefont {{Carretero}}}, \bibinfo {author} {\bibfnamefont
  {S.}~\bibnamefont {{Celli}}}, \bibinfo {author} {\bibfnamefont
  {M.}~\bibnamefont {{Chabab}}}, \bibinfo {author} {\bibfnamefont {T.~N.}\
  \bibnamefont {{Chau}}}, \bibinfo {author} {\bibfnamefont {R.}~\bibnamefont
  {{Cherkaoui El Moursli}}}, \bibinfo {author} {\bibfnamefont {T.}~\bibnamefont
  {{Chiarusi}}}, \bibinfo {author} {\bibfnamefont {M.}~\bibnamefont
  {{Circella}}}, \bibinfo {author} {\bibfnamefont {A.}~\bibnamefont
  {{Coleiro}}}, \bibinfo {author} {\bibfnamefont {R.}~\bibnamefont
  {{Coniglione}}}, \bibinfo {author} {\bibfnamefont {P.}~\bibnamefont
  {{Coyle}}}, \bibinfo {author} {\bibfnamefont {A.}~\bibnamefont {{Creusot}}},
  \bibinfo {author} {\bibfnamefont {A.~F.}\ \bibnamefont {{D{\'\i}az}}},
  \bibinfo {author} {\bibfnamefont {G.}~\bibnamefont {{de Wasseige}}}, \bibinfo
  {author} {\bibfnamefont {C.}~\bibnamefont {{Distefano}}}, \bibinfo {author}
  {\bibfnamefont {I.}~\bibnamefont {{di Palma}}}, \bibinfo {author}
  {\bibfnamefont {A.}~\bibnamefont {{Domi}}}, \bibinfo {author} {\bibfnamefont
  {C.}~\bibnamefont {{Donzaud}}}, \bibinfo {author} {\bibfnamefont
  {D.}~\bibnamefont {{Dornic}}}, \bibinfo {author} {\bibfnamefont
  {D.}~\bibnamefont {{Drouhin}}}, \bibinfo {author} {\bibfnamefont
  {T.}~\bibnamefont {{Eberl}}}, \bibinfo {author} {\bibfnamefont
  {T.}~\bibnamefont {{van Eeden}}}, \bibinfo {author} {\bibfnamefont
  {D.}~\bibnamefont {{van Eijk}}}, \bibinfo {author} {\bibfnamefont
  {N.}~\bibnamefont {{El Khayati}}}, \bibinfo {author} {\bibfnamefont
  {A.}~\bibnamefont {{Enzenh{\"o}fer}}}, \bibinfo {author} {\bibfnamefont
  {P.}~\bibnamefont {{Fermani}}}, \bibinfo {author} {\bibfnamefont
  {G.}~\bibnamefont {{Ferrara}}}, \bibinfo {author} {\bibfnamefont
  {F.}~\bibnamefont {{Filippini}}}, \bibinfo {author} {\bibfnamefont
  {L.}~\bibnamefont {{Fusco}}}, \bibinfo {author} {\bibfnamefont
  {Y.}~\bibnamefont {{Gatelet}}}, \bibinfo {author} {\bibfnamefont
  {P.}~\bibnamefont {{Gay}}}, \bibinfo {author} {\bibfnamefont
  {H.}~\bibnamefont {{Glotin}}}, \bibinfo {author} {\bibfnamefont
  {R.}~\bibnamefont {{Gozzini}}}, \bibinfo {author} {\bibfnamefont
  {R.}~\bibnamefont {{Gracia Ruiz}}}, \bibinfo {author} {\bibfnamefont
  {K.}~\bibnamefont {{Graf}}}, \bibinfo {author} {\bibfnamefont
  {C.}~\bibnamefont {{Guidi}}}, \bibinfo {author} {\bibfnamefont
  {S.}~\bibnamefont {{Hallmann}}}, \bibinfo {author} {\bibfnamefont
  {H.}~\bibnamefont {{van Haren}}}, \bibinfo {author} {\bibfnamefont {A.~J.}\
  \bibnamefont {{Heijboer}}}, \bibinfo {author} {\bibfnamefont
  {Y.}~\bibnamefont {{Hello}}}, \bibinfo {author} {\bibfnamefont {J.~J.}\
  \bibnamefont {{Hern{\'a}ndez-Rey}}}, \bibinfo {author} {\bibfnamefont
  {J.}~\bibnamefont {{H{\"o}{\ss}l}}}, \bibinfo {author} {\bibfnamefont
  {J.}~\bibnamefont {{Hofest{\"a}dt}}}, \bibinfo {author} {\bibfnamefont
  {F.}~\bibnamefont {{Huang}}}, \bibinfo {author} {\bibfnamefont
  {G.}~\bibnamefont {{Illuminati}}}, \bibinfo {author} {\bibfnamefont {C.~W.}\
  \bibnamefont {{James}}}, \bibinfo {author} {\bibfnamefont {B.}~\bibnamefont
  {{Jisse-Jung}}}, \bibinfo {author} {\bibfnamefont {M.}~\bibnamefont {{de
  Jong}}}, \bibinfo {author} {\bibfnamefont {P.}~\bibnamefont {{de Jong}}},
  \bibinfo {author} {\bibfnamefont {M.}~\bibnamefont {{Kadler}}}, \bibinfo
  {author} {\bibfnamefont {O.}~\bibnamefont {{Kalekin}}}, \bibinfo {author}
  {\bibfnamefont {U.}~\bibnamefont {{Katz}}}, \bibinfo {author} {\bibfnamefont
  {N.~R.}\ \bibnamefont {{Khan-Chowdhury}}}, \bibinfo {author} {\bibfnamefont
  {A.}~\bibnamefont {{Kouchner}}}, \bibinfo {author} {\bibfnamefont
  {I.}~\bibnamefont {{Kreykenbohm}}}, \bibinfo {author} {\bibfnamefont
  {V.}~\bibnamefont {{Kulikovskiy}}}, \bibinfo {author} {\bibfnamefont
  {C.}~\bibnamefont {{Lagunas Gualda}}}, \bibinfo {author} {\bibfnamefont
  {R.}~\bibnamefont {{Lahmann}}}, \bibinfo {author} {\bibfnamefont
  {R.}~\bibnamefont {{Le Breton}}}, \bibinfo {author} {\bibfnamefont
  {S.}~\bibnamefont {{Lestum}}}, \bibinfo {author} {\bibfnamefont
  {D.}~\bibnamefont {{Lef{\`e}vre}}}, \bibinfo {author} {\bibfnamefont
  {E.}~\bibnamefont {{Leonora}}}, \bibinfo {author} {\bibfnamefont
  {G.}~\bibnamefont {{Levi}}}, \bibinfo {author} {\bibfnamefont
  {M.}~\bibnamefont {{Lincetto}}}, \bibinfo {author} {\bibfnamefont
  {D.}~\bibnamefont {{Lopez-Coto}}}, \bibinfo {author} {\bibfnamefont
  {S.}~\bibnamefont {{Loucatos}}}, \bibinfo {author} {\bibfnamefont
  {L.}~\bibnamefont {{Maderer}}}, \bibinfo {author} {\bibfnamefont
  {J.}~\bibnamefont {{Manczak}}}, \bibinfo {author} {\bibfnamefont
  {M.}~\bibnamefont {{Marcelin}}}, \bibinfo {author} {\bibfnamefont
  {A.}~\bibnamefont {{Margiotta}}}, \bibinfo {author} {\bibfnamefont
  {A.}~\bibnamefont {{Marinelli}}}, \bibinfo {author} {\bibfnamefont {J.~A.}\
  \bibnamefont {{Mart{\'\i}nez-Mora}}}, \bibinfo {author} {\bibfnamefont
  {B.}~\bibnamefont {{Martino}}}, \bibinfo {author} {\bibfnamefont
  {K.}~\bibnamefont {{Melis}}}, \bibinfo {author} {\bibfnamefont
  {P.}~\bibnamefont {{Migliozzi}}}, \bibinfo {author} {\bibfnamefont
  {A.}~\bibnamefont {{Moussa}}}, \bibinfo {author} {\bibfnamefont
  {R.}~\bibnamefont {{Muller}}}, \bibinfo {author} {\bibfnamefont
  {L.}~\bibnamefont {{Nauta}}}, \bibinfo {author} {\bibfnamefont
  {S.}~\bibnamefont {{Navas}}}, \bibinfo {author} {\bibfnamefont
  {E.}~\bibnamefont {{Nezri}}}, \bibinfo {author} {\bibfnamefont
  {B.}~\bibnamefont {{{\'O} Fearraigh}}}, \bibinfo {author} {\bibfnamefont
  {A.}~\bibnamefont {{P{\u{a}}un}}}, \bibinfo {author} {\bibfnamefont {G.~E.}\
  \bibnamefont {{P{\u{a}}v{\u{a}}la{\c{s}}}}}, \bibinfo {author} {\bibfnamefont
  {C.}~\bibnamefont {{Pellegrino}}}, \bibinfo {author} {\bibfnamefont
  {M.}~\bibnamefont {{Perrin-Terrin}}}, \bibinfo {author} {\bibfnamefont
  {V.}~\bibnamefont {{Pestel}}}, \bibinfo {author} {\bibfnamefont
  {P.}~\bibnamefont {{Piattelli}}}, \bibinfo {author} {\bibfnamefont
  {C.}~\bibnamefont {{Pieterse}}}, \bibinfo {author} {\bibfnamefont
  {C.}~\bibnamefont {{Poir{\`e}}}}, \bibinfo {author} {\bibfnamefont
  {V.}~\bibnamefont {{Popa}}}, \bibinfo {author} {\bibfnamefont
  {T.}~\bibnamefont {{Pradier}}}, \bibinfo {author} {\bibfnamefont
  {N.}~\bibnamefont {{Randazzo}}}, \bibinfo {author} {\bibfnamefont
  {D.}~\bibnamefont {{Real}}}, \bibinfo {author} {\bibfnamefont
  {S.}~\bibnamefont {{Reck}}}, \bibinfo {author} {\bibfnamefont
  {G.}~\bibnamefont {{Riccobene}}}, \bibinfo {author} {\bibfnamefont
  {A.}~\bibnamefont {{Romanov}}}, \bibinfo {author} {\bibfnamefont
  {F.}~\bibnamefont {{Sala}}}, \bibinfo {author} {\bibfnamefont
  {A.}~\bibnamefont {{S{\'a}nchez-Losa}}}, \bibinfo {author} {\bibfnamefont
  {F.}~\bibnamefont {{Salesa Greus}}}, \bibinfo {author} {\bibfnamefont
  {D.~F.~E.}\ \bibnamefont {{Samtleben}}}, \bibinfo {author} {\bibfnamefont
  {M.}~\bibnamefont {{Sanguineti}}}, \bibinfo {author} {\bibfnamefont
  {P.}~\bibnamefont {{Sapienza}}}, \bibinfo {author} {\bibfnamefont
  {J.}~\bibnamefont {{Schnabel}}}, \bibinfo {author} {\bibfnamefont
  {J.}~\bibnamefont {{Schumann}}}, \bibinfo {author} {\bibfnamefont
  {F.}~\bibnamefont {{Sch{\"u}ssler}}}, \bibinfo {author} {\bibfnamefont
  {J.}~\bibnamefont {{Seneca}}}, \bibinfo {author} {\bibfnamefont
  {M.}~\bibnamefont {{Spurio}}}, \bibinfo {author} {\bibfnamefont
  {T.}~\bibnamefont {{Stolarczyk}}}, \bibinfo {author} {\bibfnamefont
  {M.}~\bibnamefont {{Taiuti}}}, \bibinfo {author} {\bibfnamefont
  {Y.}~\bibnamefont {{Tayalati}}}, \bibinfo {author} {\bibfnamefont {S.~J.}\
  \bibnamefont {{Tingay}}}, \bibinfo {author} {\bibfnamefont {B.}~\bibnamefont
  {{Vallage}}}, \bibinfo {author} {\bibfnamefont {V.}~\bibnamefont {{van
  Elewyck}}}, \bibinfo {author} {\bibfnamefont {F.}~\bibnamefont {{Versari}}},
  \bibinfo {author} {\bibfnamefont {S.}~\bibnamefont {{Viola}}}, \bibinfo
  {author} {\bibfnamefont {D.}~\bibnamefont {{Vivolo}}}, \bibinfo {author}
  {\bibfnamefont {J.}~\bibnamefont {{Wilms}}}, \bibinfo {author} {\bibfnamefont
  {S.}~\bibnamefont {{Zavatarelli}}}, \bibinfo {author} {\bibfnamefont
  {A.}~\bibnamefont {{Zegarelli}}}, \bibinfo {author} {\bibfnamefont {J.~D.}\
  \bibnamefont {{Zornoza}}}, \bibinfo {author} {\bibfnamefont {J.}~\bibnamefont
  {{Z{\'u}{\~n}iga}}},\ and\ \bibinfo {author} {\bibnamefont {{ANTARES
  Collaboration}}},\ }\href {https://doi.org/10.1088/1475-7516/2022/06/028}
  {\bibfield  {journal} {\bibinfo  {journal} {\jcap}\ }\textbf {\bibinfo
  {volume} {2022}},\ \bibinfo {eid} {028} (\bibinfo {year} {2022})},\ \Eprint
  {https://arxiv.org/abs/2203.06029} {arXiv:2203.06029 [astro-ph.HE]}
  \BibitemShut {NoStop}%
\bibitem [{\citenamefont {{Cirelli}}\ \emph {et~al.}(2011)\citenamefont
  {{Cirelli}}, \citenamefont {{Corcella}}, \citenamefont {{Hektor}},
  \citenamefont {{H{\"u}tsi}}, \citenamefont {{Kadastik}}, \citenamefont
  {{Panci}}, \citenamefont {{Raidal}}, \citenamefont {{Sala}},\ and\
  \citenamefont {{Strumia}}}]{Cirelli_2011}%
  \BibitemOpen
  \bibfield  {author} {\bibinfo {author} {\bibfnamefont {M.}~\bibnamefont
  {{Cirelli}}}, \bibinfo {author} {\bibfnamefont {G.}~\bibnamefont
  {{Corcella}}}, \bibinfo {author} {\bibfnamefont {A.}~\bibnamefont
  {{Hektor}}}, \bibinfo {author} {\bibfnamefont {G.}~\bibnamefont
  {{H{\"u}tsi}}}, \bibinfo {author} {\bibfnamefont {M.}~\bibnamefont
  {{Kadastik}}}, \bibinfo {author} {\bibfnamefont {P.}~\bibnamefont {{Panci}}},
  \bibinfo {author} {\bibfnamefont {M.}~\bibnamefont {{Raidal}}}, \bibinfo
  {author} {\bibfnamefont {F.}~\bibnamefont {{Sala}}},\ and\ \bibinfo {author}
  {\bibfnamefont {A.}~\bibnamefont {{Strumia}}},\ }\href
  {https://doi.org/10.1088/1475-7516/2011/03/051} {\bibfield  {journal}
  {\bibinfo  {journal} {\jcap}\ }\textbf {\bibinfo {volume} {2011}},\ \bibinfo
  {eid} {051} (\bibinfo {year} {2011})},\ \Eprint
  {https://arxiv.org/abs/1012.4515} {arXiv:1012.4515 [hep-ph]} \BibitemShut
  {NoStop}%
\bibitem [{\citenamefont {{Bauer}}\ \emph {et~al.}(2021)\citenamefont
  {{Bauer}}, \citenamefont {{Rodd}},\ and\ \citenamefont
  {{Webber}}}]{Bauer_2021}%
  \BibitemOpen
  \bibfield  {author} {\bibinfo {author} {\bibfnamefont {C.~W.}\ \bibnamefont
  {{Bauer}}}, \bibinfo {author} {\bibfnamefont {N.~L.}\ \bibnamefont
  {{Rodd}}},\ and\ \bibinfo {author} {\bibfnamefont {B.~R.}\ \bibnamefont
  {{Webber}}},\ }\href {https://doi.org/10.1007/JHEP06(2021)121} {\bibfield
  {journal} {\bibinfo  {journal} {Journal of High Energy Physics}\ }\textbf
  {\bibinfo {volume} {2021}},\ \bibinfo {eid} {121} (\bibinfo {year} {2021})},\
  \Eprint {https://arxiv.org/abs/2007.15001} {arXiv:2007.15001 [hep-ph]}
  \BibitemShut {NoStop}%
\bibitem [{\citenamefont {{Arina}}\ \emph {et~al.}(2024)\citenamefont
  {{Arina}}, \citenamefont {{Di Mauro}}, \citenamefont {{Fornengo}},
  \citenamefont {{Heisig}}, \citenamefont {{Jueid}},\ and\ \citenamefont {{Ruiz
  de Austri}}}]{Arina2024}%
  \BibitemOpen
  \bibfield  {author} {\bibinfo {author} {\bibfnamefont {C.}~\bibnamefont
  {{Arina}}}, \bibinfo {author} {\bibfnamefont {M.}~\bibnamefont {{Di Mauro}}},
  \bibinfo {author} {\bibfnamefont {N.}~\bibnamefont {{Fornengo}}}, \bibinfo
  {author} {\bibfnamefont {J.}~\bibnamefont {{Heisig}}}, \bibinfo {author}
  {\bibfnamefont {A.}~\bibnamefont {{Jueid}}},\ and\ \bibinfo {author}
  {\bibfnamefont {R.}~\bibnamefont {{Ruiz de Austri}}},\ }\href
  {https://doi.org/10.1088/1475-7516/2024/03/035} {\bibfield  {journal}
  {\bibinfo  {journal} {\jcap}\ }\textbf {\bibinfo {volume} {2024}},\ \bibinfo
  {eid} {035} (\bibinfo {year} {2024})},\ \Eprint
  {https://arxiv.org/abs/2312.01153} {arXiv:2312.01153 [astro-ph.HE]}
  \BibitemShut {NoStop}%
\bibitem [{\citenamefont {{Circiello}}\ \emph {et~al.}(2024)\citenamefont
  {{Circiello}}, \citenamefont {{McDaniel}}, \citenamefont {{Drlica-Wagner}},
  \citenamefont {{Karwin}}, \citenamefont {{Ajello}}, \citenamefont {{Di
  Mauro}},\ and\ \citenamefont {{S{\'a}nchez-Conde}}}]{Circiello2024}%
  \BibitemOpen
  \bibfield  {author} {\bibinfo {author} {\bibfnamefont {A.}~\bibnamefont
  {{Circiello}}}, \bibinfo {author} {\bibfnamefont {A.}~\bibnamefont
  {{McDaniel}}}, \bibinfo {author} {\bibfnamefont {A.}~\bibnamefont
  {{Drlica-Wagner}}}, \bibinfo {author} {\bibfnamefont {C.}~\bibnamefont
  {{Karwin}}}, \bibinfo {author} {\bibfnamefont {M.}~\bibnamefont {{Ajello}}},
  \bibinfo {author} {\bibfnamefont {M.}~\bibnamefont {{Di Mauro}}},\ and\
  \bibinfo {author} {\bibfnamefont {M.}~\bibnamefont {{S{\'a}nchez-Conde}}},\
  }\href {https://doi.org/10.48550/arXiv.2404.01181} {\bibfield  {journal}
  {\bibinfo  {journal} {arXiv e-prints}\ ,\ \bibinfo {eid} {arXiv:2404.01181}}
  (\bibinfo {year} {2024})},\ \Eprint {https://arxiv.org/abs/2404.01181}
  {arXiv:2404.01181 [astro-ph.HE]} \BibitemShut {NoStop}%
\bibitem [{\citenamefont {{Ando}}\ \emph {et~al.}(2020)\citenamefont {{Ando}},
  \citenamefont {{Geringer-Sameth}}, \citenamefont {{Hiroshima}}, \citenamefont
  {{Hoof}}, \citenamefont {{Trotta}},\ and\ \citenamefont
  {{Walker}}}]{Ando2020}%
  \BibitemOpen
  \bibfield  {author} {\bibinfo {author} {\bibfnamefont {S.}~\bibnamefont
  {{Ando}}}, \bibinfo {author} {\bibfnamefont {A.}~\bibnamefont
  {{Geringer-Sameth}}}, \bibinfo {author} {\bibfnamefont {N.}~\bibnamefont
  {{Hiroshima}}}, \bibinfo {author} {\bibfnamefont {S.}~\bibnamefont {{Hoof}}},
  \bibinfo {author} {\bibfnamefont {R.}~\bibnamefont {{Trotta}}},\ and\
  \bibinfo {author} {\bibfnamefont {M.~G.}\ \bibnamefont {{Walker}}},\ }\href
  {https://doi.org/10.1103/PhysRevD.102.061302} {\bibfield  {journal} {\bibinfo
   {journal} {\prd}\ }\textbf {\bibinfo {volume} {102}},\ \bibinfo {eid}
  {061302} (\bibinfo {year} {2020})},\ \Eprint
  {https://arxiv.org/abs/2002.11956} {arXiv:2002.11956 [astro-ph.CO]}
  \BibitemShut {NoStop}%
\bibitem [{Note1()}]{Note1}%
  \BibitemOpen
  \bibinfo {note} {\protect \url
  {https://github.com/shinichiroando/dwarf_params}}\BibitemShut {NoStop}%
\bibitem [{\citenamefont {{Graus}}\ \emph {et~al.}(2019)\citenamefont
  {{Graus}}, \citenamefont {{Bullock}}, \citenamefont {{Kelley}}, \citenamefont
  {{Boylan-Kolchin}}, \citenamefont {{Garrison-Kimmel}},\ and\ \citenamefont
  {{Qi}}}]{Graus2019}%
  \BibitemOpen
  \bibfield  {author} {\bibinfo {author} {\bibfnamefont {A.~S.}\ \bibnamefont
  {{Graus}}}, \bibinfo {author} {\bibfnamefont {J.~S.}\ \bibnamefont
  {{Bullock}}}, \bibinfo {author} {\bibfnamefont {T.}~\bibnamefont {{Kelley}}},
  \bibinfo {author} {\bibfnamefont {M.}~\bibnamefont {{Boylan-Kolchin}}},
  \bibinfo {author} {\bibfnamefont {S.}~\bibnamefont {{Garrison-Kimmel}}},\
  and\ \bibinfo {author} {\bibfnamefont {Y.}~\bibnamefont {{Qi}}},\ }\href
  {https://doi.org/10.1093/mnras/stz1992} {\bibfield  {journal} {\bibinfo
  {journal} {\mnras}\ }\textbf {\bibinfo {volume} {488}},\ \bibinfo {pages}
  {4585} (\bibinfo {year} {2019})},\ \Eprint {https://arxiv.org/abs/1808.03654}
  {arXiv:1808.03654 [astro-ph.GA]} \BibitemShut {NoStop}%
\bibitem [{\citenamefont {{Pace}}\ and\ \citenamefont
  {{Strigari}}(2019)}]{Pace2019}%
  \BibitemOpen
  \bibfield  {author} {\bibinfo {author} {\bibfnamefont {A.~B.}\ \bibnamefont
  {{Pace}}}\ and\ \bibinfo {author} {\bibfnamefont {L.~E.}\ \bibnamefont
  {{Strigari}}},\ }\href {https://doi.org/10.1093/mnras/sty2839} {\bibfield
  {journal} {\bibinfo  {journal} {\mnras}\ }\textbf {\bibinfo {volume} {482}},\
  \bibinfo {pages} {3480} (\bibinfo {year} {2019})},\ \Eprint
  {https://arxiv.org/abs/1802.06811} {arXiv:1802.06811 [astro-ph.GA]}
  \BibitemShut {NoStop}%
\bibitem [{\citenamefont {{Strigari}}\ \emph {et~al.}(2007)\citenamefont
  {{Strigari}}, \citenamefont {{Koushiappas}}, \citenamefont {{Bullock}},\ and\
  \citenamefont {{Kaplinghat}}}]{Strigari2007}%
  \BibitemOpen
  \bibfield  {author} {\bibinfo {author} {\bibfnamefont {L.~E.}\ \bibnamefont
  {{Strigari}}}, \bibinfo {author} {\bibfnamefont {S.~M.}\ \bibnamefont
  {{Koushiappas}}}, \bibinfo {author} {\bibfnamefont {J.~S.}\ \bibnamefont
  {{Bullock}}},\ and\ \bibinfo {author} {\bibfnamefont {M.}~\bibnamefont
  {{Kaplinghat}}},\ }\href {https://doi.org/10.1103/PhysRevD.75.083526}
  {\bibfield  {journal} {\bibinfo  {journal} {\prd}\ }\textbf {\bibinfo
  {volume} {75}},\ \bibinfo {eid} {083526} (\bibinfo {year} {2007})},\ \Eprint
  {https://arxiv.org/abs/astro-ph/0611925} {arXiv:astro-ph/0611925 [astro-ph]}
  \BibitemShut {NoStop}%
\bibitem [{\citenamefont {{Bullock}}\ \emph {et~al.}(2001)\citenamefont
  {{Bullock}}, \citenamefont {{Kolatt}}, \citenamefont {{Sigad}}, \citenamefont
  {{Somerville}}, \citenamefont {{Kravtsov}}, \citenamefont {{Klypin}},
  \citenamefont {{Primack}},\ and\ \citenamefont {{Dekel}}}]{Bullock2001}%
  \BibitemOpen
  \bibfield  {author} {\bibinfo {author} {\bibfnamefont {J.~S.}\ \bibnamefont
  {{Bullock}}}, \bibinfo {author} {\bibfnamefont {T.~S.}\ \bibnamefont
  {{Kolatt}}}, \bibinfo {author} {\bibfnamefont {Y.}~\bibnamefont {{Sigad}}},
  \bibinfo {author} {\bibfnamefont {R.~S.}\ \bibnamefont {{Somerville}}},
  \bibinfo {author} {\bibfnamefont {A.~V.}\ \bibnamefont {{Kravtsov}}},
  \bibinfo {author} {\bibfnamefont {A.~A.}\ \bibnamefont {{Klypin}}}, \bibinfo
  {author} {\bibfnamefont {J.~R.}\ \bibnamefont {{Primack}}},\ and\ \bibinfo
  {author} {\bibfnamefont {A.}~\bibnamefont {{Dekel}}},\ }\href
  {https://doi.org/10.1046/j.1365-8711.2001.04068.x} {\bibfield  {journal}
  {\bibinfo  {journal} {\mnras}\ }\textbf {\bibinfo {volume} {321}},\ \bibinfo
  {pages} {559} (\bibinfo {year} {2001})},\ \Eprint
  {https://arxiv.org/abs/astro-ph/9908159} {arXiv:astro-ph/9908159 [astro-ph]}
  \BibitemShut {NoStop}%
\bibitem [{Note2()}]{Note2}%
  \BibitemOpen
  \bibinfo {note} {We note that Draco II is one of the most dark-matter rich
  dSphs, but is not included in our deep-exposure sample because, during the
  data collection phase, Ando+20 had not yet been published.}\BibitemShut
  {Stop}%
\bibitem [{\citenamefont {{Song}}\ \emph {et~al.}(2024)\citenamefont {{Song}},
  \citenamefont {{Hiroshima}},\ and\ \citenamefont {{Murase}}}]{Song2024}%
  \BibitemOpen
  \bibfield  {author} {\bibinfo {author} {\bibfnamefont {D.}~\bibnamefont
  {{Song}}}, \bibinfo {author} {\bibfnamefont {N.}~\bibnamefont
  {{Hiroshima}}},\ and\ \bibinfo {author} {\bibfnamefont {K.}~\bibnamefont
  {{Murase}}},\ }\href {https://doi.org/10.1088/1475-7516/2024/05/087}
  {\bibfield  {journal} {\bibinfo  {journal} {\jcap}\ }\textbf {\bibinfo
  {volume} {2024}},\ \bibinfo {eid} {087} (\bibinfo {year} {2024})},\ \Eprint
  {https://arxiv.org/abs/2401.15606} {arXiv:2401.15606 [astro-ph.HE]}
  \BibitemShut {NoStop}%
\bibitem [{\citenamefont {Holder}\ \emph {et~al.}(2008)\citenamefont {Holder}
  \emph {et~al.}}]{VERITASInstrument}%
  \BibitemOpen
  \bibfield  {author} {\bibinfo {author} {\bibfnamefont {J.}~\bibnamefont
  {Holder}} \emph {et~al.},\ }\href {https://doi.org/10.1063/1.3076760}
  {\bibfield  {journal} {\bibinfo  {journal} {AIP Conference Proceedings}\
  }\textbf {\bibinfo {volume} {1085}},\ \bibinfo {pages} {657} (\bibinfo {year}
  {2008})},\ \Eprint
  {https://arxiv.org/abs/https://aip.scitation.org/doi/pdf/10.1063/1.3076760}
  {https://aip.scitation.org/doi/pdf/10.1063/1.3076760} \BibitemShut {NoStop}%
\bibitem [{\citenamefont {{Park}}\ \emph {et~al.}(2015)\citenamefont {{Park}}
  \emph {et~al.}}]{Park_2015}%
  \BibitemOpen
  \bibfield  {author} {\bibinfo {author} {\bibfnamefont {N.}~\bibnamefont
  {{Park}}} \emph {et~al.},\ }in\ \href {https://doi.org/10.22323/1.236.0771}
  {\emph {\bibinfo {booktitle} {34th International Cosmic Ray Conference
  (ICRC2015)}}},\ \bibinfo {series} {International Cosmic Ray Conference},
  Vol.~\bibinfo {volume} {34}\ (\bibinfo {year} {2015})\ p.\ \bibinfo {pages}
  {771},\ \Eprint {https://arxiv.org/abs/1508.07070} {arXiv:1508.07070
  [astro-ph.IM]} \BibitemShut {NoStop}%
\bibitem [{\citenamefont {{Kieda}}\ \emph {et~al.}(2013)\citenamefont {{Kieda}}
  \emph {et~al.}}]{Kieda2013}%
  \BibitemOpen
  \bibfield  {author} {\bibinfo {author} {\bibfnamefont {D.}~\bibnamefont
  {{Kieda}}} \emph {et~al.},\ }in\ \href@noop {} {\emph {\bibinfo {booktitle}
  {33rd International Cosmic Ray Conference (ICRC2013)}}},\ \bibinfo {series}
  {International Cosmic Ray Conference}, Vol.~\bibinfo {volume} {33}\ (\bibinfo
  {year} {2013})\ p.\ \bibinfo {pages} {1124},\ \Eprint
  {https://arxiv.org/abs/1308.4849} {arXiv:1308.4849 [astro-ph.IM]}
  \BibitemShut {NoStop}%
\bibitem [{\citenamefont {{Adams}}\ \emph {et~al.}(2022)\citenamefont
  {{Adams}}, \citenamefont {{Benbow}}, \citenamefont {{Brill}}, \citenamefont
  {{Buckley}}, \citenamefont {{Christiansen}}, \citenamefont {{Falcone}},
  \citenamefont {{Feng}}, \citenamefont {{Finley}}, \citenamefont {{Foote}},
  \citenamefont {{Fortson}}, \citenamefont {{Furniss}}, \citenamefont
  {{Giuri}}, \citenamefont {{Hanna}}, \citenamefont {{Hassan}}, \citenamefont
  {{Hervet}}, \citenamefont {{Holder}}, \citenamefont {{Hona}}, \citenamefont
  {{Humensky}}, \citenamefont {{Jin}}, \citenamefont {{Kaaret}}, \citenamefont
  {{Kleiner}}, \citenamefont {{Kumar}}, \citenamefont {{Lang}}, \citenamefont
  {{Lundy}}, \citenamefont {{Maier}}, \citenamefont {{Moriarty}}, \citenamefont
  {{Mukherjee}}, \citenamefont {{Nievas Rosillo}}, \citenamefont {{O'Brien}},
  \citenamefont {{Park}}, \citenamefont {{Patel}}, \citenamefont {{Pfrang}},
  \citenamefont {{Pohl}}, \citenamefont {{Prado}}, \citenamefont {{Pueschel}},
  \citenamefont {{Quinn}}, \citenamefont {{Ragan}}, \citenamefont {{Reynolds}},
  \citenamefont {{Ribeiro}}, \citenamefont {{Roache}}, \citenamefont {{Ryan}},
  \citenamefont {{Santander}}, \citenamefont {{Weinstein}}, \citenamefont
  {{Williams}},\ and\ \citenamefont {{Williamson}}}]{NievasRosillo_2021}%
  \BibitemOpen
  \bibfield  {author} {\bibinfo {author} {\bibfnamefont {C.~B.}\ \bibnamefont
  {{Adams}}}, \bibinfo {author} {\bibfnamefont {W.}~\bibnamefont {{Benbow}}},
  \bibinfo {author} {\bibfnamefont {A.}~\bibnamefont {{Brill}}}, \bibinfo
  {author} {\bibfnamefont {J.~H.}\ \bibnamefont {{Buckley}}}, \bibinfo {author}
  {\bibfnamefont {J.~L.}\ \bibnamefont {{Christiansen}}}, \bibinfo {author}
  {\bibfnamefont {A.}~\bibnamefont {{Falcone}}}, \bibinfo {author}
  {\bibfnamefont {Q.}~\bibnamefont {{Feng}}}, \bibinfo {author} {\bibfnamefont
  {J.~P.}\ \bibnamefont {{Finley}}}, \bibinfo {author} {\bibfnamefont {G.~M.}\
  \bibnamefont {{Foote}}}, \bibinfo {author} {\bibfnamefont {L.}~\bibnamefont
  {{Fortson}}}, \bibinfo {author} {\bibfnamefont {A.}~\bibnamefont
  {{Furniss}}}, \bibinfo {author} {\bibfnamefont {C.}~\bibnamefont {{Giuri}}},
  \bibinfo {author} {\bibfnamefont {D.}~\bibnamefont {{Hanna}}}, \bibinfo
  {author} {\bibfnamefont {T.}~\bibnamefont {{Hassan}}}, \bibinfo {author}
  {\bibfnamefont {O.}~\bibnamefont {{Hervet}}}, \bibinfo {author}
  {\bibfnamefont {J.}~\bibnamefont {{Holder}}}, \bibinfo {author}
  {\bibfnamefont {B.}~\bibnamefont {{Hona}}}, \bibinfo {author} {\bibfnamefont
  {T.~B.}\ \bibnamefont {{Humensky}}}, \bibinfo {author} {\bibfnamefont
  {W.}~\bibnamefont {{Jin}}}, \bibinfo {author} {\bibfnamefont
  {P.}~\bibnamefont {{Kaaret}}}, \bibinfo {author} {\bibfnamefont {T.~K.}\
  \bibnamefont {{Kleiner}}}, \bibinfo {author} {\bibfnamefont {S.}~\bibnamefont
  {{Kumar}}}, \bibinfo {author} {\bibfnamefont {M.~J.}\ \bibnamefont {{Lang}}},
  \bibinfo {author} {\bibfnamefont {M.}~\bibnamefont {{Lundy}}}, \bibinfo
  {author} {\bibfnamefont {G.}~\bibnamefont {{Maier}}}, \bibinfo {author}
  {\bibfnamefont {P.}~\bibnamefont {{Moriarty}}}, \bibinfo {author}
  {\bibfnamefont {R.}~\bibnamefont {{Mukherjee}}}, \bibinfo {author}
  {\bibfnamefont {M.}~\bibnamefont {{Nievas Rosillo}}}, \bibinfo {author}
  {\bibfnamefont {S.}~\bibnamefont {{O'Brien}}}, \bibinfo {author}
  {\bibfnamefont {N.}~\bibnamefont {{Park}}}, \bibinfo {author} {\bibfnamefont
  {S.}~\bibnamefont {{Patel}}}, \bibinfo {author} {\bibfnamefont
  {K.}~\bibnamefont {{Pfrang}}}, \bibinfo {author} {\bibfnamefont
  {M.}~\bibnamefont {{Pohl}}}, \bibinfo {author} {\bibfnamefont {R.~R.}\
  \bibnamefont {{Prado}}}, \bibinfo {author} {\bibfnamefont {E.}~\bibnamefont
  {{Pueschel}}}, \bibinfo {author} {\bibfnamefont {J.}~\bibnamefont {{Quinn}}},
  \bibinfo {author} {\bibfnamefont {K.}~\bibnamefont {{Ragan}}}, \bibinfo
  {author} {\bibfnamefont {P.~T.}\ \bibnamefont {{Reynolds}}}, \bibinfo
  {author} {\bibfnamefont {D.}~\bibnamefont {{Ribeiro}}}, \bibinfo {author}
  {\bibfnamefont {E.}~\bibnamefont {{Roache}}}, \bibinfo {author}
  {\bibfnamefont {J.~L.}\ \bibnamefont {{Ryan}}}, \bibinfo {author}
  {\bibfnamefont {M.}~\bibnamefont {{Santander}}}, \bibinfo {author}
  {\bibfnamefont {A.}~\bibnamefont {{Weinstein}}}, \bibinfo {author}
  {\bibfnamefont {D.~A.}\ \bibnamefont {{Williams}}},\ and\ \bibinfo {author}
  {\bibfnamefont {T.~J.}\ \bibnamefont {{Williamson}}},\ }\href
  {https://doi.org/10.1051/0004-6361/202142275} {\bibfield  {journal} {\bibinfo
   {journal} {\aap}\ }\textbf {\bibinfo {volume} {658}},\ \bibinfo {eid} {A83}
  (\bibinfo {year} {2022})},\ \Eprint {https://arxiv.org/abs/2111.04676}
  {arXiv:2111.04676 [astro-ph.IM]} \BibitemShut {NoStop}%
\bibitem [{\citenamefont {{Fomin}}\ \emph {et~al.}(1994)\citenamefont
  {{Fomin}}, \citenamefont {{Stepanian}}, \citenamefont {{Lamb}}, \citenamefont
  {{Lewis}}, \citenamefont {{Punch}},\ and\ \citenamefont
  {{Weekes}}}]{Fomin1994}%
  \BibitemOpen
  \bibfield  {author} {\bibinfo {author} {\bibfnamefont {V.~P.}\ \bibnamefont
  {{Fomin}}}, \bibinfo {author} {\bibfnamefont {A.~A.}\ \bibnamefont
  {{Stepanian}}}, \bibinfo {author} {\bibfnamefont {R.~C.}\ \bibnamefont
  {{Lamb}}}, \bibinfo {author} {\bibfnamefont {D.~A.}\ \bibnamefont {{Lewis}}},
  \bibinfo {author} {\bibfnamefont {M.}~\bibnamefont {{Punch}}},\ and\ \bibinfo
  {author} {\bibfnamefont {T.~C.}\ \bibnamefont {{Weekes}}},\ }\href
  {https://doi.org/10.1016/0927-6505(94)90036-1} {\bibfield  {journal}
  {\bibinfo  {journal} {Astroparticle Physics}\ }\textbf {\bibinfo {volume}
  {2}},\ \bibinfo {pages} {137} (\bibinfo {year} {1994})}\BibitemShut {NoStop}%
\bibitem [{\citenamefont {{Cogan}}(2008)}]{VEGAS}%
  \BibitemOpen
  \bibfield  {author} {\bibinfo {author} {\bibfnamefont {P.}~\bibnamefont
  {{Cogan}}},\ }in\ \href {https://doi.org/10.48550/arXiv.0709.4233} {\emph
  {\bibinfo {booktitle} {International Cosmic Ray Conference}}},\ \bibinfo
  {series} {International Cosmic Ray Conference}, Vol.~\bibinfo {volume} {3}\
  (\bibinfo {year} {2008})\ pp.\ \bibinfo {pages} {1385--1388},\ \Eprint
  {https://arxiv.org/abs/0709.4233} {arXiv:0709.4233 [astro-ph]} \BibitemShut
  {NoStop}%
\bibitem [{\citenamefont {{Maier}}\ and\ \citenamefont
  {{Holder}}(2017)}]{eventdisplay}%
  \BibitemOpen
  \bibfield  {author} {\bibinfo {author} {\bibfnamefont {G.}~\bibnamefont
  {{Maier}}}\ and\ \bibinfo {author} {\bibfnamefont {J.}~\bibnamefont
  {{Holder}}},\ }in\ \href {https://doi.org/10.22323/1.301.0747} {\emph
  {\bibinfo {booktitle} {35th International Cosmic Ray Conference
  (ICRC2017)}}},\ \bibinfo {series} {International Cosmic Ray Conference},
  Vol.\ \bibinfo {volume} {301}\ (\bibinfo {year} {2017})\ p.\ \bibinfo {pages}
  {747},\ \Eprint {https://arxiv.org/abs/1708.04048} {arXiv:1708.04048
  [astro-ph.IM]} \BibitemShut {NoStop}%
\bibitem [{\citenamefont {Krause}\ \emph {et~al.}(2017)\citenamefont {Krause},
  \citenamefont {Pueschel},\ and\ \citenamefont {Maier}}]{BDTs}%
  \BibitemOpen
  \bibfield  {author} {\bibinfo {author} {\bibfnamefont {M.}~\bibnamefont
  {Krause}}, \bibinfo {author} {\bibfnamefont {E.}~\bibnamefont {Pueschel}},\
  and\ \bibinfo {author} {\bibfnamefont {G.}~\bibnamefont {Maier}},\ }\href
  {https://doi.org/10.1016/j.astropartphys.2017.01.004} {\bibfield  {journal}
  {\bibinfo  {journal} {Astroparticle Physics}\ }\textbf {\bibinfo {volume}
  {89}},\ \bibinfo {pages} {1} (\bibinfo {year} {2017})}\BibitemShut {NoStop}%
\bibitem [{\citenamefont {{Christiansen}}\ and\ \citenamefont {{VERITAS
  Collaboration}}(2017)}]{ITM}%
  \BibitemOpen
  \bibfield  {author} {\bibinfo {author} {\bibfnamefont {J.}~\bibnamefont
  {{Christiansen}}}\ and\ \bibinfo {author} {\bibnamefont {{VERITAS
  Collaboration}}},\ }in\ \href {https://doi.org/10.22323/1.301.0789} {\emph
  {\bibinfo {booktitle} {35th International Cosmic Ray Conference
  (ICRC2017)}}},\ \bibinfo {series} {International Cosmic Ray Conference},
  Vol.\ \bibinfo {volume} {301}\ (\bibinfo {year} {2017})\ p.\ \bibinfo {pages}
  {789},\ \Eprint {https://arxiv.org/abs/1708.05684} {arXiv:1708.05684
  [astro-ph.IM]} \BibitemShut {NoStop}%
\bibitem [{Note3()}]{Note3}%
  \BibitemOpen
  \bibinfo {note} {Specifically, a run's low-energy threshold is established as
  the higher of two values: the energy corresponding to 15\% of the maximum
  point in the effective area curve or the energy at which the deviation
  between the true energy and the reconstructed energy is less than
  20\%.}\BibitemShut {Stop}%
\bibitem [{\citenamefont {{Berge}}\ \emph {et~al.}(2007)\citenamefont
  {{Berge}}, \citenamefont {{Funk}},\ and\ \citenamefont
  {{Hinton}}}]{Berge_2007}%
  \BibitemOpen
  \bibfield  {author} {\bibinfo {author} {\bibfnamefont {D.}~\bibnamefont
  {{Berge}}}, \bibinfo {author} {\bibfnamefont {S.}~\bibnamefont {{Funk}}},\
  and\ \bibinfo {author} {\bibfnamefont {J.}~\bibnamefont {{Hinton}}},\ }\href
  {https://doi.org/10.1051/0004-6361:20066674} {\bibfield  {journal} {\bibinfo
  {journal} {\aap}\ }\textbf {\bibinfo {volume} {466}},\ \bibinfo {pages}
  {1219} (\bibinfo {year} {2007})},\ \Eprint
  {https://arxiv.org/abs/astro-ph/0610959} {arXiv:astro-ph/0610959 [astro-ph]}
  \BibitemShut {NoStop}%
\bibitem [{\citenamefont {{Aleksi{\'c}}}\ \emph {et~al.}(2012)\citenamefont
  {{Aleksi{\'c}}}, \citenamefont {{Rico}},\ and\ \citenamefont
  {{Martinez}}}]{Aleksi__2012}%
  \BibitemOpen
  \bibfield  {author} {\bibinfo {author} {\bibfnamefont {J.}~\bibnamefont
  {{Aleksi{\'c}}}}, \bibinfo {author} {\bibfnamefont {J.}~\bibnamefont
  {{Rico}}},\ and\ \bibinfo {author} {\bibfnamefont {M.}~\bibnamefont
  {{Martinez}}},\ }\href {https://doi.org/10.1088/1475-7516/2012/10/032}
  {\bibfield  {journal} {\bibinfo  {journal} {\jcap}\ }\textbf {\bibinfo
  {volume} {2012}},\ \bibinfo {eid} {032} (\bibinfo {year} {2012})},\ \Eprint
  {https://arxiv.org/abs/1209.5589} {arXiv:1209.5589 [astro-ph.HE]}
  \BibitemShut {NoStop}%
\bibitem [{\citenamefont {{Aleksi{\'c}}}\ \emph {et~al.}(2014)\citenamefont
  {{Aleksi{\'c}}} \emph {et~al.}}]{Aleksi__2014}%
  \BibitemOpen
  \bibfield  {author} {\bibinfo {author} {\bibfnamefont {J.}~\bibnamefont
  {{Aleksi{\'c}}}} \emph {et~al.},\ }\href
  {https://doi.org/10.1088/1475-7516/2014/02/008} {\bibfield  {journal}
  {\bibinfo  {journal} {\jcap}\ }\textbf {\bibinfo {volume} {2014}},\ \bibinfo
  {eid} {008} (\bibinfo {year} {2014})},\ \Eprint
  {https://arxiv.org/abs/1312.1535} {arXiv:1312.1535 [hep-ph]} \BibitemShut
  {NoStop}%
\bibitem [{\citenamefont {Li}\ and\ \citenamefont {Ma}(1983)}]{LiMa}%
  \BibitemOpen
  \bibfield  {author} {\bibinfo {author} {\bibfnamefont {T.}~\bibnamefont
  {Li}}\ and\ \bibinfo {author} {\bibfnamefont {Y.}~\bibnamefont {Ma}},\ }\href
  {https://doi.org/10.1086/161295} {\bibfield  {journal} {\bibinfo  {journal}
  {Astrophysical Journal}\ }\textbf {\bibinfo {volume} {272}},\ \bibinfo
  {pages} {317} (\bibinfo {year} {1983})}\BibitemShut {NoStop}%
\bibitem [{\citenamefont {{Bergstr{\"o}m}}\ and\ \citenamefont
  {{Kaplan}}(1994)}]{Bergstrom1994}%
  \BibitemOpen
  \bibfield  {author} {\bibinfo {author} {\bibfnamefont {L.}~\bibnamefont
  {{Bergstr{\"o}m}}}\ and\ \bibinfo {author} {\bibfnamefont {J.}~\bibnamefont
  {{Kaplan}}},\ }\href {https://doi.org/10.1016/0927-6505(94)90005-1}
  {\bibfield  {journal} {\bibinfo  {journal} {Astroparticle Physics}\ }\textbf
  {\bibinfo {volume} {2}},\ \bibinfo {pages} {261} (\bibinfo {year} {1994})},\
  \Eprint {https://arxiv.org/abs/hep-ph/9403239} {arXiv:hep-ph/9403239
  [hep-ph]} \BibitemShut {NoStop}%
\bibitem [{\citenamefont {{Lin}}\ \emph {et~al.}(2022)\citenamefont {{Lin}},
  \citenamefont {{Hammer}},\ and\ \citenamefont {{Mei{\ss}ner}}}]{Lin2022}%
  \BibitemOpen
  \bibfield  {author} {\bibinfo {author} {\bibfnamefont {Y.-H.}\ \bibnamefont
  {{Lin}}}, \bibinfo {author} {\bibfnamefont {H.-W.}\ \bibnamefont
  {{Hammer}}},\ and\ \bibinfo {author} {\bibfnamefont {U.-G.}\ \bibnamefont
  {{Mei{\ss}ner}}},\ }\href {https://doi.org/10.1103/PhysRevLett.128.052002}
  {\bibfield  {journal} {\bibinfo  {journal} {\prl}\ }\textbf {\bibinfo
  {volume} {128}},\ \bibinfo {eid} {052002} (\bibinfo {year} {2022})},\ \Eprint
  {https://arxiv.org/abs/2109.12961} {arXiv:2109.12961 [hep-ph]} \BibitemShut
  {NoStop}%
\bibitem [{\citenamefont {Martinez}\ \emph {et~al.}(2011)\citenamefont
  {Martinez}, \citenamefont {Minor}, \citenamefont {Bullock}, \citenamefont
  {Kaplinghat}, \citenamefont {Simon},\ and\ \citenamefont
  {Geha}}]{Martinez:2010xn}%
  \BibitemOpen
  \bibfield  {author} {\bibinfo {author} {\bibfnamefont {G.~D.}\ \bibnamefont
  {Martinez}}, \bibinfo {author} {\bibfnamefont {Q.~E.}\ \bibnamefont {Minor}},
  \bibinfo {author} {\bibfnamefont {J.}~\bibnamefont {Bullock}}, \bibinfo
  {author} {\bibfnamefont {M.}~\bibnamefont {Kaplinghat}}, \bibinfo {author}
  {\bibfnamefont {J.~D.}\ \bibnamefont {Simon}},\ and\ \bibinfo {author}
  {\bibfnamefont {M.}~\bibnamefont {Geha}},\ }\href
  {https://doi.org/10.1088/0004-637X/738/1/55} {\bibfield  {journal} {\bibinfo
  {journal} {Astrophys. J.}\ }\textbf {\bibinfo {volume} {738}},\ \bibinfo
  {pages} {55} (\bibinfo {year} {2011})},\ \Eprint
  {https://arxiv.org/abs/1008.4585} {arXiv:1008.4585 [astro-ph.GA]}
  \BibitemShut {NoStop}%
\bibitem [{\citenamefont {McGaugh}\ \emph {et~al.}(2021)\citenamefont
  {McGaugh}, \citenamefont {Lelli}, \citenamefont {Schombert}, \citenamefont
  {Li}, \citenamefont {Visgaitis}, \citenamefont {Parker},\ and\ \citenamefont
  {Pawlowski}}]{McGaugh:2021tyj}%
  \BibitemOpen
  \bibfield  {author} {\bibinfo {author} {\bibfnamefont {S.~S.}\ \bibnamefont
  {McGaugh}}, \bibinfo {author} {\bibfnamefont {F.}~\bibnamefont {Lelli}},
  \bibinfo {author} {\bibfnamefont {J.~M.}\ \bibnamefont {Schombert}}, \bibinfo
  {author} {\bibfnamefont {P.}~\bibnamefont {Li}}, \bibinfo {author}
  {\bibfnamefont {T.}~\bibnamefont {Visgaitis}}, \bibinfo {author}
  {\bibfnamefont {K.~S.}\ \bibnamefont {Parker}},\ and\ \bibinfo {author}
  {\bibfnamefont {M.~S.}\ \bibnamefont {Pawlowski}},\ }\href
  {https://doi.org/10.3847/1538-3881/ac2502} {\bibfield  {journal} {\bibinfo
  {journal} {Astron. J.}\ }\textbf {\bibinfo {volume} {162}},\ \bibinfo {pages}
  {202} (\bibinfo {year} {2021})},\ \Eprint {https://arxiv.org/abs/2109.03251}
  {arXiv:2109.03251 [astro-ph.GA]} \BibitemShut {NoStop}%
\bibitem [{\citenamefont {{Geringer-Sameth}}\ \emph {et~al.}(2015)\citenamefont
  {{Geringer-Sameth}}, \citenamefont {{Koushiappas}},\ and\ \citenamefont
  {{Walker}}}]{GS2015}%
  \BibitemOpen
  \bibfield  {author} {\bibinfo {author} {\bibfnamefont {A.}~\bibnamefont
  {{Geringer-Sameth}}}, \bibinfo {author} {\bibfnamefont {S.~M.}\ \bibnamefont
  {{Koushiappas}}},\ and\ \bibinfo {author} {\bibfnamefont {M.}~\bibnamefont
  {{Walker}}},\ }\href {https://doi.org/10.1088/0004-637X/801/2/74} {\bibfield
  {journal} {\bibinfo  {journal} {\apj}\ }\textbf {\bibinfo {volume} {801}},\
  \bibinfo {eid} {74} (\bibinfo {year} {2015})},\ \Eprint
  {https://arxiv.org/abs/1408.0002} {arXiv:1408.0002 [astro-ph.CO]}
  \BibitemShut {NoStop}%
\end{thebibliography}%

\appendix

\section{Deep exposure dwarfs}\label{app:deepexp}

\begin{figure*}[t!]
    \centering
    \subfigure{\includegraphics[width=0.329\textwidth]{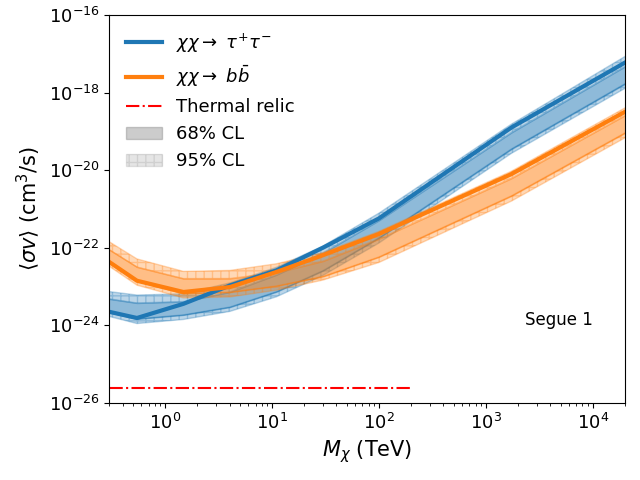}}
    \subfigure{\includegraphics[width=0.329\textwidth]{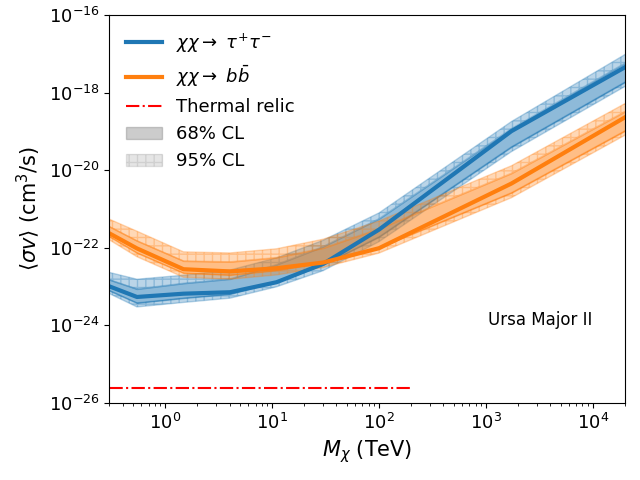}}
    \subfigure{\includegraphics[width=0.329\textwidth]{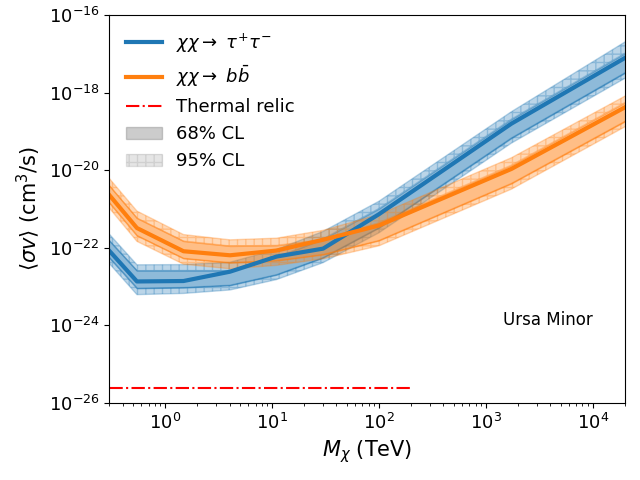}}
    \caption{VERITAS upper limit curves for three deep-exposure dSphs in two annihilation channels. The results for Segue 1, Ursa Major II, and Ursa Minor are displayed in the left, middle, and right panels, respectively. The results from $\tau^+\tau^-$ and $b\bar{b}$ channels are depicted in blue and orange, respectively. A solid curve is produced from each dSph observation, while shaded and hatched uncertainty bands depict the 68\% and 95\% containment intervals obtained from 300 realizations of random fluctuations of the background. The red dotted-dashed line is the expected velocity-weighted annihilation cross section for a thermal-relic dark matter scenario.}
    \label{fig:deep_exp}
\end{figure*}

Based on their J-factors, we have observed three targets with deeper exposures ($>$100 hours): Segue 1, Ursa Major II, and Ursa Minor. For these three dSphs, we compute the UL curves on the dark matter annihilation cross section and the null hypothesis bands for two annihilation channels, $\tau^+\tau^-$ and $b\bar{b}$. As shown in Fig.~\ref{fig:deep_exp}, Segue 1 exhibits a deviation between the observed UL curve and the null hypothesis band. This deviation arises from Poisson fluctuations in the high-energy regime (above 10 TeV). The dark matter analysis without events beyond 10 TeV eliminates this deviation. As Segue 1 heavily influences the joint-fit UL curve, a similar deviation shows up in the joint-fit result (Fig.~\ref{fig:ul_exp}). In the case of other two dSphs, the observed UL curve is consistent with the null hypothesis band. 

\section{Comparison Ando+20 profile with GS+15 profile}\label{app:comp}

The J-factors estimated in GS+15 and Ando+20 show noticeable differences. Table~\ref{tab:j_comp} summarises the J-factor at 0.5$^\circ$ for dSphs from the two studies. They differ by a factor of 0.5 to 4.5, depending on the dSph. Generally, GS+15 tends to have much higher J-factors for the dark matter-dense dSphs, compared to Ando+20. Conversely, in the case of dSphs with lower dark matter density, GS+15 reports smaller J-factors than those reported by Ando+20. This disparity is particularly decisive in our study because the J-factors of dSphs with deeper exposure (Segue 1, Ursa Major II, and Ursa Minor) from Ando+20 are approximately 2.8 and 4.3 times lower than values from GS+15. 

As we lack access to the J-factor parameters from GS+15 for our complete set of 17 dSphs, an accurate computation of the extent to which this discrepancy impacts the upper limit curve for the dark matter cross section cannot be achieved. Nonetheless, assuming the two J-factors vary across all angles by a constant factor for each dSph, a rough estimation of the dark matter upper limit curve can be obtained; i.e., for each dSph, the number of expected events from the dark matter signal (as described in Equation~\ref{equation:expectedEvnts}) is multiplied by a constant factor based on Table~\ref{tab:j_comp}. Note that for Draco II and Triangulum II, we assume the ratio is 1 because GS+15 did not present the J-factors for those dSphs. As a result, we find that the upper limit is lowered by a factor of 3, consistent with the J-factor difference between GS+15 and Ando+20 for the dSphs for which VERITAS has collected deep exposures.

\begin{table}[t!]
	\centering 
	\begin{tabular}{ c | c | c | c }
    \hline\hline
     Dwarf& \multicolumn{2}{c|}{log$_{10}J(0.5^o)$ [GeV$^2$/cm$^5$]} & ratio \\ \cline{2-3}
    & Ando+20 & GS+15 & GS+15 : Ando+20  \\ \hline
    B\"ootes& $17.77_{-0.24}^{+0.23}$ & $18.24^{+0.40}_{-0.37}$ & 2.95\\
    Coma Berenices& $18.37_{-0.33}^{+0.30}$ & $19.02^{+0.37}_{-0.41}$ & 4.47\\
    CVn I& $17.38_{-0.11}^{+0.11}$ & $17.44^{+0.37}_{-0.28}$ & 1.15\\
    CVn II& $17.19_{-0.47}^{+0.37}$ & $17.65^{+0.45}_{-0.43}$ & 2.88\\
    Hercules I& $16.93_{-0.39}^{+0.34}$ & $16.86^{+0.74}_{-0.68}$ & 0.85\\
    Leo I& $17.70_{-0.08}^{+0.07}$ & $17.84^{+0.20}_{-0.16}$ & 1.38\\
    Leo II& $17.54_{-0.10}^{+0.10}$ & $17.97^{+0.20}_{-0.18}$ & 2.69\\
    Leo IV& $16.56_{-0.66}^{+0.57}$ & $16.32^{+1.06}_{-1.69}$ & 0.58\\
    Leo V& $16.58_{-0.69}^{+0.60}$ & $16.37^{+0.94}_{-0.87}$ & 0.62\\
    Segue 1& $18.91_{-0.48}^{+0.39}$ & $19.36^{+0.32}_{-0.35}$ & 2.82\\
    Segue 2& $17.23_{-0.99}^{+0.58}$ & $16.21^{+1.06}_{-0.98}$ & 0.10\\
    Sextans I& $18.05_{-0.29}^{+0.25}$ & $17.92^{+0.35}_{-0.29}$ & 0.74\\
    Ursa Major I& $18.19_{-0.25}^{+0.22}$ & $17.87^{+0.56}_{-0.33}$ & 0.48\\
    Ursa Major II& $18.79_{-0.48}^{+0.36}$ & $19.42^{+0.44}_{-0.42}$ & 4.27\\
    Ursa Minor& $18.47_{-0.22}^{+0.20}$ & $18.95^{+0.26}_{-0.18}$ & 3.02\\
    \hline\hline
    \end{tabular}
    \caption{The J-factor comparison between \cite{Ando2020} and \cite{GS2015}. Note that the J-factors for Draco II and Triangulum II are not included here due to their absence in \cite{GS2015}.}
    \label{tab:j_comp}
\end{table}

\end{document}